\documentclass[11pt,a4paper]{article}
\usepackage{amssymb,amsmath}
\usepackage{graphicx}

\numberwithin{equation}{section}

\def\be{\begin{equation}}
\def\ee{\end{equation}}

\newcommand{\bz}{{\bar{z}}}
\newcommand{\bw}{{\bar{w}}}

\newcommand{\bxi}{{\bar{\xi}}}
\newcommand{\bh}{{\bar{h}}}
\newcommand{\bJ}{{\bar{J}}}

\DeclareMathOperator{\str}{\text{str}}
\DeclareMathOperator{\Ad}{{\text{Ad}}}
\DeclareMathOperator{\ad}{{\text{ad}}}
\DeclareMathOperator{\Res}{{\text{Res}}}

\newcommand{\Complex}{\mathbb{C}}
\newcommand{\bartial}{{\bar{\partial}}}
\newcommand{\g}{\mathfrak{g}}

\newcommand{\cO}{\mathcal{O}}

\newcommand{\cS}{\mathcal{S}}

\title{Non-chiral current algebras for deformed supergroup WZW models}
\author{Anatoly Konechny$^{1,2}$ and Thomas Quella$^{3,4}$\\[5mm]
$^1$ Department of Mathematics, Heriot-Watt University,\\
EH14 4AS Edinburgh, United Kingdom\\[3mm]
$^2$ Maxwell Institute for Mathematical Sciences, Edinburgh, United Kingdom\\[3mm]
$^3$ Institut f\"ur Theoretische Physik, Universit\"at zu K\"oln,\\
Z\"ulpicher Stra\ss{}e 77, 50937 Cologne, Germany\\[3mm]
$^4$ Korteweg de Vries Instituut voor Wiskunde, Universiteit van Amsterdam,\\
 PO Box 94248, 1090 GE Amsterdam, The Netherlands\\[5mm]
{\small E-mail: }{\small\tt
  A.Konechny@hw.ac.uk, Thomas.Quella@uni-koeln.de}
}

\date{}

\begin{document}

\begin{titlepage}
  \maketitle
  \vspace*{-9.3cm} {\tt {EMPG-10-23}} 
  \hfill {\tt
    arXiv:\,{1011.4813}}
  \vspace{9cm}
\begin{abstract}
  We study deformed WZW models on supergroups with vanishing Killing
  form. The deformation is generated by the isotropic current-current
  perturbation which is exactly marginal under these assumptions. It
  breaks half of the global isometries of the original supergroup. The
  current corresponding to the remaining symmetry is conserved but its
  components are neither holomorphic nor anti-holomorphic. We obtain
  the exact two- and three-point functions of this current and a
  four-point function in the first two leading orders of a $1/k$
  expansion but to all orders in the deformation parameter. We further
  study the operator product algebra of the currents, the equal time
  commutators and the quantum equations of motion. The form of the
  equations of motion suggests the existence of non-local charges
  which generate a Yangian. Possible applications to string theory on
  Anti-de\,Sitter spaces and to condensed matter problems are briefly
  discussed.
\end{abstract}

\end{titlepage}

  \tableofcontents

\newpage
%************************************************************************
%************************************************************************
%************************************************************************
\section{Introduction}

  Conformally invariant  $\sigma$-models with superspace target play a
  prominent role in various branches of
  mathematical physics. In the context of condensed matter theory they
  arise as means for an effective description of disordered systems
  \cite{Efetov1983:MR708812}. They are also an essential ingredient in
  the covariant quantisation of superstrings, especially in
  backgrounds involving Ramond-Ramond fields
  \cite{Berkovits:2002zk}. In this context one of the main
  applications concerns the AdS/CFT correspondence where
  $\sigma$-models on supersymmetric versions of $AdS_d\times S^d$ and
  related spaces are used to describe the string theory side of the
  duality \cite{Metsaev:1998it,Berkovits:1999im,Berkovits:1999zq,
    Arutyunov:2008if, Stefanski:2008ik, Babichenko:2009dk,Zarembo:2010sg}.

  The string background $AdS_3\times S^3$ with pure Neveu-Schwarz flux
  can be formulated in terms of a WZW model on the supergroup
  $PSU(1,1|2)$ \cite{Berkovits:1999im,Gotz:2006qp}. In such a
  formulation one can describe deformations corresponding to switching
  on a mixture of Neveu-Schwarz and Ramond-Ramond fluxes which
  preserves the full isometry of $PSU(1,1|2)$. Using $G$ as an
  abbreviation for the supergroup $PSU(1,1|2)$, the isometry is
  $G\times G$. In the Lagrangian description the term describing such
  a deformation is the kinetic term of the WZW theory: $\str(Jg\bJ g^{-1})$
  \cite{Berkovits:1999im,Gotz:2006qp}. This corresponds to an operator
  in the WZW theory which can be written as
  $:\!J^a\phi_{ab}\bJ^b\!:(z,\bz)$. Here $J^{a}(z)$ and $\bar
  J^{b}(\bz)$ are the left and right WZW currents and
  $\phi_{ab}(z,\bz)$ is the full primary field corresponding to the
  adjoint representation of the supergroup $G$. The presence of the
  field $\phi_{ab}$ ensures that the perturbation is invariant under
  the full isometry $G\times G$. It compensates the non-trivial
  transformation behaviour of $J$ under left multiplication and of
  $\bar{J}$ under right multiplication by elements from $G$. The
  perturbing field is exactly marginal due to the remarkable fact that 
  the supergroup $PSU(1,1|2)$ has a vanishing Killing form. The
  vanishing of the Killing form in particular implies that the Casimir element
  is trivial in the adjoint representation and hence that the field
  $\phi_{ab}$ has zero conformal dimension.
   
  The deformation above is described geometrically as a principal
  chiral model on the supergroup $G$ with a Wess-Zumino term. It was first
  argued in \cite{Bershadsky:1999hk} that such principal chiral models
  (in the absence of a Wess-Zumino term) are conformal when the supergroup has
  a vanishing Killing form. The focus of that paper was on  the
  particular series of supergroups: $PSL(n|n)$ which includes the
  $PSU(1,1|2)$ case.  The considerations of \cite{Bershadsky:1999hk}
  were further extended to  general supergroups with vanishing Killing
  form and to their cosets in \cite{Kagan:2005wt, Babichenko:2006uc}
  (see also \cite{Metsaev:1998it,Berkovits:1999im,Berkovits:1999zq,
    Arutyunov:2008if, Stefanski:2008ik,Babichenko:2009dk,Zarembo:2010sg} for more
  specific models).

  Another interesting deformation of the $PSU(1,1|2)$ WZW theory is
  realised by the perturbing operator $\str(J\bJ)$. Such deformations
  are exactly marginal again due to the vanishing of the Killing
  form. However, they preserve only the diagonal part of the global
  symmetry group. Their global symmetry is thus isomorphic to one copy
  of $G$. As in the previous kind of deformation we expect a string
  theory interpretation in terms of a non-trivial background with
  mixtures of RR and NS flux. Although it is straightforward to
  compute the metric and $B$-field in the deformed sigma model, the
  extraction of physical background fields requires more work and the
  corresponding calculations have not been carried out so far.

  Both of the deformations above easily generalise to WZW theories on
  arbitrary supergroups with vanishing Killing form. These theories
  are typically logarithmic, admitting the existence of fields with
  zero conformal dimension besides the identity. For the
  supergroups with trivial Killing form one of these operators is the
  adjoint representation primary field $\phi_{ab}(z,\bz)$. Supergroup
  WZW theories are well understood, at least at the conceptual level
  \cite{Rozansky:1992rx,Maassarani:1996jn,
    Schomerus:2005bf,Gotz:2006qp,Saleur:2006tf,Quella:2007hr}. Even
  though the logarithmic structure leads to many complications, the
  models can be solved due to the existence of two copies of an affine
  Kac-Moody superalgebra symmetry. The latter are realised in terms of
  two current algebras, one being holomorphic, the other
  anti-holomorphic.

  For supergroups with vanishing Killing form, deformed by one of the
  two types of deformations described above, the global symmetry also
  implies the existence of conserved currents, however now with a much
  more complicated operator product expansion. In particular, the
  conserved local currents no longer split into a holomorphic and an
  antiholomorphic component which are separately conserved. These
  statements are well-known for conserved currents belonging to global
  symmetries in massive theories. The general structure of the
  operator product algebra generated by conserved currents in massive
  2D theories was first considered in \cite{Luscher:1977uq}. It was
  demonstrated in \cite{Luscher:1977uq} that for the massive $O(n)$
  $\sigma$-models the OPE algebra of conserved currents allows one to
  construct an infinite tower of non-local conserved charges. These
  results were later generalised and reinterpreted in terms of Yangian
  symmetries in \cite{Bernard:1990jw}. In \cite{Luscher:1977uq} the
  OPE algebras generated by currents were called ``massive current
  algebras''. As we will work with similar algebras in the context of
  conformal theories we prefer to call them ``non-chiral current
  algebras''.

  It is well-known that the $G\times G$-preserving deformations
  discussed above are integrable, at least on a classical level, see
  \cite{Schwarz:1995td, Lu:2008kb} and references therein. 
  Our deformations are thus also very likely to be both conformal
  and integrable. It should be noted that the integrability is associated
  with the global symmetry. More precisely, it can be understood as
  being a consequence of current conservation and the existence of a
  Maurer-Cartan equation. While a priori there is no geometric reason
  to expect that the second, $G$-preserving, deformation considered in
  this note leads to an integrable theory, we will show that the
  algebraic structure of the theory allows one to use the construction
  of \cite{Luscher:1977uq, Bernard:1990jw} to obtain natural
  candidates for Yangian charges directly at the quantum level. More
  technically, we argue that the quantum equation of motion for the
  $G$-preserving deformation can be rephrased as a Maurer-Cartan
  equation for the conserved current associated with the global
  $G$-symmetry.

  To summarise our considerations so far, we see  that there are at
  least two good reasons to study the non-chiral current algebras
  generated by the above two deformations. Firstly one can use the
  methods of  \cite{Luscher:1977uq,Bernard:1990jw} to
  construct an infinite  tower of non-local conserved charges and to
  prove the integrability of such models. And secondly one may hope
  that such algebras will be useful for organising the spectrum of
  such conformal models. Besides potential applications such algebras
  are also interesting in their own right. In particular it would be
  interesting to understand in detail how the conformal symmetry is
  interrelated with integrability.

  The non-chiral current algebra for the $G\times G$-preserving
  deformation was recently studied in \cite{Ashok:2009xx,Ashok:2009jw,
    Benichou:2010rk}. In those papers the operator product algebra of
  currents was investigated using perturbation theory both around the
  WZW point and in the classical (large level) limit. As a starting
  point, the authors postulated a quantum version of the Maurer-Cartan
  equation. Furthermore an interesting bootstrap approach using the
  Maurer-Cartan equation was put forward to obtain the OPE algebra of
  the non-chiral currents and primary fields. While very inspiring,
  that approach however hinges on the crucial assumption that the OPE
  algebra of the deformed currents closes on itself. This assumption
  is quite strong in view of the fact that the dimension zero field
  $\phi_{ab}(z,\bz)$ is involved in the deformation which could appear
  in various combinations in the OPE of the currents. We provide a
  quantitative discussion of this question by deriving two relations
  involving the operator $\phi_{ab}(z,\bz)$ that are necessary (but
  maybe not sufficient) for the closure of the current-current
  OPEs. Unfortunately further analysis of this issue is stalled due
  to the lack of knowledge of the OPE of the operator
  $\phi_{ab}(z,\bz)$ with itself. As this question lies outside the
  main scope of this paper the details of that computation are
  relegated to appendix \ref{ap:GG}.

  In the present paper we study  the $G$-preserving current-current
  deformation. In this case the technical complications related to the
  field $\phi_{ab}(z,\bz)$ are absent and the situation is under very
  good control. The OPE closure of the deformed currents is easily
  established. Furthermore, despite the fact that for these
  deformations there is no Maurer-Cartan equation of geometric origin,
  the quantum equation of motion does take up essentially the same
  form (see section \ref{sc:EOM}). We believe that this property
  allows one to prove the integrability of the model similarly to how
  it was done in \cite{Bernard:1990jw} for the massive models.
    
  Besides having potential applications in the context of the AdS/CFT
  correspondence the $G$-preserving deformation is also relevant for
  the study of Gross-Neveu models with supergroup symmetries. In the
  case of $OSP(2S+2|2S)$ the latter have been argued to be dual to
  $S^{2S+1|2S}$ supersphere $\sigma$-models
  \cite{Candu:2008yw,Mitev:2008yt}. This correspondence is very
  interesting from a conceptual point of view, since superspheres
  belong to the class of (one-sided) supercosets for which a genuine
  CFT description is currently beyond reach. One would hence hope that
  the same type of correspondence can be established for other
  supercoset spaces, in particular for those appearing in the
  AdS/CFT correspondence.
  
  The main technical tool to be employed in the present paper is
  abelian conformal perturbation theory, used in conjunction with
  certain representation-theoretic assumptions valid for the
  supergroups of interest. The basic ideas of this method first
  appeared in \cite{Bershadsky:1999hk} where it was shown that
  two-point functions of closed string vertex operators and
  three-point functions of currents can be determined exactly by
  extrapolation from the semi-classical (flat) limit. The vanishing of
  the Killing form and some classification of the low rank invariant
  tensors on the supergroups $PSU(N|N)$ were used to argue for the
  absence of corrections involving the structure constants for certain
  correlators. One can then use the abelian conformal perturbation
  theory (metric and $B$-field perturbations in toroidal theories) to
  obtain those correlators.\footnote{The abelian perturbation series
    did not explicitly appear in \cite{Bershadsky:1999hk}. Presumably,
    in the context of that paper such series appeared only as overall
    factors in the correlators and were absorbed in the normalisation
    conventions.} Using essentially the same method the exact open
  string spectra for certain D-branes in deformed WZW theories on
  supergroups were obtained in \cite{Quella:2007sg,Mitev:2008yt} (see
  also \cite{Obuse:2008nc} for a similar calculation in a condensed
  matter context).

  In the present paper we apply the same method to bulk correlation
  functions of currents in the current-current deformed model. We are
  able to determine all the two- and three-point functions of currents
  exactly to all orders in the deformation parameter and we calculate
  their OPE at leading order in perturbation theory. Furthermore we
  compute the first non-trivial terms in the inverse level expansion
  for a four-point function of the currents, again to all orders in
  the coupling constant. Both the OPE algebra and the four-point
  function of currents exhibit logarithms, just as expected for this
  type of theories. Using the exact two- and three-point functions of
  the currents we compute the quantum equation of motion and the equal
  time commutators to all orders in the coupling constant. The
  equation of motion takes up the form of the Maurer-Cartan equation,
  giving strong indications for the integrability of the model
  following the arguments of \cite{Bernard:1990jw}. The main body of
  the paper is organised as follows. In section \ref{sc:Defo} we
  introduce supergroup WZW models and discuss the available conformal
  deformations in some detail. In section \ref{sc:TwoThree} we first
  calculate the exact two- and three-point functions of the deformed
  theory. Our calculation is valid to all orders in the deformation
  parameter and  shows a non-trivial coupling between holomorphic and
  anti-holomorphic currents away from the WZW point. In section
  \ref{sc:OPEs} we elaborate on the precise form of the full OPE
  between the currents to lowest order in perturbation theory. We find
  again that the holomorphicity is spoiled and that there are
  logarithmic contributions in the OPE between currents of opposite
  chirality. Section \ref{sc:FourPT} contains a calculation of the
  leading terms of the current four-point function in an expansion in
  $1/k$ but to all orders in the deformation parameter. In that result
  the logarithmic nature of the CFT is clearly visible. In section
  \ref{sc:CR} we use the exact knowledge of the singular terms in the
  OPE of currents to compute their equal time commutators. Remarkably
  the commutator algebra is isomorphic to the direct sum of two copies
  of the current algebra. This means that the phase space of the model
  is isomorphic to two copies of an affine Kac-Moody superalgebra. The
  equation of motion, however, exhibits significant differences from
  that of WZW models. It is derived in section \ref{sc:EOM} to all
  orders in the coupling constant.

  The appendices \ref{ap:Conventions}, \ref{ap:Distributions},
  \ref{ap:1storder}, and \ref{ap:4pt} contain our conventions for Lie
  superalgebras, a thorough review of the abelian conformal
  perturbation theory and some details of calculations which are
  particularly cumbersome. Some technical details pertaining to our
  discussion of the $G\times G$-preserving deformation have been put
  in appendix \ref{ap:GG}.

%************************************************************************
%************************************************************************
%************************************************************************
\section{\label{sc:Defo}Supergroup WZW models and their deformations}

  Supergroup WZW models exhibit a number of peculiar features that
  their bosonic cousins are lacking. One of them is the occurrence of
  logarithmic correlation functions which is intimately connected to
  the supergeometry and the non-factorisation of the state space into
  left and right movers
  \cite{Schomerus:2005bf,Gotz:2006qp,Saleur:2006tf,Quella:2007hr}.

  While the presence of the logarithms is a common
  feature,\footnote{Possible exceptions are WZW models at low levels
    which admit a free field description (see
    e.g.~\cite{Saleur:2006tf,Mitev:2008yt}). In that case realisations
    of logarithmic and non-logarithmic theories both exist.} special
  phenomena arise if the Killing form of the underlying supergroup
  is vanishing \cite{Bershadsky:1999hk}. In that case the supergroup
  WZW model admits marginal perturbations of current-current type
  which would otherwise  break conformal invariance. In this section
  we review the construction and the symmetries of WZW models and
  discuss different types of marginal deformations and their
  implications.

%************************************************************************
%************************************************************************
\subsection{Supergroup WZW models}

  Let us fix a supergroup $G$ and a non-degenerate invariant form
  $\langle\cdot,\cdot\rangle$. We assume the supergroup to be simple
  and simply-connected and the invariant form to be normalised in the
  standard way (see below). The supergroup WZW model is a
  two-dimensional $\sigma$-model describing the propagation of strings
  on $G$. The action functional is given by
\begin{align}
  \label{eq:WZW}
  \cS^{\text{WZW}}[g]
  = -\frac{ik}{4\pi}\int_\Sigma\,
       \bigl\langle g^{-1}\partial g,g^{-1}\bartial g\bigr\rangle\,dz\wedge d\bz
       -\frac{ik}{24\pi}\,\int_B
       \bigl\langle g^{-1}dg,[g^{-1}dg,g^{-1}dg]\bigr\rangle\, ,
\end{align}
  where $\Sigma$ is a closed Riemann surface and $B$ is a
  three-dimensional extension of this surface such that $\partial
  B=\Sigma$. The form $\langle\cdot,\cdot\rangle$ is supposed to be
  normalised such that the topological Wess-Zumino term is
  well-defined up to multiples of $2\pi i$ as long as $k$ is an
  integer. The level $k$ is thus the only parameter of the
  model.

  By construction, every WZW model has a global symmetry $G\times G$
  corresponding to multiplying the field $g(z,\bz)$ by arbitrary group
  elements from the left and from the right. In fact, this symmetry is
  elevated to an affine Kac-Moody algebra symmetry
\begin{align}
  \label{eq:WZWOPE}
  J^a(z)\,J^b(w)
  = \frac{k\,\kappa^{ab}}{(z-w)^2}+\frac{i{f^{ab}}_c\,J^c(w)}{z-w}
  \ ,\qquad\qquad\qquad
  \bJ^a(\bz)\,\bJ^b(\bw)
  =
  \frac{k\,\kappa^{ab}}{(\bz-\bw)^2}+\frac{i{f^{ab}}_c\,\bJ^c(\bw)}{\bz-\bw}\, ,
\end{align}
  if one allows these group elements to depend holomorphically and
  antiholomorphically on $z$, respectively. In the last formula, the
  currents are defined by
\begin{align}
  \label{eq:WZWCurrents}
  J= -k\partial gg^{-1}
  \ ,\qquad
  \bJ= kg^{-1}\bartial g\, ,
\end{align}
  and the equations of motion guarantee that they are holomorphic and
  antiholomorphic, respectively. The tensor $\kappa^{ab}$ refers to
  the non-degenerate invariant form, see appendix \ref{ap:Conventions}
  for the details of our Lie superalgebra conventions.

  Supergroup WZW models are a very exciting subject by themselves
  but for the purpose of this paper it is not necessary to introduce
  further details. The interested reader is referred to
  \cite{Quella:2007hr} where a more comprehensive discussion can be
  found.

%************************************************************************
%************************************************************************
\subsection{Marginal deformations}

  WZW models allow for a  number of local deformations which
  preserve conformal invariance. On an abstract level, such
  deformations have to be of the form
\begin{align}
  \label{eq:Action}
  \cS_\lambda[g]
  = \cS^{\text{WZW}}[g]+\cS_{\text{def}}[g]
  \qquad\text{ with }\qquad
  \cS_{\text{def}}
  = \lambda\int\!d^2z\,\cO_{\text{def}}(z,\bz)\, ,
\end{align}
  where the perturbing field $\cO_{\text{def}}(z,\bz)$ has conformal weights
  $(h,\bh)=(1,1)$ in order to render the coupling $\lambda$
  dimensionless. Consequently, a canonical candidate for the
  perturbing field is an arbitrary bilinear in the currents $J^a$ and
  $\bJ^a$, e.g.
\begin{align}
  \label{eq:GenDef}
  \cO_{\text{def}}(z,\bz)
  = m_{ab}\,:\!J^a\bJ^b\!:(z,\bz)\, .
\end{align}
  A simple calculation, however, implies that perturbations of this
  form are generically marginally relevant, i.e.\ that conformal
  invariance is spoiled at higher orders in perturbation
  theory. Roughly speaking, marginality requires that the currents of
  one specific chirality which enter eq.~\eqref{eq:GenDef} mutually
  commute (up to central terms). The only exception are supergroups
  with vanishing Killing form for which also deformations induced by
  the perturbing field
\begin{align}
  \label{eq:DefField}
  \cO_{\text{def}}(z,\bz)
  =\ :\!\bigl\langle J(z),\bJ(\bz)\bigr\rangle\!:\ 
  = \kappa_{ba}:\!J^a\bJ^b\!:(z,\bz) 
\end{align}
  are marginal to all orders in perturbation theory. Here
  $\kappa_{ba}$ denotes the inverse of the invariant form
  $\kappa^{ab}$. The argument for the marginality will be reviewed
  below. The deformation just discussed has obvious generalisations
  such as
\begin{align}
  \label{eq:AutoDefField}
  \cO_{\text{def}}(z,\bz)
  =\ :\!\bigl\langle J(z),\Omega\bigl(\bJ(\bz)\bigr)\bigr\rangle\!:\, ,
\end{align}
  where $\Omega$ is an arbitrary (but constant) automorphism of
  $G$. The insertion of the automorphism merely corresponds to a
  reinterpretation of the current $\bJ$ and will hence not be
  considered in this note.\footnote{In other words, the usual right
    action of $G$ on itself is replaced by an $\Omega$-twisted right
    action of $G$ on itself.}

  Let us now analyse what kind of symmetries are preserved by
  current-current perturbations. Under the isometry $g\mapsto
  lgr^{-1}$ the currents transform as
\begin{align}
  J\mapsto lJl^{-1}
  \ ,\qquad\qquad
  \bJ\mapsto r\bJ r^{-1}\, .
\end{align}
  This in particular implies that none of the deforming fields
  \eqref{eq:GenDef} is invariant under the full $G\times G$
  symmetry. For a general matrix $m_{ab}$ hardly any of the global
  symmetries will remain, even if the deformation preserves conformal
  invariance. The special choice \eqref{eq:DefField} for instance
  preserves one copy of $G$ and that is the maximal possible global
  symmetry for an ansatz of the form \eqref{eq:GenDef}.

  The full symmetry can only be preserved if one of the currents is
  conjugated by group elements as in $g\bJ g^{-1}$.\footnote{Or a
    twisted version of this if a twisted $G\times G$ symmetry is to be
  preserved.} In operator language this
  requires the use of a non-chiral primary field $\phi_{ab}(z,\bz)$ that
  transforms in the representation $\ad\otimes\ad$ with respect to the
  global $G\times G$ symmetry \cite{Gotz:2006qp}. It is obvious from
  the construction that the resulting perturbing field
\begin{align}
  \label{eq:DefFullSym}
  \cO_{\text{def}}(z,\bz)
  &=\ :\!\bigl\langle J,g\bJ g^{-1}\bigr\rangle\!:\ 
   =\ :\!J^a\phi_{ab}\bJ^b\!:(z,\bz)
\end{align}
  can only be marginal if $\phi_{ab}(z,\bz)$ is a primary field with
  conformal dimensions $h=\bh=0$. Since the conformal dimension of
  $\phi_{ab}(z,\bz)$ is proportional to the quadratic Casimir element
  $C_{\ad}$ evaluated in the adjoint representation, this is possible
  precisely when the underlying supergroup has a vanishing Killing
  form.

  It should be noted that a composite, normal ordered operator as the
  one in \eqref{eq:DefFullSym} has complicated properties and leads to
  a variety of subtleties in perturbation theory, even more so since
  it belongs to a non-unitary representation and has the same
  conformal dimension as the identity operator.\footnote{One should
    also bear in mind that for non-compact models the representation
    $\ad\otimes\ad$ is not part of the spectrum. A priori this makes
    it difficult to make sense out of correlation functions involving
    the field $\phi_{ab}(z,\bz)$.} For this reason, we restrict
  ourselves to pure current-current perturbations which preserve the
  diagonal $G$ symmetry. As explained above, this is the maximal
  symmetry which can be preserved under such circumstances.
  
  Throughout the rest of the paper we will only treat the deformation
  by the perturbing field \eqref{eq:DefField}.\footnote{Note that
    $\kappa_{ba}:\!J^a\bJ^b\!:$ and $\kappa_{ab}:\!J^a\bJ^b\!:$ are
    basically the same perturbing fields since they result from each
    other by application of the automorphism
    $\Omega(T^a)=(-1)^{a}T^a$ to one of the two currents.} More
  concretely, the Lagrangian we are considering corresponds to the
  deformation
\begin{align}
  \label{eq:Def}
  \cS_{\text{def}}[g]
  &=\frac{\lambda}{\pi k}\int_\Sigma d^2z\,\bigl\langle J,\bar{J}\bigr\rangle
   = -\frac{k\lambda}{\pi}\int_\Sigma d^2z\,\bigl\langle\partial
   gg^{-1},g^{-1}\bartial g\bigr\rangle\, ,
\end{align}
  where we have included a convenient normalisation factor
  $1/k\pi$. From the geometrical construction it is evident that this
  deformation preserves the diagonal $G$ symmetry. Conformal
  invariance, however, can only be checked on the level of
  operators. As is well known, a necessary condition for the
  marginality of the deformation is that the OPE
  $\cO_{\text{def}}(z,\bz)\cO_{\text{def}}(w,\bw)$ does not contain a
  copy of $\cO_{\text{def}}(w,\bw)$ again \cite{Cardy:1989da}. This
  condition arises at first order in perturbation theory. A quick
  calculation keeping only the relevant terms yields
\begin{align}
  \cO_{\text{def}}(z,\bz)\,\cO_{\text{def}}(w,\bw)
  &= \kappa_{ba}\,\kappa_{dc}\,(-1)^{bc}\,J^a(z)J^c(w)\,\bJ^b(\bz)\bJ^d(\bw)\\[2mm]
  &= \cdots-\kappa_{ba}\,\kappa_{dc}\,(-1)^{bc}\,{f^{ac}}_e\,{f^{bd}}_f\,\frac{:\!J^e\bJ^f\!:(w,\bw)}{|z-w|^2}
        +\cdots\, .
\end{align}
  The coefficient of the current bilinear $J^e\bJ^f$ can easily be
  shown to be proportional to $C_{\ad}\,\kappa_{fe}$. It hence
  vanishes for supergroups with vanishing Killing form, thus proving
  our assertion.

  Conformal invariance can also be shown at higher orders in
  perturbation theory. For that purpose we review an argument of
  Bershadsky et al \cite{Bershadsky:1999hk} that in this or a similar
  form will be used frequently throughout the text. The basic idea is
  the following: We know that the perturbing field (and hence also the
  corresponding coupling) transforms trivially under a certain action
  of the supergroup $G$. As a consequence, the associated
  $\beta$-function will also be invariant with respect to $G$. Let us
  then look at all possible Feynman diagrams that can contribute to
  the perturbative expansion of the $\beta$-function. Since the
  deformation can be fully expressed in terms of currents whose OPE is
  given in \eqref{eq:WZWOPE}, the $\beta$-function  will be a sum over
  diagrams made up from trivalent vertices (corresponding to the
  structure constants). These diagrams all cannot have external legs
  since otherwise the $\beta$-function would not transform trivially
  under $G$.

  Let us first consider an arbitrary diagram which contains
  at least one trivalent vertex and let us separate it from the rest
  of the diagram. We now have a vertex connected by three lines to a
  blob containing the rest of the diagram. Assuming further that the
  structure constants are the unique invariant rank three tensor of
  $G$ (this is known to be true in all cases of interest
  \cite{Bershadsky:1999hk,Babichenko:2006uc,Quella:2007sg}), we
  conclude that the diagram is proportional to
  $(-1)^{d_d}\kappa_{ba}{f^{ac}}_d{f^{bd}}_c$. This tensor is identically
  zero due to the vanishing of the dual Coxeter
  number. Hence only diagrams without trivalent vertices remain. Under
  these circumstances, however, the perturbative series reduces to
  that of a multi-component free boson which is known to possess
  marginal current-current perturbations changing the respective
  radii. In our context this also proves conformal invariance for our
  deformed non-abelian model.

  One may wonder whether the moduli space is one-dimensional or
  whether one can combine two marginal deformations in order to
  explore additional directions. Let us analyse this question by
  switching on an additional perturbing field of the form
  \eqref{eq:AutoDefField} with automorphism
  $\Omega(T^a)=(-1)^{a}T^a$. We focus on this choice since $\Omega$ is
  the only automorphism available for all supergroups. Denoting the
  second deforming operator by
  $\tilde{\cO}_{\text{def}}(z,\bz)=\kappa_{ab}J^a\bJ^b$ one finds
\begin{equation}
  \begin{split}
    \cO_{\text{def}}(z,\bz)\,\tilde{\cO}_{\text{def}}(w,\bw)
    &= \kappa_{ba}\,\kappa_{cd}\,(-1)^{bc}\,J^a(z)J^c(w)\,\bJ^b(\bz)\bJ^d(\bw)\\[2mm]
    &= \cdots-\kappa_{ba}\,\kappa_{cd}\,(-1)^{bc}\,{f^{ac}}_e\,{f^{bd}}_f\,\frac{:\!J^e\bJ^f\!:(w,\bw)}{|z-w|^2}
          +\cdots\, .
  \end{split}
\end{equation}
  A closer investigation shows that there is no reason for the
  coefficient of $J^e\bJ^f$ to vanish.
  As a result we conclude that the two deformations $\cO_{\text{def}}$
  and $\tilde{\cO}_{\text{def}}$ are incompatible in
  the sense that conformal invariance is broken as soon as both fields
  are switched on, even though each of them preserves conformal
  invariance separately. This is not too surprising since the two
  deformations preserve different $G$-symmetries, hence destroying
  supersymmetry (but not the bosonic symmetry) when both couplings are
  non-zero.
  
  The operator formalism also allows us to verify our earlier statement
  that the diagonal $G$-symmetry is preserved. This is the case
  precisely when the current $\cO_{\text{def}}(z,\bz)$ is invariant with respect to
  the diagonal action of $G$. In algebraic terms this amounts to the
  statement
\begin{align}
  \Biggl[\oint_w\frac{dz}{2\pi i}\,J^a(z)
  +\oint_{\bw}\frac{d\bz}{2\pi i}\,\bJ^a(\bz)\Biggr]\cO_{\text{def}}(w,\bw)
  = 0
\end{align}
  in the undeformed theory. The validity of
  this equation follows immediately from the OPEs
\begin{align}
  \label{eq:OPEJO}
  J^a(z)\,\cO_{\text{def}}(w,\bw)
  &= \frac{k\bJ^a(\bw)}{(z-w)^2}+\frac{i{f^a}_{dc}:\!J^c\bJ^d\!:(w,\bw)}{z-w}\\[2mm]
  \bJ^a(\bz)\,\cO_{\text{def}}(w,\bw)
  &= \frac{kJ^a(w)}{(\bz-\bw)^2}
        -\frac{i{f^a}_{dc}:\!J^c\bJ^d\!:(w,\bw)}{\bz-\bw}\, .
\end{align}
  An argument similar to the previous one for the $\beta$-function, but now involving
  diagrams with precisely one external leg, can be used to show that a
  quantity with a single $G$-index does not receive corrections at
  higher orders in perturbation theory. Hence we conclude that
  $\cO_{\text{def}}(z,\bz)$ is $G$-invariant to all orders in perturbation
  theory. 
 
%************************************************************************
%************************************************************************
%************************************************************************
\section{\label{sc:TwoThree}Exact two- and three-point functions of currents}

  In this section we will compute the deformed two- and three-point
  functions of the currents to all orders in $\lambda$. To this end we use
  certain algebraic properties of the supergroups at hand (similar to
  the ones employed in the previous section) combined with the abelian
  conformal perturbation theory of \cite{Moore:1993zc}.
 
%************************************************************************
%************************************************************************
\subsection{\label{22}Two-point functions}
 
  Following \cite{Bernard:1990jw} we will assume that the ultraviolet
  limit in the deformed theory is smooth and the perturbed operators
  are in one-to-one correspondence with the operators at the WZW
  point. In particular this concerns the currents which for the
  deformed theory will be denoted by the same symbol but now including
  both holomorphic and antiholomorphic arguments: $J^{a}(z, \bar z)$
  and $\bar J^{a}(z, \bar z)$. On general grounds we expect the
  deformed currents $J^{a}$, $\bar J^{a}$ to remain Virasoro primary
  fields of conformal weights $(1,0)$ and $(0,1)$ respectively. This
  is because the perturbing operator preserves the spin, the currents
  are conserved (and thus their scaling dimension is one) and there
  are no fields that  could form a Jordan block with the currents. The
  last statement follows from the fact that the perturbing fields are
  built from the currents only and thus cannot generate anything but
  the fields in the vacuum representation of the underlying affine Lie
  superalgebra.

  A general remark must be made regarding  correlation functions. In a
  WZW theory for a supergroup the expectation value of the identity
  operator is typically zero: $\langle {\bf 1} \rangle = 0$. This is
  because the identity field in such models belongs to the bottom of a
  Jordan block (the socle of a projective cover). However, this effect
  is absent for the free field realisations of the WZW models
  encountered in \cite{Mitev:2008yt} (giving rise to Gross-Neveu
  models) which allow to establish non-logarithmic theories. Thus, at
  least in that case, the computations of correlation functions which
  are done below  with the convention $\langle {\bf 1} \rangle = 1$
  are fully justified. For other WZW theories on supergroups the
  computations below should be understood more formally as means of
  obtaining OPE coefficients (see section \ref{sc:OPEs}).

  The two-point functions of currents in the deformed theory can be
  obtained by summing up the perturbation theory expansion:
\begin{align}
  \label{JJ} 
  \bigl\langle J^{a}(z_{1}, \bar z_{1}) J^{b}(z_{2}, \bar z_{2})\bigr\rangle_{\lambda}
  &= \Bigl\langle J^{a}(z_{1}) J^{b}(z_{2}) 
\exp\Bigl(-\frac{\lambda}{k\pi}\int\!\! d^{2}w\, :\! J^{e}\bar J^{r}\! : \kappa_{re}\Bigr) 
\Bigr\rangle_{0}
\end{align}
  and analogously for the $\bigl\langle J^{a} \bar J^{b}\bigr\rangle_{\lambda}$
  correlator. Here and elsewhere $\bigl\langle\cdots\bigr\rangle_{0}$
  stands for a WZW theory correlator.

  Each term in the perturbation series is given in terms of integrals
  of correlators evaluated at the WZW point. To compute such integrals
  we need to have these correlators defined in the distributional
  sense (and not merely as functions defined for finite separation of
  variables as is customarily done in CFT). Such distributional
  correlators  in general contain contact term ambiguities related by
  reparameterisations of the coupling constant $\lambda$. A particular
  choice of these contact terms should be considered as part of the
  definition of the composite operator $:\! J^{e}\bar J^{r} \!:$
  coupling to $\lambda$.

  We will fix the distributional correlators of the currents extending
  the prescription of \cite{Moore:1993zc}. In appendix D of that paper,
  G.~Moore gave a prescription for partial integrals of distributions
  arising in conformal perturbation theory of free bosons. That
  prescription is easily adopted for  integrals of  correlators
  containing currents in WZW theory. Moreover in Appendix
  \ref{ap:Distributions} we show how one can define the correlation
  functions at hand as distributions so that the prescription of
  \cite{Moore:1993zc} holds and we justify, based on that definition,
  various manipulations with such integrals.
 
  Specialising to a particular class of supergroups crucially
  simplifies summing up the perturbation series \eqref{JJ}. In
  addition to the vanishing of the adjoint Casimir element we will
  assume that the structure constants $f^{abc}$ are the only invariant
  3-tensor. Under these assumptions the terms containing the structure
  constants in the perturbative series  \eqref{JJ} drop out
  \cite{Bershadsky:1999hk}. For completeness we repeat here the
  argument. Since a simple Lie superalgebra has a unique
  non-degenerate invariant bilinear form, the general form of the
  deformed correlator is
\begin{align}
 \bigl\langle J^{a}(z_{1}, \bar z_{1}) J^{b}(z_{2}, \bar z_{2})\bigr\rangle_{\lambda}
 = \kappa^{ab}g(\lambda; z_{12}, \bar z_{12})
\end{align}
  where $g$ transforms trivially under the global symmetry group. Thus
  if $g$ contains terms dependent on $f^{abc}$ those terms must be of
  the form $f^{abc}C_{abc}$ where $C_{abc}$ is an invariant
  tensor. But since the only invariant 3-tensor is given by the
  structure constants, and since $C_{\ad}=0$ for the quadratic Casimir
  of the adjoint representation, such contributions vanish. The same
  argument goes through for the $\bigl\langle J^{a} \bar
  J^{b}\bigr\rangle_{\lambda}$ correlator. The remaining perturbation series
  is effectively that of the free boson theory. Using the results of
  \cite{Moore:1993zc} (see Appendix \ref{ap:Distributions} for
  details) we obtain
\begin{align}
  \label{JJ2}
 \bigl\langle J^{a}(z_{1}, \bar z_{1}) J^{b}(z_{2}, \bar z_{2})\bigr\rangle_{\lambda}
  &=  
 \frac{k \kappa^{ab}}{(1-\lambda^{2})z_{12}^{2}} \ , &
 \bigl\langle \bar J^{a}(z_{1}, \bar z_{1}) \bar J^{b}(z_{2}, \bar
 z_{2})\bigr\rangle_{\lambda}
 &=  
 \frac{k \kappa^{ab}}{(1-\lambda^{2})\bar z_{12}^{2}} \, , \\[2mm]
  &&
 \bigl\langle J^{a}(z_{1}, \bar z_{1}) \bar J^{b}(z_{2}, \bar
 z_{2})\bigr\rangle_{\lambda}
 &= 0\, ,
\end{align}
  where the correlators are taken at finite separation.
   
%************************************************************************
%************************************************************************
\subsection{Three-point functions} 

  Consider next the deformed three-point functions  
\begin{equation}\label{JJJ}
\bigl\langle J^{a}(z_1, \bar z_1)J^{b}(z_2, \bar z_2)J^{c}(z_3, \bar
z_3)\bigr\rangle_{\lambda}= \Bigl\langle J^{a}(z_1)J^{b}(z_2)J^{c}(z_3)
\exp\Bigl(-\frac{\lambda}{k\pi}\int\!\! d^{2}w\, J^{e}\bar
J^{r}\kappa_{re}\Bigr) \,
\Bigr\rangle_{0}\, .
\end{equation}
  It was argued in \cite{Bershadsky:1999hk} that the terms in the
  perturbation series \eqref{JJJ} containing two or more factors of
  the structure constants vanish. For the argument of
  \cite{Bershadsky:1999hk} to work one needs to assume that there are
  only three traceless invariant tensors of rank 4:
\begin{equation}
  \label{4t}
{f_{ab}}^{e}f_{cde}\, , \quad {f_{ac}}^{e}f_{bde} \, , \quad \kappa_{ab}\kappa_{cd} 
+ (-1)^{bc}\kappa_{ac}\kappa_{bd} + \kappa_{ad}\kappa_{bc} 
\end{equation}
  Any 3-tensor resulting from a contraction of more than two structure
  constants can be represented diagrammatically as in figure
  \ref{fig:3pt} below.
\begin{figure}[!h]
\centering
\includegraphics[width=6cm]{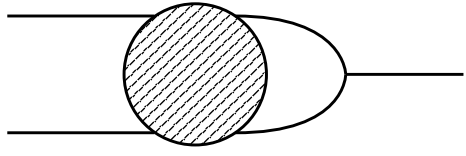}
\caption{\label{fig:3pt}Contraction of tensors for a three-point function}
\end{figure}
%\begin{figure}[t]
%\begin{center}
%\includegraphics[scale=1]{PaperDiagram.jpg}
%\end{center}
%\end{figure}
\noindent 
  Every structure constant corresponds to a three-vertex in the
  diagram and any contraction to a link. The blob containing four
  external lines must correspond to a traceless tensor as there are no
  corrections to the invariant metric. The desired result follows now
  from the fact that every traceless four-tensor listed in \eqref{4t}
  vanishes upon contracting any two indices with the structure
  constants.
 
  We therefore only need to extract all terms containing a single
  factor of the structure constants in \eqref{JJJ}. To this end we
  first extract terms that are singular as  $J^{a}$ approaches the
  other insertions. Singularities with  one of the external fields
  give contributions proportional to the deformed two-point functions
  \eqref{JJ2} while contractions with the perturbing fields can be
  rearranged again into correlators of the perturbed theory. We obtain
\begin{eqnarray}
  \label{JJJ2}
  && \Bigl\langle J^{a}(z_1,\bar z_1)J^{b}(z_2, \bar z_2)J^{c}(z_3, \bar z_3)\Bigr\rangle_{\lambda}
   =\ \frac{1}{1-\lambda^{2}}\,\Bigl\langle J^{a}(z_1)J^{b}(z_2)J^{c}(z_3)\Bigr\rangle_{0} \nonumber \\
 &&  - (-1)^{a(b+c)}\left(\frac{\lambda}{k\pi}\right)i {f^{a}}_{rs}\int \frac{d^2 w}{z_1 - w}
\,\Bigl\langle J^{b}(z_2, \bar z_2)J^{c}(z_3, \bar z_3)J^{s}(w, \bar w)\bar J^{r}(w, \bar w)\Bigr\rangle_{\lambda} \nonumber \\
&& - (-1)^{a(b+c)}\left(\frac{\lambda}{\pi}\right)\int \frac{d^2 w_1}{(z_1-w_1)^2}\,\Bigl\langle J^b(z_2, \bar z_2)
J^{c}(z_3, \bar z_3)\bar J^{a}(w_1,\bar w_1)
\Bigr\rangle_{\lambda}\, .
\end{eqnarray}
  The correlators in the second line can be evaluated in the abelian
  theory that is dropping the structure constants. Using the technique
  of abelian conformal perturbation explained in Appendix
  \ref{ap:Distributions} we obtain
\begin{align}
  \label{Jinterim}
 - (-1)^{a(b+c)}\left(\frac{\lambda}{k\pi}\right)i {f^{a}}_{rs}\int \frac{d^2 w}{z_1 - w}
\,\Bigl\langle J^{b}(z_2, \bar z_2)J^{c}(z_3, \bar z_3)J^{s}(w, \bar w)\bar
J^{r}(w,\bar w)\Bigr\rangle_{\lambda}\\[2mm]
  =\frac{\lambda^{2}}{(1-\lambda^{2})^{2}}\,\Bigl\langle
  J^{a}(z_1)J^{b}(z_2)J^{c}(z_3)\Bigr\rangle_{0} \, .
\end{align}
  We next take up the correlators in the third line of
  \eqref{JJJ2}. Extracting the singularities of $\bar J^{a}(\bar w_1)$
  with other fields we obtain
\begin{align}
  \label{JJJ3}
 & - (-1)^{a(b+c)}\left(\frac{\lambda}{\pi}\right)\int \frac{d^2 w_1}{(z_1-w_1)^2}
 \,\Bigl\langle J^b(z_2, \bar z_2)J^{c}(z_3, \bar z_3)\bar J^{a}(w_1 , \bar
 w_1)\Bigr\rangle_{\lambda} 
 \nonumber \\
 = & -\frac{\lambda^{3}}{(1-\lambda^{2})^{2}}\,\Bigl\langle J^{a}(z_1)J^{b}(z_2)J^{c}(z_3)\Bigr\rangle_{0} \nonumber \\
 & + (-1)^{a(b+c)}\left(\frac{\lambda}{\pi}\right)^2  \iint
\frac{d^{2} w_1 d^{2}w_2}{(z_1-w_1)^{2}(\bar w_1 - \bar w_2)^2}
\,\Bigl\langle J^{b}(z_2, \bar z_2)J^{c}(z_3, \bar z_3)J^{a}(w_2, \bar
w_2)\Bigr\rangle_{\lambda}\, .
\end{align}
 
  Using the integral \eqref{int1}  the integral in \eqref{JJJ3} yields
  back the original three-point function $\langle J^a
  J^bJ^c\rangle_{\lambda}$. Collecting together \eqref{JJJ2},
  \eqref{Jinterim} and \eqref{JJJ3} we finally obtain
\begin{equation}
  \label{JJJanswer}
 \Bigl\langle J^{a}(z_1, \bar z_1)J^{b}(z_2, \bar z_2)J^{c}(z_3, \bar
 z_3)\Bigr\rangle_{\lambda}=  
 \frac{1-\lambda^{3}}{(1-\lambda^{2})^{3}}\left[\,\frac{-ikf^{abc}}{z_{12}z_{23}z_{31}}\,\right]\, . 
\end{equation}
 
  Consider next the mixed three-point function 
\begin{equation} 
 \Bigl\langle J^{a}(z_1, \bar z_1)J^{b}(z_2, \bar z_2)\bar J^{c}(z_3, \bar
 z_3)\Bigr\rangle_{\lambda}=  
\Bigl\langle J^{a}(z_1)J^{b}(z_2)\bar J^{c}(\bar z_3)
\exp\Bigl(-\frac{\lambda}{k\pi}\int\!\! d^{2}w\, J^{e}\bar J^{r}\kappa_{re}\Bigr) \, 
\Bigr\rangle_{0}\, .
\end{equation}
  Extracting the singularities of $\bar J^{c}$ with the perturbing
  fields we obtain
\begin{align}
 & \Bigl\langle J^{a}(z_1, \bar z_1)J^{b}(z_2, \bar z_2)\bar J^{c}(z_3, \bar z_3)\Bigr\rangle_{\lambda} 
 \nonumber \\[2mm]
 = & -\left(\frac{\lambda}{k\pi}\right) 
i{f^{c}}_{es}(-1)^{ce+e} \int \frac{d^{2}w}{\bar z_{3}-\bar w}\,\Bigl\langle J^{a}(z_{1},\bar z_1) J^{b}(z_{2}, \bar z_2 )
:\!J^{e}\bar J^{s}\!:(w,\bar w)\Bigr\rangle_{\lambda} \nonumber \\[2mm]
&- \left(\frac{\lambda}{\pi}\right)  \int \frac{d^{2}w}{(\bar z_{3}-\bar w)^{2}}\,
\Bigl\langle J^{a}(z_{1}, \bar z_1) J^{b}(z_{2}, \bar z_2 ) J^{c}(w, \bar w)\Bigr\rangle_{\lambda} \, .
\end{align}
  The first term in the above equation is easily evaluated using
  \eqref{dr_conts} while for the second term we use
  \eqref{JJJanswer}. Altogether we find
\begin{equation}
  \label{JJbJanswer}
 \Bigl\langle J^{a}(z_1, \bar z_1)J^{b}(z_2, \bar z_2)\bar J^{c}(z_3,
 \bar z_3)\Bigr\rangle_{\lambda}=  
 \frac{\lambda(1-\lambda)}{(1-\lambda^{2})^{3}}\left[\,\frac{-ik f^{abc}\bar z_{12}}{z_{12}^{2}\bar z_{23}\bar z_{31}}\,\right] \, .
\end{equation}
  We see that, as was expected on general grounds, the position dependence of 
  the three-point functions \eqref{JJJanswer} and \eqref{JJbJanswer} 
  matches with the fields $J^{a}(z,\bar z)$ and $\bar J^{a}(z, \bar z)$ 
  being primaries of conformal weights $(1,0)$ and $(0,1)$ respectively. 

%************************************************************************
%************************************************************************
%************************************************************************
\section{\label{sc:OPEs}The OPE algebra of currents}

  In this section we compute the first order contributions to the
  current-current OPEs in the deformed theory. Already at this order
  we find several characteristic features expected from such OPEs
  \cite{Luscher:1977uq,Bernard:1990jw,Ashok:2009xx,Benichou:2010rk}.
  The OPE of $J^{a}$ with $J^{b}$ acquires non-holomorphic
  contributions and the OPE of $J^{a}$ with $\bJ^{b}$ ceases to be
  vanishing. In fact, it the latter even receives a logarithmic
  correction. While we restrict our attention to first order
  perturbation theory, some of the structure constants can be
  determined to all orders using the exact knowledge about the two-
  and three-point functions obtained in section
  \ref{sc:TwoThree}. These coefficients are written out in subsection
  \ref{sec:exOPE}.

%************************************************************************
%************************************************************************
\subsection{The method}

  Our calculation of the OPE between two currents will be based on the
  method developed in \cite{Guida:1995kc}. The starting point is an
  abstract CFT which we assume to be under
  complete control, at least conceptually if not calculationally. 
  This CFT will then be
  perturbed by a deformation term
  $\cS_{\text{def}}=\lambda\int\!d^2z\,\cO_{\text{def}}(z,\bz)$ which could,
  in principle, be either marginal or relevant. Below, we will focus
  on one of the marginal current-current deformations considered in section
  \ref{sc:Defo}. Let us use the symbols $\Phi^a$ to denote a basis
   of  operators in the undeformed
  theory. 
  
  The deformed theory has an OPE of the form
\begin{align}
  \Phi^a(z_1,\bz_1)\,\Phi^b(z_2,\bz_2)
  = \sum_{c}C_c^{ab}(z_{12},\bz_{12}|\lambda)\,\Phi^c(z_2,\bz_2)\, .
\end{align}
  In this equation, the OPE coefficients $C_c^{ab}(z_1-z_2|\lambda)$ depend
  on the deformation parameter $\lambda$. Up to potential logarithmic
  contributions, the explicit coordinate dependence on the right hand
  side of the equation is completely determined by the conformal
  dimensions of the operators $\Phi^c$.

  The fundamental ingredient in our perturbative evaluation of the
  OPE coefficients $C_c^{ab}(z_1-z_2|\lambda)$ is the action
  principle (see e.g. references in \cite{Guida:1995kc}). 
  At the leading order in the deformation parameter $\lambda$
  the action principle reduces to
\begin{multline}
  \label{eq:ActionPrinciple}
    \Bigl\langle\Bigl[\Phi^a(z_1,\bz_1)\,\Phi^b(z_2,\bz_2)
    -\sum_cC_c^{ab}(z_{12},\bz_{12}|0)\,\Phi^c(z_2,\bz_2)\Bigr]
    X(z_3,\bz_3,\cdots)\,\int\!\!d^2z\,\cO_{\text{def}}(z,\bz)\,\Bigr\rangle_0\\[2mm]
    = -\sum_i\partial_\lambda C_c^{ab}(z_{12},\bz_{12}|0)\,
          \bigl\langle\Phi^c(z_2,\bz_2)X(z_3,\bz_3,\cdots)\bigr\rangle_0\, ,
\end{multline}
  where all correlators are evaluated in the unperturbed theory. Using
  a suitable sequence of choices for the multi-local operator
  $X(z_3,\bz_3,\cdots)$ we can therefore determine the derivatives
  $\partial_\lambda C_c^{ab}(z_{12},\bz_{12}|0)$ one after
  another. Together with the knowledge of the unperturbed theory we
  can then easily reconstruct the OPE coefficients to first order in
  the deformation parameter $\lambda$. It is worth noting that in general 
  infrared divergences are present in perturbation expansion around
  massless theories.
  Under certain regularity assumptions the action principle 
  allows one to confine all such divergences to  one-point functions  so that the 
  deformed OPE coefficients are infrared finite \cite{Guida:1995kc}.

  Let us now specialise our considerations to the current-current
  perturbation of WZW models that have been introduced in section
  \ref{sc:Defo}. The perturbation operator in
  \eqref{eq:ActionPrinciple} now becomes
  ${\cal O}_{\rm def}= \frac{1}{\pi k}\,\kappa_{ab}:\! J^{b}\bar J^{a}\! :$.
  The perturbation theory of such current-current
  perturbations is infrared finite since the currents fall off as
  $1/z^2$ and $1/\bz^2$ at infinity, respectively, leaving an integral
  over $1/|z|^4$ which is integrable. For this reason we will omit all
  infrared regulators in what follows. However, in order to apply the
  general formalism presented above we need to discuss the basis of
  operators we are using. As was already mentioned in section
  \ref{sc:TwoThree} we  assume that  a basis of operators in the
  deformed theory can be labelled by the elements of a basis in the
  undeformed theory and that the limit $\lambda\to0$ is smooth. We
  will therefore denote the deformed operators by the corresponding
  bare operators. To insert the undeformed operator into the
  perturbation series one needs to fix the correlation functions as
  distributions. The distributional correlators are defined up to
  contact terms. Any particular choice of such contact terms is part
  of the definition of the composite operator. We will stick to the
  definition of distributional correlators of currents generalising
  the prescription of \cite{Moore:1993zc} as described in Appendix
  \ref{ap:Distributions}.

  In the case of the deformed currents we explicitly put both holomorphic and 
  antiholomorphic coordinates in the notation $J^{a}(z, \bar z)$,
  $\bar J^{b}(w, \bar w)$ to emphasise that these are the deformed
  currents. Since we are only interested in the OPEs between the
  currents and since the deformation term itself only contains
  currents we will only encounter composite operators made up from
  normal ordered products of currents and their derivatives. Thus we
  can take a basis labelled by operators in the vacuum sector of the
  WZW theory. A basis of such operators is built from the composites
  of currents and their derivatives. Let us express a word of caution
  regarding the meaning of such composite operators in the deformed
  theory. The operator $:\!\! J^{a}\bar J^{b}\!\!:(z, \bar z)$
  denotes an operator of the deformed theory whose correlators are
  obtained  by inserting the bare composite $:\!\! J^{a}J^{b}\!\!:$ in
  the perturbation theory series. Such an operator in general will not
  coincide with the normal ordered product of the deformed currents
  $J^a(z,\bz)$ and $\bJ^b(w,\bw)$, i.e.\ with an operator of the
  deformed theory obtained by taking the limit $z \to w$ in the
  deformed OPE of the currents $J^{a}(z,\bar z)$ and $\bar J^{b}(w,
  \bar w)$ and subtracting the singular terms.

  In the undeformed theory the OPE of currents will only contain
  composites which are built of no more than two currents. This will
  no longer be the case for the deformed OPEs which will contain on
  the right hand side composites containing an arbitrarily large
  number of currents. There is a simple rule to be noted for the
  appearance of such composites: the composite built of $N$ currents
  may  appears at the orders $\lambda^{M}$ with $M\ge N-1$. Thus at
  the order $\lambda$ we do not need to include  the composite
  operators beyond bilinears in the currents.

%************************************************************************
%************************************************************************
\subsection{The OPE of $J$ with itself}

  We first discuss the deformed OPE of $J^{a}$ and $J^{b}$. Our
  calculations show that at the leading order in $\lambda$ it gets
  deformed as
\begin{align}
  \label{eq:JJOPEresult}
  J^a(z_1,\bz_1)\,J^b(z_2,\bz_2)
  &= \frac{k\,\kappa^{ab}}{(z_1-z_2)^2}
        +\frac{i{f^{ab}}_c\,J^c(z_2,\bz_2)}{z_1-z_2}
        +\frac{i}{2}{f^{ab}}_c\,\partial J^c(z_2,\bz_2) + \frac{1}{2}:\!(J^{a}J^{b} + J^{b}J^{a})\!:(z_{2}, \bar z_{2})\nonumber\\[2mm]
  &\qquad-\frac{\bz_1-\bz_2}{z_1-z_2}\,\frac{\lambda}{k}(-1)^{bd}{f^a}_{dg}{f^{gb}}_c\,:\!J^c\bJ^d\!:(z_2,\bz_2)
        -\frac{\bz_1-\bz_2}{(z_1-z_2)^2}\,i\lambda \,{f^{ab}}_c\,\bJ^c(z_2,\bz_2)\nonumber\\[2mm]
  &\qquad-\frac{(\bz_1-\bz_2)^2}{(z_1-z_2)^2}\,\frac{i\lambda}{2}\,{f^{ab}}_c\,\bartial\bJ^c(z_2,\bz_2)
        +\cO(\lambda^2) + \cdots\, .
\end{align}
  The only terms of order $\lambda$ in this OPE which are not explicitly written on the right hand side 
  are the terms vanishing as $z_1 \to z_2$.

%************************************************************************
\subsubsection{General ansatz and initial conditions}

  Since we are only interested in terms which are singular or constant
  as $z_1\to z_2$ we can pick a basis of bare operators from the set
  containing $1$, $J^{a}$, $\bar J^{b}$, $\partial J^{a}$, $\bar
  \partial \bar J^{b}$, $:\!\! J^{a}J^{b}\!\!:$, $:\!\! \bar J^{a}\bar
  J^{b}\!\!:$, $:\!\! J^{a}\bar J^{b}\!\!:$. In choosing a linear
  independent set among these operators one needs to take into account
  the following relation between operators in the  WZW model
\begin{align}
  \label{eq:JJdJ}
  :\!\bigl[J^a,J^b\bigr]\!:\ = i{f^{ab}}_c\,\partial J^c
\end{align}
  and a similar one for the antiholomorphic currents. Usually one can
  choose to form a basis using either the derivatives  or the
  antisymmetric bilinears, but for supergroups with a vanishing dual
  Coxeter number $g^\vee$ the usual inversion of equation
  \eqref{eq:JJdJ},
\begin{align}
  \partial J^a
  = \frac{i}{2g^\vee}\,{f^a}_{cb}\,:\!\bigl[J^b,J^c\bigr]\!:\, ,
\end{align}
  does not work. Thus for the subspace at hand we pick the basis
  containing the operators $1$, $J^{a}$, $\bar J^{b}$, $\partial
  J^{a}$, $\bar \partial \bar J^{b}$, $:\!\! J^{a}\bar J^{b}\!\!:$ and
  the symmetric bilinears $:\!\! (J^{a}J^{b}+ J^{b}J^{a})\!\!:$,
  $:\!\! (\bar J^{a}\bar J^{b}+ \bar J^{b}\bar J^{a})\!\!:$.

  We thus write the  following ansatz for the deformed OPE
\begin{align}
  \label{eq:JJOPEansatz}
  J^a(z_1,\bz_1)\,J^b(z_2,\bz_2)
  &= \frac{k\kappa^{ab}(\lambda)}{(z_1-z_2)^2}
        +\frac{i{f^{ab}}_c(\lambda)J^c(z_2,\bz_2)}{z_1-z_2}
        +g^{ab}_{cd}(\lambda):\!J^cJ^d\!:(z_2,\bz_2)
        +h_c^{ab}(\lambda)\partial J^c(z_2,\bz_2)\nonumber\\[2mm]
  &\qquad+\frac{\bz_1-\bz_2}{z_1-z_2}t^{ab}_{cd}(\lambda)\,:\!J^c\bJ^d\!:(z_2,\bz_2)
        +\frac{\bz_1-\bz_2}{(z_1-z_2)^2}u^{ab}_c(\lambda)\bJ^c(z_2,\bz_2)\nonumber\\[2mm]
  &\qquad+\frac{(\bz_1-\bz_2)^2}{(z_1-z_2)^2}v^{ab}_c(\lambda)\bartial\bJ^c(z_2,\bz_2)
        +\frac{(\bz_1-\bz_2)^2}{(z_1-z_2)^2}w^{ab}_{cd}(\lambda):\!\bJ^c\bJ^d\!:(z_2,\bz_2)
        \, ,
\end{align}
  where the coefficients $g$ and $w$ are symmetric in the lower indices:
\begin{align}
  g^{ab}_{cd}(\lambda)
  = (-1)^{cd}g^{ab}_{dc}(\lambda)
  \ ,\qquad\qquad
  w^{ab}_{cd}(\lambda)
  = (-1)^{cd}w^{ab}_{dc}(\lambda)\, .
\end{align}
  The explicit form of the coordinate dependence could be in principle
  modified by logarithms but explicit computations below show that
  this does not happen to first order in $\lambda$.

  Further constraints on the OPE coefficients are obtained by exchanging the order
  of the two currents in the OPE and re-expanding around
  $z_1=z_2$. Working out the details we find
\begin{align}
  \kappa^{ab}(\lambda)
  &= (-1)^{ab}\kappa^{ba}(\lambda)&
  {f^{ab}}_c(\lambda)
  &= -(-1)^{ab}{f^{ba}}_c(\lambda)\\[2mm]
  g^{ab}_{cd}(\lambda)
  &= (-1)^{ab}g^{ba}_{cd}(\lambda)&
  h_c^{ab}(\lambda)
  &= (-1)^{ab}h_c^{ba}(\lambda)+i{f^{ab}}_c(\lambda)\\[2mm]
  u_c^{ab}(\lambda)
  &= -(-1)^{ab}u_c^{ba}(\lambda)&
  v_c^{ab}(\lambda)
  &= (-1)^{ab}v_c^{ba}(\lambda)\\[2mm]
  &&w_{cd}^{ab}(\lambda)
  &= (-1)^{ab}w_{cd}^{ba}\, .  
\end{align}
  More interestingly, we also obtain the equation
\begin{align}
  \Bigl[t_{cd}^{ab}(\lambda)-(-1)^{ab}t_{cd}^{ba}(\lambda)\Bigr]:\!J^c\bJ^d\!:
  \ = u_c^{ab}(\lambda)\,\partial\bJ^c\, .
\end{align}
  When considering the OPE $\bJ^a(z_1,\bz_1)\bJ^b(z_2,\bz_2)$, a
  similar equation is obtained for $\bartial J^c$.
  Additional constraints may arise from the associativity of the
  OPE. Since we will determine the coefficients perturbatively using
  an underlying Lagrangian description, associativity should
  automatically be satisfied.

  It remains to write down the undeformed values for the structure
  constants as they appear in the WZW model. In the conventions chosen
  for the ansatz \eqref{eq:JJOPEansatz}, the non-trivial values are
\begin{align}
  \kappa^{ab}(0)
  &= k\kappa^{ab}\ ,&
  {f^{ab}}_c(0)
  &= {f^{ab}}_c\ ,&
  g_{cd}^{ab}(0)
  &=\frac{1}{2}\Bigl[\delta_c^a\delta_d^b+\delta_d^a\delta_c^b\Bigr]\ ,&
  h_c^{ab}(0)
  &= \frac{i}{2}{f^{ab}}_c\, .
\end{align}
  All these relations are straightforward to see except for the last
  two, which follow from
\begin{align}
  :\!J^aJ^b\!:
  \ = \frac{1}{2}:\!\Bigl[J^aJ^b+J^bJ^a\Bigr]\!:
       +\frac{1}{2}:\!\Bigl[J^aJ^b-J^bJ^a\Bigr]\!:
  \ = \frac{1}{2}\Bigl[\delta_c^a\delta_d^b+\delta_d^a\delta_c^b\Bigr]:\!J^cJ^d\!:
       +\frac{i}{2}{f^{ab}}_c\,\partial J^c\, .
\end{align}

%************************************************************************
\subsubsection{Further details}

  Starting from the ansatz \eqref{eq:JJOPEansatz} for the deformed OPE
  we will now successively determine the individual coefficients using
  the action principle \eqref{eq:ActionPrinciple}. For later
  convenience and in order to enable a systematic evaluation of the
  individual contributions we introduce a number of abbreviations. For
  the first term on the left hand side of
  eq.~\eqref{eq:ActionPrinciple} we use the symbol $A(X)$,
\begin{align}
  A(X)= \kappa_{fe}\Bigl\langle J^a(z_1)J^b(z_2)\,X(\cdots)
        \frac{1}{\pi k}\int\!\! d^2z\,:\!J^e\bJ^f\!:(z,\bz)\Bigr\rangle_{0}\, .
\end{align}
  For the terms originating from the unperturbed OPE we use the
  symbols $B_i(X)$
\begin{align}
  B_1(X)
  &= \kappa_{fe}\biggl\langle\frac{k\kappa^{ab}}{(z_1-z_2)^2}\,X(\cdots)
       \frac{1}{\pi k} \int\!\! d^2z\,:\!J^e\bJ^f\!:(z,\bz)\biggr\rangle_{0}\, , \\[2mm]
  B_2(X)
  &= \frac{i{f^{ab}}_d}{z_1-z_2}\kappa_{fe}\Bigl\langle J^d(z_2)\,X(\cdots)
       \frac{1}{\pi k} \int\!\! d^2z\,:\!J^e\bJ^f\!:(z,\bz)\Bigr\rangle_{0}\, , \\[2mm]
  B_3(X)
  &= \kappa_{fe}\Bigl\langle :\! J^aJ^b\! :(z_2)\,X(\cdots)
        \frac{1}{\pi k} \int\!\! d^2z\,:\!J^e\bJ^f\!:(z,\bz)\Bigr\rangle_{0}\, .
\end{align}
  Here we assume $X(\cdots)$ to be a fixed multi-local field.

\paragraph{Determination of $\kappa$, $f$, $h$ and $g$.}

  We start by  checking that $\kappa^{ab}(\lambda)$ does not
  receive any corrections at first order in perturbation theory (it does so
   at higher orders, cf.\ our exact result in
  eq.~\eqref{JJ2}). Indeed, one can easily check that $A(X)=B_i(X)=0$
  if one chooses $X=1$. Consistency then requires $\partial_\lambda
  k^{ab}(0)=0$. Similar remarks apply to the coefficients
  ${f^{ab}}_c(\lambda)$ and $h_c^{ab}(\lambda)$ which both remain
  undeformed up to $\cO(\lambda^2)$. In this case one has to choose
  $X(\xi)=J^c(\xi)$.

  It is also easy to argue that the coefficient $g^{ab}_{cd}$ must not
  receive any corrections at this order. For that purpose we only note
  that the operator content of 
\begin{align}
  J^{a}(z_{1})J^{b}(z_{2}) \int\!\! d^{2}z\, :\!J^{e}\bar J^{f}\!:(z, \bar z) \kappa_{fe}
\end{align}
  must contain operators with a non-trivial $\bar J$
  component.\footnote{It may look like a contact term between $J^{a}$
    and $\bar J^{b}$ proportional to a delta function can spoil this
    argument, but as formula \eqref{dr_conts} shows such contact
    terms are of order $\lambda$ and can thus be discarded at the leading
    order.}
  
  A more formal proof of this statement would require more information
  on the supergroup. Here we briefly point out the problem. In order
  to determine $g$ it is natural to choose $X(\xi)=:\!J^cJ^d\!:(\xi)$. The
  resulting correlators for $A(X)$ and $B_i(X)$  contain precisely one
  insertion of $\bJ$, implying that $A(X)=B_i(X)=0$. This implies
\begin{align}
  \partial_\lambda g_{rs}^{ab}(0)\,\Bigl\langle:\!J^rJ^s\!:(z_2):\!J^cJ^d\!:(\xi)\Bigr\rangle
  &= \frac{\partial_\lambda
        g_{rs}^{ab}(0)}{(z_2-\xi)^4}\Bigl\{k^2(-1)^{rs}\kappa^{rc}\kappa^{sd}+k^2\kappa^{rd}\kappa^{sc}\\[2mm]
  &\qquad-2k(-1)^{rs}{f^{rc}}_ef^{sed}-k(-1)^{r(s+c)}{f^{rd}}_ef^{sce}\Bigr\}
  =0 \, . 
\end{align}
  One may worry that there exists a non-vanishing tensor such that the
  contraction above yields zero. However, even if this does
  happen this merely means that the $X$ we chose was not a good
  choice to determine $g_{rs}^{ab}(0)$ and we have to consider other
  $X$'s. The presence of a non-trivial kernel for the above 4-tensor
  depends on a particular group. As long as the states $:\!(J^rJ^s+
  J^{s}J^{r})\!:$ are not of zero norm there will be another $X$ with a
  nonzero overlap which by the above simple argument must detect
  $g_{rs}^{ab}(0)=0$. We will not pursue this   more formal line of
  argument any further.

\paragraph{Determination of $u$ and $v$.}

  In the next step we select $X(\bxi)=\bJ^c(\bxi)$. This will allow us to
  determine the coefficients $u$ and $v$. A straightforward
  calculation using the explicit form of the undeformed two- and
  three-point functions as well as a decomposition into partial
  fractions yields
\begin{align}
  A(X)
  &= \frac{1}{\pi k}\kappa_{fe}(-1)^{ce}\int d^2z\,\frac{-ikf^{abe}}{(z_1-z_2)(z_2-z)(z-z_1)}\frac{k\kappa^{cf}}{(\bxi-\bz)^2}\\[2mm]
  &= -\frac{\bz_1-\bz_2}{(z_1-z_2)^2}\,ikf^{abc}\,
        \,\frac{1}{(\bxi-\bz_1)(\bxi-\bz_2)}\, .
\end{align}
  Here the integral was evaluated using  formulas
  \eqref{int1}-\eqref{int3}. Later we will also need the expansion of
  this expression in terms of inverse powers of $\bxi$,
\begin{align}
  A(X)
  = -\frac{\bz_1-\bz_2}{(z_1-z_2)^2}\,ikf^{abc}\,
        \,\frac{1}{\bxi^2}\,\biggl[1+\frac{\bz_1+\bz_2}{\bxi}+\cdots\biggr]\, .
\end{align}
  One also finds $B_i(X)=0$ here, even though
  this time it is a result of integration. The total contribution that
  has to be matched is thus given by $A(X)$ itself.

  According to eq.~\eqref{eq:ActionPrinciple} this result should be
  compared to the unperturbed correlation functions. With the present
  choice $X=\bJ^c$, the most singular contribution arises from
\begin{align}
  \frac{\bz_1-\bz_2}{(z_1-z_2)^2}\,\partial_\lambda u^{ab}_d(0)
  \,\Bigl\langle\bJ^d(\bz_2)\bJ^c(\bxi)\Bigr\rangle
  &=   \frac{\bz_1-\bz_2}{(z_1-z_2)^2}\,\partial_\lambda u^{ab}_d(0)
  \,\frac{k\kappa^{dc}}{(\bz_2-\bxi)^2}\\[2mm]
  &= \frac{\bz_1-\bz_2}{(z_1-z_2)^2}
        \,k\kappa^{dc}\partial_\lambda u^{ab}_d(0)
        \,\frac{1}{\bxi^2}\biggl[1+\frac{2\bz_2}{\bxi}+\cdots\biggr]\, .
\end{align}
  Comparison of the leading terms yields
\begin{align}
  \label{eq:solu}
  u^{ab}_c(\lambda)
  = -i\lambda{f^{ab}}_c+\cO(\lambda^2)\, .
\end{align}
  While the leading contribution allows to determine $u$, the
  subleading contribution provides enough information to calculate
  $v$. Plugging our finding for $u$ back into
  eq.~\eqref{eq:ActionPrinciple} leads to the expression
\begin{align}
  A(X)+\frac{\bz_1-\bz_2}{(z_1-z_2)^2}\,\partial_\lambda u^{ab}_d(0)
  \,\Bigl\langle\bJ^d(\bz_2)\bJ^c(\bxi)\Bigr\rangle
  &= -\frac{(\bz_1-\bz_2)^2}{(z_1-z_2)^2}\,
        \frac{1}{\bxi^3}\,ikf^{abc}+\cdots\ , 
\end{align}
  which has to be matched by a linear combination of the following two
  terms:
\begin{align}
  -\frac{(\bz_1-\bz_2)^2}{(z_1-z_2)^2}\,\partial_\lambda v^{ab}_d(0)
  \Bigl\langle\bartial\bJ^d(\bz_2)\bJ^c(\bxi)\Bigr\rangle
  &= \frac{(\bz_1-\bz_2)^2}{(z_1-z_2)^2}\,\partial_\lambda v^{ab}_d(0)
        \frac{2k\kappa^{dc}}{(\bz_2-\bxi)^3}\\[2mm]
  &= -\frac{(\bz_1-\bz_2)^2}{(z_1-z_2)^2}\,\frac{1}{\bxi^3}
        \,2k\kappa^{dc}\partial_\lambda v^{ab}_d(0)
        \Bigl[1+\cdots\Bigr]\, , \\[2mm]
  -\frac{(\bz_1-\bz_2)^2}{(z_1-z_2)^2}\,\partial_\lambda w^{ab}_{de}(0)
  \Bigl\langle:\!\bJ^d\bJ^e\!:(\bz_2)\bJ^c(\bxi)\Bigr\rangle
  &= \frac{(\bz_1-\bz_2)^2}{(z_1-z_2)^2}
        \,\partial_\lambda w^{ab}_{de}(0)
        \,\frac{ikf^{dec}}{(\bz_2-\bxi)^3}
   = 0\, .
\end{align}
  Again the last term vanishes since we assumed $w$ to be
  antisymmetric in the lower indices. We can therefore easily solve for
  $v(\lambda)$, obtaining
\begin{align}
  \label{eq:solv}
  v^{ab}_c(\lambda)
  =-\frac{i\lambda}{2}\,{f^{ab}}_c+\cO(\lambda^2)\, .
\end{align}

\paragraph{Determination of $t$ and $w$.}
These coefficients are determined in a similar fashion.
  In order to streamline the presentation in the main text, these
  calculations have been moved to appendix \ref{ap:1storder}.

%************************************************************************
%************************************************************************
\subsection{The OPE of $J$ with $\bJ$}

  The next goal is to determine the mixed OPE between $J$ and
  $\bJ$. In the undeformed theory this OPE vanishes
  identically. However, the calculations below imply that the deformed
  OPE acquires a correction that is given by
\begin{align}
  J^a(z_1,\bz_1)\,\bJ^b(z_2,\bz_2)
  &= \frac{i\lambda{f^{ab}}_c}{z_1-z_2}\,\bJ^c(z_2,\bz_2)
        + \frac{i\lambda{f^{ab}}_c}{\bz_1-\bz_2}\,J^c(z_2,\bz_2)\\[2mm]
  &\qquad-\frac{z_1-z_2}{\bz_1-\bz_2}\,i\lambda
  {f^{ab}}_c\,\partial J^c(z_2,\bz_2)\\[2mm]
  &\qquad-(-1)^{fs}\,\frac{\lambda}{k}\kappa_{ef}{f^{ae}}_c{f^{bf}}_d
        \ln\frac{|z_1-z_2|^2}{\epsilon^2}\,:\!J^c\bJ^d\!:(z_2,\bz_2)+\cO(\lambda^2)\, .
\end{align}
  at first order in the coupling constant. Here, the constant
  $\epsilon$ in the logarithmic contribution is a UV regulator which can
  be thought of as a normal ordering ambiguity. We start with
  discussing the general ansatz for the OPE
  $J^a(z_1,\bz_1)\bJ^b(z_2,\bz_2)$. Instead of showing all the
  individual steps leading to our result, we then present the
  calculation of three coefficients in some detail. The determination
  of the remaining ones can be found in appendix \ref{ap:1storder}.

%************************************************************************
\subsubsection{General ansatz and initial conditions}

  As in the previous section we start with a general ansatz which in
  this case reduces to
\begin{align}
  \label{eq:JbJOPEansatz}
  J^a(z_1,\bz_1)\,\bJ^b(z_2,\bz_2)
  &= \frac{A^{ab}(\lambda)}{|z_1-z_2|^2}
        +\frac{1}{z_1-z_2}\,B^{ab}_c(\lambda)\,\bJ^c(z_2,\bz_2)
        +\frac{\bz_1-\bz_2}{z_1-z_2}\,\tilde{B}^{ab}_c(\lambda)\,\bartial\bJ^c(z_2,\bz_2)\\[2mm]
  &\qquad+\frac{\bz_1-\bz_2}{z_1-z_2}\,\hat{B}^{ab}_{cd}(\lambda)\,:\!\bJ^c\bJ^d\!:(z_2,\bz_2)
        +\frac{1}{\bz_1-\bz_2}\,C^{ab}_c(\lambda)\,J^c(z_2,\bz_2)\\[2mm]
  &\qquad+\frac{z_1-z_2}{\bz_1-\bz_2}\,\tilde{C}^{ab}_c(\lambda)\,\partial
        J^c(z_2,\bz_2)
        +\frac{z_1-z_2}{\bz_1-\bz_2}\,\hat{C}^{ab}_{cd}(\lambda)\,
        :\!J^cJ^d\!:(z_2,\bz_2)\\[2mm]
  &\qquad+D_{cd}^{ab}(\lambda)\,:\!J^c\bJ^d\!:(z_2,\bz_2)\, .
\end{align}
  Without loss of generality we assume that the coefficients
  $\hat{B}^{ab}_{cd}(\lambda)$ and $\hat{C}^{ab}_{cd}(\lambda)$ are
  symmetric in the lower indices, i.e.
\begin{align}
  \hat{B}^{ab}_{cd}(\lambda)
  = (-1)^{cd}\hat{B}^{ab}_{dc}(\lambda)
  \ ,\qquad\qquad
  \hat{C}^{ab}_{cd}(\lambda)
  = (-1)^{cd}\hat{C}^{ab}_{dc}(\lambda)\, .
\end{align}
  It should also be noted that the explicit coordinate dependence only
  reflects the contributions needed to account for the proper scaling
  dimension of the terms on the right hand side. The coefficients may
  carry an additional implicit logarithmic (and hence dimensionless)
  coordinate dependence. Indeed, we will soon recognise that such
  logarithmic contributions arise in the coefficient
  $D_{cd}^{ab}(\lambda)$.

  In the current setting, the initial conditions at $\lambda=0$ are almost trivial
  since $J(z)$ and $\bJ(w)$ commute in the undeformed theory. We only
  need to keep track of the non-singular term
\begin{align}
  J^a(z_1)\,\bJ^b(\bz_2)
  =\ :\!J^a\bJ^b\!:(z_2,\bz_2) + \cdots \, .
\end{align}
  Consequently we have $D_{cd}^{ab}(0)=\delta_c^a\delta_d^b$,
%\begin{align}
%  D_{cd}^{ab}(0)
%  = \delta_c^a\delta_d^b\, ,
%\end{align}
  with all other coefficients vanishing.

%************************************************************************
\subsubsection{Further details}

  The general procedure outlined above instructs us to determine and
  to compare the following two quantities for suitable choices of the
  field $X(\cdots)$. The first one is
\begin{align}
  A(X)= \kappa_{fe}\Bigl\langle
  J^a(z_1)\bJ^b(z_2)X(\cdots)\frac{1}{k\pi}\int\! \!d^2z\,J^e\bJ^f(z,\bz)\Bigr\rangle_{0}
\end{align}
  and the second one is
\begin{align}
  B(X)
  = \lim_{:z_1\to z_2:}A(X)
  = \underset{z_1\to z_2}{\Res}\:\frac{A(X)}{z_1-z_2}\, ,
\end{align}
  that corresponds to the non-singular part of $A(X)$ as $z_1$
  approaches $z_2$. Practically, the limit extracts the constant part
  if $A(X)$ is considered as a Laurent series in $z_1-z_2$.

\paragraph{Determination of $A^{ab}$.}

  The result $A^{ab}(\lambda)=0$ for the
  central term follows immediately -- and even to all orders -- from
  our knowledge of the exact two-point function \eqref{JJ2}. It can
  also be understood using the general framework using $X=1$. 

\paragraph{Determination of $D_{cd}^{ab}$.}

  From a physical perspective, the most interesting coefficient in the
  ansatz \eqref{eq:JbJOPEansatz} is certainly $D_{cd}^{ab}(\lambda)$
  since it turns out to contain logarithmic contributions. This
  coefficient can be determined by setting $X(\xi, \bar \xi)=:\!J^c\bJ^d\! :(\xi, \bar \xi)$. We  first
  evaluate
\begin{align}
  A(X)
  &= (-1)^{bc}\frac{k\kappa_{ef}f^{ace}f^{bdf}}{(z_1-\xi)^2(\bz_2-\bxi)^2}
        \ln\frac{\epsilon^2|z_1-z_2|^2}{|\xi-z_2|^2|z_1-\xi|^2}\, .
\end{align}
  During the calculation we followed the standard recipe of replacing
  $(\xi-\xi)^2$ by a regulator $\epsilon^2$. In a similar fashion we
  then evaluate
\begin{align}
  B(X)
  &= \lim_{:z_1\to z_2:}A(X)
   = (-1)^{bc}\frac{k\kappa_{ef}f^{ace}f^{bdf}}{|z_2-\xi|^4}
        \ln\frac{\epsilon^4}{|z_2-\xi|^4}\, .
\end{align}
  For large values of the variable $\xi$ one obtains
\begin{align}
  A(X)-B(X)
  &= (-1)^{bc}\frac{k\kappa_{ef}f^{ace}f^{bdf}}{|\xi|^4}
        \ln\frac{|z_1-z_2|^2|z_2-\xi|^4}{\epsilon^2|\xi-z_2|^2|z_1-\xi|^2}
        +\cdots\\[2mm]
  &= (-1)^{bc}\frac{k\kappa_{ef}f^{ace}f^{bdf}}{|\xi|^4}
        \ln\frac{|z_1-z_2|^2}{\epsilon^2}+\cdots\, .
\end{align}
  This difference has to be compared to
\begin{align}
 - \partial_\lambda D^{ab}_{rs}(0)
  \,\Bigl\langle J^r(z_2)\bJ^s(\bz_2):\!J^c\bJ^d\!:(\xi,\bxi)\Bigr\rangle
  &= -\frac{k^2(-1)^{cs}\kappa^{rc}\kappa^{sd}}{(z_2-\xi)^2(\bz_2-\bxi)^2}\,\partial_\lambda D^{ab}_{rs}(0)\\[2mm]
  &= -k^2(-1)^{cs}\kappa^{rc}\kappa^{sd}
        \partial_\lambda D^{ab}_{rs}(0)\,\frac{1}{|\xi|^4}
        \Bigl[1+\cdots\Bigr]\, .
\end{align}
  The comparison yields a logarithmic dependence of the structure
  constants on the difference $z_1-z_2$,
\begin{align}
  D^{ab}_{rs}(\lambda)
  &= -(-1)^{fs}\,\frac{\lambda}{k}\kappa_{ef}{f^{ae}}_r{f^{bf}}_s
        \ln\frac{|z_1-z_2|^2}{\epsilon^2}+\cO(\lambda^2)\, .
\end{align}
  The presence of the regulator $\epsilon$ can be interpreted as a normal
  ordering ambiguity \footnote{At the level of correlators in the bare theory this is 
  an ambiguity in defining  distributional three-point functions $\langle :\!J^{a}\bar J^{b}\!: \, \, 
  :\!J^{c}\bar J^{d} \!: \, \, :\! J^{e}\bar J^{f}\!:\rangle_{0}\kappa_{fe}$. Such distributions are 
  not considered in appendix B. This can be considered as a further ambiguity in defining 
  the normal ordering in the deforming operator. }.

\paragraph{Determination of $C_c^{ab}$ and $\tilde{C}_c^{ab}$.}

  We next explain how to  determine the coefficients $C$ and
  $\tilde{C}$. Both can be obtained from one single calculation using
  the choice $X=J^c$. Like before we first evaluate
\begin{align}
  A(X)
  &= -\frac{ikf^{abc}}{(z_1-\xi)^2}\biggl[\frac{1}{\bz_1-\bz_2}-\frac{1}{\bxi-\bz_2}\biggr]
   = -\frac{ikf^{abc}}{\xi^2}\biggl[\frac{1}{\bz_1-\bz_2}+\frac{2z_1}{(\bz_1-\bz_2)\xi}-\frac{1}{\bxi}+\cdots\biggr]\, .
\end{align}
  Taking the limit $:\!z_1\to z_2\!:$ leads to the expression
\begin{align}
  B(X)
  &= \frac{ikf^{abc}}{(z_2-\xi)^2(\bxi-\bz_2)}
   = \frac{ikf^{abc}}{\xi^2\bxi}\biggl[1+\frac{2z_2}{\xi}+\frac{\bz_2}{\bxi}+\cdots\biggr]\, .
\end{align}
  Putting these together we obtain
\begin{align}
  A(X)-B(X)
  = -\frac{ikf^{abc}}{(\bz_1-\bz_2)\xi^2}\biggl[1+\frac{2z_1}{\xi}+\cdots\biggr]\, .
\end{align}
  The leading term in this expression can be accounted for by the term
\begin{align}
  -\frac{1}{\bz_1-\bz_2}\,\partial_\lambda C^{ab}_d(0)
  \,\Bigl\langle J^d(z_2)J^c(\xi)\Bigr\rangle
  &= -\frac{1}{\bz_1-\bz_2}\,\frac{k\kappa^{dc}}{(z_2-\xi)^2}\,\partial_\lambda C^{ab}_d(0)\\[2mm]
  &= -\frac{k\kappa^{dc}\partial_\lambda
     C^{ab}_d(0)}{\bz_1-\bz_2}\,\frac{1}{\xi^2}
        \biggl[1+\frac{2z_2}{\xi}+\cdots\biggr]\, .
\end{align}
  A comparison of the most singular terms yields
\begin{align}
  C^{ab}_c(\lambda)
  = i\lambda{f^{ab}}_c\, .
\end{align}
  In order to fix the subleading contributions we analyse
\begin{align}
  A(X)-B(X)+\frac{\partial_\lambda C^{ab}_d(0)}{\bz_1-\bz_2}
  \,\Bigl\langle J^d(z_2)J^c(\xi)\Bigr\rangle
  = -\frac{z_1-z_2}{\bz_1-\bz_2}\frac{1}{\xi^3}\,2
  ikf^{abc}\Bigl[1+\cdots\Bigr]\, .
\end{align}
  There are in principle two different correlation functions which
  could give rise to such a contribution,
\begin{align}
  -\frac{z_1-z_2}{\bz_1-\bz_2}\,\partial_\lambda\tilde{C}^{ab}_d(0)
  \,\Bigl\langle\partial J^d(z_2)J^c(\xi)\Bigr\rangle
  &= \frac{z_1-z_2}{\bz_1-\bz_2}\,\partial_\lambda\tilde{C}^{ab}_d(0)
        \,\frac{2k\kappa^{dc}}{(z_2-\xi)^3}\\[2mm]
  -\frac{z_1-z_2}{\bz_1-\bz_2}\,\partial_\lambda\hat{C}^{ab}_{de}(0)
  \,\Bigl\langle:\!J^dJ^e\!:(z_2)J^c(\xi)\Bigr\rangle
  &= \frac{z_1-z_2}{\bz_1-\bz_2}\,\partial_\lambda\hat{C}^{ab}_{de}(0)
        \,\frac{ikf^{dec}}{(z_2-\xi)^3}
   = 0\, .
\end{align}
  The latter vanishes due to our symmetry assumptions. Hence we find
\begin{align}
  \tilde{C}^{ab}_c(\lambda)
  = -i\lambda{f^{ab}}_c+\cO(\lambda^2)\, .
\end{align}

\paragraph{Determination of the other coefficients.}
  The computational details related to the remaining OPE coefficients
  are put into appendix \ref{ap:1storder}.

%************************************************************************
%************************************************************************
\subsection{\label{sec:exOPE}Exact OPE coefficients}

  The exact two- and three-point functions of the currents obtained in
  section \ref{sc:TwoThree} allow us to compute the OPE coefficients
  $\kappa^{ab}(\lambda)$, ${f^{ab}}_{c}(\lambda)$,
  $u^{ab}_{c}(\lambda)$ from \eqref{eq:JJOPEansatz} and
  $A^{ab}(\lambda)$, $B^{ab}_{c}(\lambda)$, $C^{ab}_{c}(\lambda)$ from
  \eqref{eq:JbJOPEansatz} to all orders in $\lambda$. We obtain 
\begin{equation}
  \label{exOPE1}
  \kappa^{ab}(\lambda) = \frac{\kappa^{ab}}{1-\lambda^{2}}\, , \quad A^{ab}(\lambda) = 0\, , \quad  {f^{ab}}_{c}(\lambda)=\frac{1-\lambda^3}{(1-\lambda^2)^{2}}{f^{ab}}_{c}\, , 
\end{equation}
\begin{equation}
  \label{exOPE2}
u^{ab}_{c}(\lambda) = C^{ab}_{c}(\lambda)=B^{ab}_{c}(\lambda)=
i{f^{ab}}_{c}\frac{\lambda(1-\lambda)}{(1-\lambda^{2})^{2}} \, .
\end{equation}

%************************************************************************
%************************************************************************
%************************************************************************
\section{\label{sc:FourPT} Four-point function of currents in a $1/k$ expansion}

  In this section we will use the methods of section \ref{sc:TwoThree} to
  obtain an approximation to a four-point function of the
  currents. Consider the perturbation series expansion 
\begin{align}
 A_{4}(\lambda,k)
  &= \Bigl\langle J^{a}(z_1, \bar z_1)J^{b}(z_2, \bar z_2)J^{c}(z_3, \bar z_3)J^{d}(z_4, \bar z_4) \Bigr\rangle_{\lambda} 
 \nonumber \\
 &=  \Bigl\langle J^{a}(z_1)J^{b}(z_2)J^{c}(z_3)J^{d}(z_4) 
 \exp\Bigl(-\frac{\lambda}{k\pi}\int d^{2}w\, J^{e}\bar J^{r}\kappa_{re}\Bigr) \Bigr\rangle_{0} \, .
\end{align}
Every contraction 
of the bare currents in the perturbative integrals either comes with the metric 
multiplied by the level $k$ or with the structure constants. Thus  the number of 
the structure constants appearing in the perturbative expansion effectively
measures the power of $k$. For fixed $\lambda$ we have an expansion in powers of $1/k$
of the form
\begin{align}
  \label{kexp}
  A_{4}(\lambda,k)
  &= k^{2}\Bigl[ A_{4}^{(0)}(\lambda) +
  \frac{1}{k}A_{4}^{(1)}(\lambda) + \dots +
  \frac{1}{k^{p}}A_{4}^{(p)}(\lambda) + \cdots \Bigr]\, ,
\end{align}
  where the term $A^{(p)}(\lambda)$ comes from all terms in the perturbation
  series with $2p$ structure constant contractions. We compute the
  functions $A_{4}^{(0)}(\lambda)$ and $A_{4}^{(1)}(\lambda)$ to all orders in
  $\lambda$. In this section we sketch the main steps in the computation
  relegating more details to Appendix \ref{ap:4pt}.

  We begin by extracting all singularities of the bare current
  $J^{a}(z_{1})$ in the perturbative integrands. As in section \ref{sc:TwoThree}
  contractions with the perturbing operators are rearranged into
  integrated correlators in the deformed theory. Using \eqref{JJ2},
  \eqref{JJJanswer}, \eqref{JJbJanswer} we obtain
\begin{align}
 \label{step1}
 A_{4}
 &= a_{4}  -\left(\frac{\lambda}{k\pi}\right) \int \frac{d^{2}w}{z_1-w}\, 
\Bigl\langle J^{b}(z_2, \bar z_2)J^{c}(z_3, \bar z_3)J^{d}(z_4, \bar z_4)\, i{f^{a}}_{re}
 :\!J^{e}\bar J^{r}\!:(w,\bar w)\Bigr\rangle_{\lambda} (-1)^{a(b+c+d)} \nonumber \\[2mm]
&\qquad - \left(\frac{\lambda}{\pi}\right) \int \frac{d^{2}w}{(z_1-w)^2}\,
\Bigl\langle  J^{b}(z_2)J^{c}(z_3)J^{d}(z_4) \bar J^{a}(\bar w)\Bigr\rangle_{\lambda}(-1)^{a(b+c+d)}
\end{align}
where 
\begin{align}
  a_{4}
  &= \frac{k^{2}}{1-\lambda^2}\biggl[\,\frac{\kappa^{ab}\kappa^{cd}}{z_{12}^{2}z_{34}^{2}}  
+ \frac{(-1)^{ab}\kappa^{ac}\kappa^{bd}}{z_{13}^{2}z_{24}^{2}} 
+ \frac{\kappa^{ad}\kappa^{bc}}{z_{14}^{2}z_{23}^{2}}\,\biggr] \nonumber \\[2mm]
&\qquad  + \frac{1-\lambda^{3}}{(1-\lambda^2)^{3}}\biggl[\, \frac{{f^{ab}}_{s}f^{scd}}{z_{12}} + 
\frac{(-1)^{ab}{f^{ac}}_{s}f^{bsd}}{z_{13}} + 
\frac{(-1)^{a(b+c)}{f^{ad}}_{s}f^{bcs}}{z_{14}}\,\biggr] \frac{k}{z_{23}z_{34}z_{42}}\, .
\end{align}
We next extract the singularities of $\bar J^{a}(\bar w)$ in the second term on the 
right hand side of \eqref{step1}. This way we obtain 
\begin{align}
  \label{step2}
 (1-\lambda^{2})A_{4}
  &= a_{4} + (1-\lambda) \left(\frac{-\mu}{2\pi}\right)(-1)^{a(b+c+d)} \nonumber \\[2mm]
&\qquad  \times \int\!\! \frac{d^{2}w}{z_1-w}\,i{f^{a}}_{re}\Bigl\langle J^{b}(z_2, \bar z_2)J^{c}(z_3, \bar z_3)
J^{d}(z_4, \bar z_4) :\!J^{e}\bar J^{r}\!:(w,\bar w)
\Bigr\rangle_{\lambda}\, . 
\end{align}
Notice that  a factor of the structure constants stands in front of the second term on the right hand side 
of the above expression. Thus,  in our approximation, we need to compute all perturbative terms with a single 
contraction $JJ\to J$ in the correlation function inside the integral. 
Extracting singularities of $J^{b}(z_2)$ in that correlator we obtain 
terms  all containing a single factor of the structure constant except for the 
contribution 
\begin{equation}
-(-1)^{b(c+d+e+r)}\left(\frac{\lambda}{\pi}\right)\int\!\!\frac{d^{2}w_2}{(z_2-w_2)^2}
\,\Bigl\langle J^{c}(z_3, \bar z_3)J^{d}(z_4, \bar z_4) :\!J^{e}\bar J^{r}\!:(w,\bar w)\,\bar J^{b}(w_2, \bar w_2) \Bigr\rangle_{\lambda}
\end{equation}
that comes from the abelian contraction of $J^{b}(z_2)$ with perturbing operators. 
Extracting the singularities of $\bar J^{b}(\bar w_2)$ in the above correlator 
gives terms which are explicitly evaluated by the methods of Appendix
\ref{ap:Distributions} plus a term proportional to 
\begin{equation}
\lambda^{2}\Bigl\langle J^{b}(z_2, \bar z_2)J^{c}(z_3,\bar z_3)J^{d}(z_4,\bar z_4):\!J^{e}\bar J^{r}\!:(w,\bar w)\Bigr\rangle_{\lambda}
\end{equation}
which closes the system of equations. Collecting all contributions we obtain 
\begin{align}
  \label{corr2}
  (1-\lambda^{2})\Bigl\langle J^{b}(z_2, \bar z_2)J^{c}(z_3, \bar z_3)J^{d}(z_4, \bar z_4) :\!J^{e}\bar J^{r}\!:(w,\bar w)
\Bigr\rangle_{\lambda}  = \sum_{i=1}^{7} a^{(i)} 
\end{align} 
where 
\begin{align}
  a^{(1)}
  &=(-1)^{b(c+d)}\frac{k\kappa^{be}}{(z_2-w)^{2}}
    \bigl\langle J^{c}(z_3, \bar z_3)J^{d}(z_4, \bar z_4) \bar J^{r}(w, \bar w)\bigr\rangle_{\lambda} \, , \\[2mm]
  a^{(2)}
  &= (-1)^{b(c+d+e+r)}k\pi (-\lambda)\kappa^{rb}\delta(z_2-w)\bigl\langle J^{c}(z_3, \bar z_3)J^{d}(z_4, \bar z_4)J^{e}(w, \bar w)\bigr\rangle_{\lambda}  \, , \\[2mm]
  a^{(3)}
  &= (-1)^{b(c+d)}\frac{i{f^{be}}_{s}}{z_2-w}\, \bigl\langle J^{c}(z_3, \bar z_3)J^{d}(z_4, \bar z_4) 
:\!J^{s}\bar J^{r}\!:(w,\bar w)\bigr\rangle_{\lambda}  \, , \\[2mm]
  a^{(4)}
  &=  \lambda(-1)^{b(c+d + e)}\frac{i{f^{br}}_{s}}{z_2-w}\, \bigl\langle J^{c}(z_3, \bar z_3)J^{d}(z_4, \bar z_4) 
:\!J^{e}\bar J^{s}\!:(w,\bar w)\bigr\rangle_{\lambda} \, , \\[2mm]
  a^{(5)}
  &=  (1-\lambda)\left(\frac{-\mu}{2\pi}\right)(-1)^{b(c+d+e+r)}i{f^{b}}_{qs}\int\!\!\frac{d^{2}w_2}{z_2-w_2}
\nonumber \\
&\qquad\bigl\langle J^{c}(z_3, \bar z_3)J^{d}(z_4, \bar z_4) :\!J^{e}\bar J^{r}\!:(w,\bar w)\, :\!J^{s}\bar J^{q}\!:(w_2,\bar w_2)\bigr\rangle_{\lambda}\, , \\[2mm]
  a^{(6)}
  &=  \frac{i{f^{bc}}_{s}}{z_{23}}\bigl\langle J^{s}(z_{3}, \bar z_3)J^{d}(z_{4}, \bar z_4)\, :\!\!J^{e}\bar J^{r}\!\!:(w,\bar w)\bigr\rangle_{\lambda}\, , \\[2mm]
  a^{(7)}
  &= \frac{i(-1)^{bc}{f^{bd}}_{s}}{z_{24}}\bigl\langle J^{c}(z_{3}, \bar z_3)J^{s}(z_{4},\bar z_4)\, :\!\!J^{e}\bar J^{r}\!\!:(w,\bar w)\bigr\rangle_{\lambda} \, .
\end{align}
  The first two terms on the right hand side of \eqref{corr2} can be
  calculated using the exact three-point functions. The remaining
  terms all contain a factor of the structure constants. In our
  approximation they can be computed using abelian perturbation theory.
  Substituting the result into \eqref{step2} we finally obtain
%% Final result
\begin{align}
  \label{4pt_final}
& \Bigl\langle J^{a}(z_1, \bar z_1)J^{b}(z_2, \bar z_2)J^{c}(z_3, \bar
z_3)J^{d}(z_4, \bar z_4) \Bigr\rangle_{\lambda}
  =\frac{k^{2}}{(1-\lambda^2)^{2}}\biggl[\frac{\kappa^{ab}\kappa^{cd}}{z_{12}^{2}z_{34}^{2}}  
+ \frac{(-1)^{ab}\kappa^{ac}\kappa^{bd}}{z_{13}^{2}z_{24}^{2}} 
+ \frac{\kappa^{ad}\kappa^{bc}}{z_{14}^{2}z_{23}^{2}} \biggr] \nonumber \\[2mm]
& +\frac{1-\lambda^{4}}{(1-\lambda^2)^{4}}\biggl[ \frac{{f^{ab}}_{s}f^{scd}}{z_{12}} + 
\frac{(-1)^{ab}{f^{ac}}_{s}f^{bsd}}{z_{13}} + 
\frac{(-1)^{a(b+c)}{f^{ad}}_{s}f^{bcs}}{z_{14}} \biggr] \frac{k}{z_{23}z_{34}z_{42}} \nonumber \\[2mm]
&+ k\frac{\lambda^{2}(1-\lambda)^{2}}{(1-\lambda^2)^{5}}\biggl[ -{f^{ab}}_{r}f^{rcd}
 \frac{1}{z_{34}^{2}z_{12}^{2}}\ln\left| \frac{z_{13}z_{24}}{z_{23}z_{14}}\right|^{2} + (-1)^{a(b+c)}{f^{ad}}_{r}f^{bcr}\frac{1}{z_{14}^{2}z_{23}^{2}}
\ln\left|\frac{z_{24}z_{13}}{z_{12}z_{34}}\right|^{2} \nonumber \\[2mm]
 & 
-(-1)^{ab}{f^{ac}}_{r}f^{brd}\frac{1}{z_{13}^{2}z_{24}^{2}}\ln\left|\frac{z_{23}z_{14}}{z_{12}z_{34}}\right|^{2} 
\biggr] +  {\cal O}(k^{0})
\end{align}
  The above result is crossing symmetric and  exhibits logarithms as
  expected for this type of theories.

  Note also that the form of the terms that we obtained does not
  depend on the particular supergroup we have. This will not be so for
  further terms in the expansion \eqref{kexp} in which we would need
  to use specific group theoretic identities reducing combinations of
  four structure constants to those built from two structure constants
  and the metric.

  It is possible to extend the above computation  to  obtaining higher
  order corrections $A^{(p)}_{4}$ as well as to obtaining
  approximations to four-point functions of the form $\langle
  J^{a}J^{b}J^{c}\bar J^{d}\rangle_{\lambda}$ and  $\langle
  J^{a}J^{b}\bar J^{c}\bar J^{d}\rangle_{\lambda}$. However, with the
  above computation being already quite laborious, it is clear that
  the method we use quickly stops being efficient. One has to search
  for more sophisticated methods, perhaps using integrability
  techniques.

%************************************************************************
%************************************************************************
%************************************************************************
\section{\label{sc:CR}Equal time commutators}
  
  The OPE for the currents which we analysed in the previous sections
  appears to have quite a complex structure. One may hope to reveal a
  simpler structure looking at the equal time commutators of the
  currents. This indeed turns out to be the case as we show below.

  The equal time commutator algebra of the currents can be obtained
  from the most singular terms in the current's OPEs via the
  Bjorken-Johnson-Low limit
\begin{equation}
\bigl[J_{\mu}^{a}(\sigma_{1} ,0), J^{b}_{\nu}(\sigma_2,0) \bigr]
  = \lim_{\epsilon \to 0}\Bigl(J_{\mu}^{a}(\sigma_{1},i\epsilon)
J^{b}_{\nu}(\sigma_{2},0) - J^{b}_{\nu}(\sigma_{2},i\epsilon)J_{\mu}^{a}(\sigma_{1},0)\Bigr) 
\end{equation}
  Using the exact OPE coefficients \eqref{exOPE1} and \eqref{exOPE2}
  we obtain
\begin{align}
 \bigl[J^{a}(\sigma_{1},0), J^{b}(\sigma_2,0)\bigr]
  &= 2\pi i\Bigl\{-\frac{k \kappa^{ab}}{1-\lambda^2}\delta'(\sigma_1-\sigma_2) + i{f^{ab}}_{c}\delta(\sigma_1-\sigma_2)
(F_{1}(\lambda)J^{c}(\sigma_2,0)  + F_{2}(\lambda)\bar J^{c}(\sigma_2,0))  \Bigr\} \nonumber \\[2mm]
 \bigl[J^{a}(\sigma_1,0), \bar J^{b}(\sigma_2,0)\bigr]
  &= 2\pi {f^{ab}}_{c}\delta(\sigma_1-\sigma_2)F_{2}(\lambda)\bigl[J^{c}(\sigma_2,0) - \bar J^{c}(\sigma_2,0)\bigr]
\end{align}
where 
\begin{equation}
F_{1}(\lambda) = \frac{1-\lambda^{3}}{(1-\lambda^{2})^{2}}  \, , \qquad F_{2}(\lambda)=\frac{\lambda}{(1+\lambda)(1-\lambda^{2})} \, . 
\end{equation}
Compactifying the spatial direction on a circle: $\sigma \sim \sigma+ 2\pi$ we introduce the Fourier modes
 \begin{align}
 J^{a}(\sigma, \tau)
 &= i \sum_{n\in {\mathbb Z}}e^{-in\sigma}J_{n}(\tau)\, , &
 \bar J^{a}(\sigma, \tau)
 &= -i \sum_{n\in {\mathbb Z}}e^{-in\sigma}\bar J_{n}(\tau) \, . 
\end{align}
For these modes we obtain the equal time commutation relations (ETC)
\begin{align}
  \label{ETC}
 \bigl[J_{n}^{a}(\tau), J_{m}^{b}(\tau)\bigr]
  &= \frac{k}{1-\lambda^2} \kappa^{ab} n\delta_{n,-m} + i{f^{ab}}_{c}\bigl[F_{1}(\lambda)J^{c}_{n+m}(\tau) - F_{2}(\lambda)\bar J_{n+m}^{c}(\tau)\bigr]
\, , \nonumber \\[2mm]
  \bigl[\bar J_{n}^{a}(\tau), \bar J_{m}^{b}(\tau)\bigr]
  &= -\frac{k}{1-\lambda^2} \kappa^{ab} n\delta_{n,-m} + i{f^{ab}}_{c}\bigl[F_{1}(\lambda)\bar J^{c}_{n+m}(\tau) - F_{2}(\lambda) J_{n+m}^{c}(\tau)\bigr]
\, , \nonumber \\[2mm]
  \bigl[ J_{n}^{a}(\tau), \bar J_{m}^{b}(\tau)\bigr]
  &= i{f^{ab}}_{c}F_{2}(\lambda)\bigl[\bar J^{c}_{n+m}(\tau) + J_{n+m}^{c}(\tau)\bigr]\, . 
\end{align}
 In terms of the modes 
\begin{align}
  l_{n}^{a}(\tau)
  = J_{n}^{a}(\tau) - \lambda\bar J^{a}_{n}(\tau)
  \ ,\qquad\qquad
  r_{n}^{a}(\tau)
  = \bar J_{n}^{a}(\tau) - \lambda J^{a}_{n}(\tau)
\end{align}
the ETC algebra takes the form 
\begin{align}\label{ETC2}
  \bigl[l_{n}^{a}(\tau), l_{m}^{b}(\tau)\bigr]
  &= k \kappa^{ab} n\delta_{n,-m} + i{f^{ab}}_{c}l^{c}_{n+m}(\tau) \, , \\[2mm]
  \bigl[r_{n}^{a}(\tau), r_{m}^{b}(\tau)\bigr]
  &= -k \kappa^{ab} n\delta_{n,-m} + i{f^{ab}}_{c}r^{c}_{n+m}(\tau)\, , \\[2mm]
  \bigl[r_{n}^{a}(\tau), l_{m}^{b}(\tau)\bigr]
  &= 0\, .
\end{align}
  We see that the phase space of our model is isomorphic to two
  commuting copies of the affine current algebra with opposite central
  extensions. We hope that this simple result will be useful in the
  further analysis of the model. It is interesting to notice that  the
  phase space of a principal chiral model has exactly the same
  description (see \cite{Faddeev:1987ph}). The Hamiltonian governing
  the $\tau$-evolution of the modes is however different. It is
  derived in the next section.

%************************************************************************
%************************************************************************
%************************************************************************
\section{\label{sc:EOM}Equations of motion}

  At the WZW point the equations of motion are the conditions for the
  (anti-)holomorphicity of the current components: $\partial \bar
  J^{a}=\bar \partial J^{a} =0$. In the perturbed theory these
  equations get deformed. Since the perturbing operator is made only
  of the currents and the vacuum sector closes on itself via OPEs we
  expect the additional term in the deformed equations of motion to be
  built from the currents. Based on the current conservation, spin
  conservation and for dimensional reasons the deformed equations
  must be of the form
\begin{align}
  \label{eqmo1}
  \bar \partial J^{a}(z, \bar z)
  =- \partial \bar J^{a}(z, \bar z)
  =iG(\lambda) {t^{a}}_{bc}:\! J^{c}\bar J^{b}\!:(z, \bar z)
\end{align}
  where ${t^{a}}_{bc}$ is some invariant group tensor and $G(\lambda)$ is
  some function of the coupling constant. For the supergroups at hand
  the structure constants give a unique invariant three tensor so that
  we can set ${t^{a}}_{bc}={f^{a}}_{bc} $. The function $G(\lambda)$ in
  general depends on the particular definition of the normal ordering
  in $:\! J^{c}\bar J^{b}\!:$. We defined such an operator following
  Moore's assignment of contact terms in the abelian conformal
  perturbation theory. A different choice of the composite operator
  would in general result in  a different function $G(\lambda)$. In our prescription the function
  $G(\lambda)$ can be computed by matching the leading singular terms in the
  OPE of \eqref{eqmo1} with the currents. For computing the OPE of
  ${f^{a}}_{bc}:\! J^{c}\bar J^{b}\!:(z, \bar z)$ with one of the
  currents the abelian conformal perturbation theory can be used. 
  Using the exact OPE coefficients \eqref{exOPE1} and \eqref{exOPE2} we obtain
  \begin{align}
  \label{eqmo2}
  G(\lambda){t^{a}}_{bc}
  =- \frac{\lambda}{(1+\lambda)k}{f^{a}}_{bc}\, . 
\end{align}  
The equation of motion thus reads 
\begin{align} \label{eqmo_full}
 \bar \partial J^{a}(z, \bar z)
  =- \partial \bar J^{a}(z, \bar z)
  =-i\frac{\lambda}{(1+\lambda)k}\,{f^{a}}_{bc}:\! J^{c}\bar J^{b}\!:(z, \bar z) \, . 
\end{align}

  As a consistency check we can  compare the quantum equation of
  motion \eqref{eqmo_full} with the classical one that
  follows from the Lagrangian \eqref{eq:Action}, \eqref{eq:Def}. Both
  equations should match in form at the leading order in
  perturbation. The classical equation  can be written as 
\begin{align}
  \label{eqm_class}
  \partial \bar J 
  =- \bar \partial J
  =\lambda\bigl(\bar \partial J_{0} - \partial \bar J_{0}\bigr)
       + \frac{\lambda}{k}\bigl[\bar J_{0}, J_{0}\bigr] 
\end{align}
  where $J$ and $\bar J$ are the components of the conserved Noether
  current
\begin{align}
  J&= (1-\lambda)J_{0} + \lambda {\rm Ad}_{g} (J_{0})  \, , \\
 \bar J&=  (1-\lambda)\bar J_{0} + \lambda {\rm Ad}_{g^{-1}}(\bar J_{0})  \, . 
\end{align}
  and 
\begin{align}
  J_{0}
  = -k\partial g g^{-1}
  \ ,\qquad\qquad
  \bar J_{0}
  = kg^{-1}\bar \partial g \, . 
\end{align}
  At the leading order in $\lambda$ the classical equation of motion
  \eqref{eqm_class} takes the form
\begin{align}
  \partial \bar J
  =- \bar \partial J \approx   \frac{\lambda}{k} \bigl[\bar J, J\bigr] 
\end{align}
  that matches with the leading order term in the quantum equation of
  motion \eqref{eqmo_full}.

  Another consistency check concerns the time evolution. It is easily
  verified that the ETC algebra \eqref{ETC} is preserved by the time
  evolution. Moreover, one can check that the Hamiltonian densities
  giving the equation of motion \eqref{eqmo_full} are
\begin{align}
  \label{Virops}
  T(\sigma, \tau)
  &= \left(\frac{1-\lambda^2}{2  k}\right) \kappa_{dc} :\! J^{c}J^{d}\!:(\sigma, \tau)\, , &
  \bar T(\sigma, \tau)
  &= \left(\frac{1-\lambda^2}{2  k}\right) \kappa_{dc} :\! \bar J^{c} \bar J^{d}\!:(\sigma, \tau) 
\end{align}
  so that 
\begin{align}
  \bar \partial J^{a}(z, \bar z)
  &= \frac{i}{2\pi}\Bigl[\,\int\!\! d\sigma\,\bar T(\sigma, \tau), J^{a}(z, \bar z)\Bigr]\, , &
 \partial \bar J^{a}(z, \bar z)
 &= \frac{i}{2\pi}\Bigl[\,\int\!\! d\sigma\,T(\sigma, \tau), \bar
 J^{a}(z, \bar z)\Bigr]\, .
\end{align}
  In verifying these relations we used\footnote{These relations are
    true for every term in perturbation series and thus hold in the
    deformed theory.}
\begin{align}\label{comm1}
  {f^{c}}_{ba}:\!J^{a}J^{b}\!:
  \ = \frac{i}{2}{f^{c}}_{ba}{f^{ab}}_{d}\,\partial J^{d}=0
  \ ,\qquad\qquad 
  :\! J^{a}\bar J^{b}\!:
  \ = (-1)^{ab}:\! \bar J^{b}J^{a}\!: \, . 
\end{align}

  The operators \eqref{Virops} are the Virasoro generators of the
  deformed CFT. Unlike the deformed currents $J^{a}(z, \bar z)$, $\bar
  J^{b}(z,\bar z)$ these generators  remain holomorphic and
  antiholomorphic. On general grounds this follows from the vanishing
  of the beta function. However it is instructive to show this more
  directly using the equation of motion \eqref{eqmo_full}. To that end
  we write $T(z, \bar z)$ as
\begin{align}
  \label{Tlimit}
  T(z, \bar z)
  =C \left(\frac{1-\lambda^2}{2  k}\right) \kappa_{dc} \lim_{:z\to w:} J^{c}(z, \bar z)J^{d}(w,\bar w)
\end{align}
  where $\lim_{:z\to w:}$ stands for taking the limit and subtracting
  the singular terms in the OPE. In \eqref{Tlimit} $C$ is a constant
  of proportionality which may depend on $\lambda$ and, if logarithms are
  present in the OPE of the deformed currents, on the  subtraction
  scale. The proportionality of the two definitions of the composite
  field follows from the fact that in the WZW theory $\kappa_{dc} :\!
  J^{c}J^{d}\!:$ is the unique group-invariant operator of its
  conformal weights. Using \eqref{eqmo_full} and\eqref{Tlimit} we
  obtain
\begin{align}\label{holT}
  \bar \partial T(w,\bar w)
  &= -i\frac{C\lambda(1-\lambda)}{2k^{2}}f_{dca}\lim_{:z\to w:}   
  J^{a}(z,\bar z) :\!\!J^{c}\bar J^{d}\!\!:(w, \bar w) \, . 
\end{align}
Based on the spin and scaling dimension of the operator on the left hand side and the global 
symmetry conservation the limit on the right hand side
must be a linear combination of the operators $f_{cba}:\! J^{a}J^{b}\bar J^{c}\!:$ and 
$\kappa_{ab}:\! \partial J^{b}\bar J^{a}\!:$. However since the right hand side of (\ref{holT}) 
already contains a factor of the structure constants and the metric tensor $\kappa_{ab}$ does not 
receive any corrections the second operator cannot appear. Therefore  
  %Due to the presence of $f_{dca}$ in the above expression we note
  %that the subtracted limit $:\,z\to w\,:$ can be taken in the abelian
  %theory with the structure constants set to zero. With the definition
  %of composite operators that we have, the right hand side of
  \eqref{holT} can then be rewritten as
\begin{align}
  \bar \partial T(w,\bar w)
  =i\tilde C f_{dca}:\!\! J^{a}J^{c}\bar J^{d}\!\!:(w, \bar w)
\end{align}
  that vanishes due to %the symmetry properties of the structure
  %constant tensor and 
   our definition of the operators $:\!J^{a}J^{c}\bar J^{d}\!:$ 
   (see formulae (\ref{comm1})).

  We conclude this section by noting that the form of equation
  \eqref{eqmo_full} is, up to rescaling of the currents, the same as
  that discussed in \cite{Bernard:1990jw}. We thus expect that our
  model is integrable and possesses the same Yangian symmetries as
  defined in  \cite{Bernard:1990jw}. We leave the detailed
  investigation of these symmetries to future work.

%************************************************************************
%************************************************************************
%************************************************************************
\section{\label{sc:Conclusions}Conclusions }

  In this paper we have considered current-current perturbations of
  WZW models on supergroups $G$. These perturbations break the global
  symmetry $G\times G$ down to the diagonal action of $G$ but preserve
  conformal invariance if $G$ has vanishing Killing form. Perturbative
  calculations provided a number of explicit results regarding the
  OPEs and correlation functions of currents as well as the quantum
  equation of motion.

  More specifically we were able to determine the most singular terms
  in the deformed OPE of WZW currents exactly to all orders in the
  coupling. In turn, this allowed us to obtain the exact quantum
  equations of motion \eqref{eqmo_full} and the equal time
  commutators of currents \eqref{ETC}. These exact results provide a
  non-perturbative Hamiltonian reformulation of the model. In view of
  the simple form of \eqref{ETC} and \eqref{eqmo_full} (see also
  \eqref{ETC2}) we expect this reformulation to be useful in
  understanding the structure of the operator product expansion. One
  of the consequences of our results that could already be seen
  at first order in perturbation theory is the occurrence of
  logarithmic contributions in the mixed OPE between the two
  components of the conserved current.

  The full operator product algebra of currents contains an infinite
  tower of operators which are composites built
  of arbitrarily many currents. Our results in section \ref{sc:OPEs} 
  on the first order OPEs  and on the four-point function of currents
  (see section \ref{sc:FourPT}) only give very limited information
  about these terms. Some organising principle is needed to understand
  the full OPE algebra, perhaps related to the conjectured Yangian
  structure. We plan to return to this question in the future.

  In this paper we have focused on the deformed current algebra. Our
  main method -- quasi-abelian conformal perturbation theory -- can
  also be applied to obtain precise analytical information about the
  deformed spectrum of conformal dimensions. In \cite{Mitev:2008yt}
  this was achieved for the boundary spectrum on symmetry preserving
  D-branes. As will be reported in \cite{AKTQ2}, similar considerations
  apply to the deformed bulk spectrum. Our findings will enable
  additional checks on the conjectured equivalence between supersphere
  $\sigma$-models and $OSP(2S+2|2S)$-symmetric Gross-Neveu models
  \cite{Candu:2008yw,Mitev:2008yt}. Our results may also shed light
  onto open questions related to the parabolic paradigm for
  multifractality spectra in quantum Hall systems
  \cite{Evers:2008,Obuse:2008nc}.

  Finally we comment on some potential applications of our
  results in string theory. First of all, referring to the example
  mentioned in the previous paragraph one might hope for further
  examples of dualities between conformal supercoset $\sigma$-models
  and deformed WZW
  models. It would be particularly interesting to investigate
  the deformations of the $PSU(2,2|4)_1$ WZW model and to see whether
  it can be related to $AdS_5\times S^5$ string theory which -- in the
  Green-Schwarz formalism -- is known to be described by a
  $\sigma$-model on the coset superspace $PSU(2,2|4)/SO(1,4)\times SO(5)$
  \cite{Metsaev:1998it}. A more obvious connection of our
    deformations to string theory exists in the case of the
    supergroup $PSU(1,1|2)$ which is known to describe the string
    background $AdS_3\times S^3$ with mixtures of Neveu-Schwarz and
    Ramond-Ramond fluxes \cite{Berkovits:1999im}. Since the metric
    and the fluxes preserve the full isometry of $AdS_3\times S^3$,
    the $G\times G$-preserving deformations of the $PSU(1,1|2)$ WZW
    model have to be used for their description. On the other hand,
    the $G$-preserving deformations discussed here should also
    correspond to some string background, possibly to some squashed
    version of $AdS_3\times S^3$ with fluxes. Extracting the precise form of the
    metric and the fluxes from the deformed WZW Lagrangian is
    left to future work.
  \bigskip

  \noindent{\em Note added:} While this paper was nearing completion a
  new preprint has appeared  \cite{Benichou:2010ts} in which the
  integrability of the $G\times G$-preserving deformations is
  discussed.

%************************************************************************
\subsubsection*{Acknowledgements}

  The authors thank Matthias Gaberdiel,  Volker Schomerus and Raphael Benichou for
  useful discussions. T.Q. acknowledges the warm hospitality at
  Heriot-Watt University and the financial support during a visit in
  the initial phase of this project. The work of A.K. was supported in
  part by grant ST/G000514/1 ``String Theory Scotland'' from the UK
  Science and Technology Facilities Council.

\appendix
%************************************************************************
%************************************************************************
%************************************************************************
\section{\label{ap:Conventions}Lie superalgebra conventions}

  A Lie superalgebra is a graded vector space $\g=\g_{0}\oplus\g_{1}$
  equipped with a bracket $[\cdot,\cdot]:\g\otimes\g\to\g$. The
  bracket is required to be bilinear, grade-preserving and graded
  antisymmetric. Moreover it has to satisfy a graded version of the
  Jacobi identity. For the definition of the physical action
  functional it is essential to have a non-degenerate,
  grade-preserving and graded symmetric bilinear form
  $\langle\cdot,\cdot\rangle:\g\otimes\g\to\Complex$ which plays the
  role of a metric. For a comprehensive introduction into Lie
  superalgebras we refer the reader to
  \cite{Kac:1977em,Frappat:1996pb}.

  For concrete calculations it is convenient to fix a basis $T^a$ of
  homogeneous generators, with the generator $T^a$ having degree
  $d_a$, i.e.\ $T^a\in\g_{d_a}$. In most of the paper we will actually
  use the abbreviation $d_a\equiv a$, hoping that no misunderstandings
  arise. The structure constants ${f^{ab}}_c$
  are defined in terms of the commutation relation
\begin{align}
  [T^a,T^b]=i{f^{ab}}_c\,T^c\, .
\end{align}
  They inherit a number of properties from the requirement that the
  bracket so defined gives rise to a Lie superalgebra. In particular,
  in terms of the structure constants, the graded antisymmetry and the
  Jacobi identity can be written as
\begin{align}
  \label{eq:Jacobif}
  {f^{ba}}_c
  &= -(-1)^{ab}\,{f^{ab}}_c&
  &,&
  {f^{ab}}_d\,{f^{dc}}_e+(-1)^{c(a+b)}\,{f^{ca}}_d\,{f^{db}}_e+(-1)^{a(b+c)}\,{f^{bc}}_d\,{f^{da}}_e
  &= 0\, .
\end{align}
  For a simple Lie superalgebra the desired non-degenerate metric can
  be obtained through the definition\footnote{We restrict our
    attention to simple Lie superalgebras which admit an even
    non-degenerate form. The other Lie superalgebras are not relevant
    in our context.}
\begin{align}
  \kappa^{ab}
  =\langle T^a,T^b\rangle
  =\str_R(T^aT^b)
\end{align}
  by evaluating the supertrace in a suitable representation
  $R$.\footnote{For a simple Lie superalgebra all metrics are
    identical up to a scalar factor. Hence the concrete choice
    of representation $R$ is not essential. It is however important
    to note that for certain Lie superalgebras the supertrace vanishes
    identically, for instance in the adjoint representation. }
  Its inverse is defined by
\begin{align}
  \label{eq:DefinitionInverse}
  \kappa_{ba}\,\kappa^{bc}=\delta_a^c\, .
\end{align}
  Indices are raised and lowered according to the rule
\begin{align}
  A^a&= A_b\,\kappa^{ba}&
  A_a&= \kappa_{ab}\,A^b\, .
\end{align}
  The last convention in particular implies
\begin{align}
  {\kappa^b}_a&= \delta_a^b&
  {\kappa_a}^b&= (-1)^{a}\,\delta_a^b\, .
\end{align}
  One also obtains
\begin{align}
  A^aB_a
  =A_b\,\kappa^{ba}\,B_a
  =(-1)^{b}\,A_b\,\kappa^{ab}\,B_a
  =(-1)^{a}\,A_aB^a\, .
\end{align}

  In the main part of the paper, a crucial role will be played by the
  quadratic Casimir element, defined by
\begin{align}
  \label{eq:Casimir}
  C_2
  =\kappa_{ba}T^aT^b
  =(-1)^{a}\,\kappa_{ab}\,T^aT^b\, .
\end{align}
  If evaluated in the adjoint representation, the quadratic Casimir element
  provides a simple way of determining whether the Killing form
  vanishes or not. Indeed, in the former case one finds
\begin{align}
  \label{eq:CasAdjoint}
  C_{\ad}\,\kappa^{ab}
  =-(-1)^{d}\,{f^{ac}}_d{f^{bd}}_c
  =\str_{\ad}(T^a,T^b)
  =0\, .
\end{align}
  This relation will be at the heart of many of the special features
  that WZW models exhibit for supergroups with vanishing Killing form.

%************************************************************************
%************************************************************************
%************************************************************************
\section{\label{ap:Distributions}Abelian conformal perturbation
  theory}

  In Appendix A of \cite{Moore:1993zc} a prescription was given for
  computing the integrals arising in conformal perturbation theory of
  free bosons when changing the metric and $B$-field. We refer to
  these perturbations as  abelian conformal perturbations. They are
  equivalent to current-current perturbations in toroidal WZW
  theories. The  perturbation of the Euclidean action is
\begin{equation}
\Delta S = \frac{\lambda}{k\pi}\int\!\! d^{2}w\, :\! J^{e}\bar J^{r}\! : \kappa_{re}
\end{equation}
  where the  OPEs of the currents are 
\begin{equation}
  \label{bare_ope}
J^{a}(z)J^{b}(w) \sim \frac{k\kappa^{ab}}{(z-w)^{2}} + \text{ non-sing. } \, , \quad 
\bar J^{a}(\bar z)\bar J^{b}(\bar w) \sim \frac{k\kappa^{ab}}{(\bar z-\bar w)^{2}} + \text{ non-sing. .}
\end{equation}

  One is interested in computing perturbation theory integrals
\begin{equation}
  \label{gen_int}
\int \dots \int d^{2}w_{1}\dots  d^{2} w_{k} \, 
\bigl\langle V_{1}(z_{1}, \bar z_{1})\dots V_{n}(z_{n}, \bar z_{n}) O(w_{1}, \bar w_{1})\dots O(w_{k}, \bar w_{k})\bigr\rangle
\end{equation}
  where $O =\ :\! J^{e}\bar J^{r}\! : \kappa_{re}$ and  $V_{1}, \dots,
  V_{n}$ stand for external insertions. Without loss of generality
  such insertions can be taken to be affine primaries and their
  descendants. The correlator entering \eqref{gen_int} taken at finite
  separations can be computed using Wick's theorem and the
  contractions \eqref{bare_ope}. Each correlator is a sum over all
  contraction schemes. Each contraction scheme can be represented as a
  collection of chains of pairwise contractions. In general a chain of
  contractions that starts and ends on the same external operator
  contributes only to the renormalisation of the corresponding
  operator (change of normal ordering prescription). Also chains that
  start and end on one of the operators $O(w_{i}, \bar w_{i})$
  contribute only to the overall normalisation factor. For all
  computations done in the present paper such contributions are not
  needed and thus such contraction schemes are assumed to be dropped
  everywhere where we use abelian conformal perturbation. The
  remaining contractions each produce functions with non-integrable
  singularities. A consistent prescription is needed for integrals of
  such functions.

  For the case when the external operators $V_{i}$ are currents and
  their composites the prescription of \cite{Moore:1993zc} for the
  integrals defining the deformed correlation functions  can be
  succinctly summarised by specifying  dressed contractions of
  currents
\begin{align}
  \label{dr_conts}
 J^{a}(z_{1})J^{b}(z_{2}) &\sim  \frac{k \kappa^{ab}}{1-\lambda^{2}}\,\frac{1}{(z_1 - z_2)^{2}}\, ,&
\bar J^{a}(\bar z_{1})\bar J^{b}(\bar z_{2}) &\sim  \frac{k \kappa^{ab}}{1-\lambda^{2}}\,
\frac{1}{(\bar z_1 - \bar z_2)^{2}}\, , \nonumber \\
 J^{a}(z_{1})\bar J^{b}(\bar z_{2}) &\sim -\frac{\pi \lambda k \kappa^{ab}}{1-\lambda^{2}}\,\delta(z_{1}-z_{2})\, . 
\end{align}
  The deformed correlators of currents and their composites are then
  obtained using Wick's theorem with the above contractions. Strictly
  speaking this prescription works for the external fields taken at
  finite separation: $|z_{i}-z_{j}|>0$ for $i\ne j$. This avoids the
  appearance of meaningless expressions such as squares of delta
  functions, etc. However for some correlators the distributional
  answers obtained using \eqref{dr_conts} are correct. In particular
  this holds for correlators with a single external field being the
  composite $:\! J^{a}\bar J^{b}\! :$ and the rest of the external
  fields being currents. For example
\begin{align} \label{example}
& \bigl\langle J^{c}(z_{1})J^{d}(z_2):\! J^{e}\bar J^{f}\!:(w, \bar w)\bigr\rangle_{\lambda}= 
\biggl(-\frac{\pi \lambda k \kappa^{cf}}{1-\lambda^{2}}\delta(z_{1}-w)\biggr)
\biggl( \frac{k \kappa^{de}}{1-\lambda^{2}}\frac{1}{(z_2 - w)^{2}}\biggr) \nonumber \\
& + (-1)^{cd}\biggl(\frac{k \kappa^{ce}}{1-\lambda^{2}}\frac{1}{(z_1 - w)^{2}}   \biggr)
\biggl(-\frac{\pi \lambda k \kappa^{df}}{1-\lambda^{2}}\delta(z_{2}-w)
\biggr) \, .
\end{align}
  In our computations such distributional correlators appear inside
  integrals and may give non-vanishing contributions at finite
  separation upon integration. In each case one needs to be careful
  applying \eqref{dr_conts} to obtain well-defined distributions.

  For any prescription of the kind introduced in  \cite{Moore:1993zc}
  to be consistent it must come from some {\it distributional}
  correlation functions defined in the undeformed theory. Also,
  besides taking the integrals one sometimes is interested in doing
  other manipulations with the correlators arising in perturbation
  series (e.g. of the type we do in sections \ref{sc:TwoThree} and
  \ref{sc:FourPT}). To justify such manipulations one needs a rigorous
  definition of the arising distributions. Below we give such a
  definition and discuss some properties of these distributions.

  There are two basic classes of functions arising from abelian chains
  of contractions of currents
\begin{equation}
  \label{1}
  \frac{1}{(z_{1}-w_{1})^{2}} \frac{1}{(\bar w_{1}-\bar w_{2})^{2}} \frac{1}{ (w_{2}- w_{3})^{2}} \cdot \dots \cdot 
  \frac{1}{(w_{2k}- z_{2})^{2}} \, , 
\end{equation}
\begin{equation}
  \label{2}
\frac{1}{(z_{1}-w_{1})^{2}} \frac{1}{(\bar w_{1}-\bar w_{2})^{2}} \frac{1}{ (w_{2}- w_{3})^{2}} \cdot \dots \cdot 
\frac{1}{(\bar w_{2k+1}- \bar z_{2})^{2}} \, . 
\end{equation}
  These functions correspond to a chain of contractions running from
  an insertion in $z_{1}$ to an insertion in $z_{2}$. We want to
  promote the first function to a distribution on ${\mathbb
    R}^{2(2k+2)}$ and the second one to a distribution on ${\mathbb
    R}^{2(2k+3)}$ so that $z_{1}$ and $z_{2}$ are distribution
  variables as well.
  
  Before regularising the above functions we make a couple of  general
  remarks. Note that if $D(z_{2}-z_{3}, z_{3}-z_{4}, \dots ,
  z_{n-1}-z_{n})$ is a translation invariant distribution on
  $({\mathbb R}^{2})^{n-1}$ then there is a natural definition of the
  product
\begin{equation}
d\delta(z_{1}-z_{2}) D 
\end{equation}
  where $d$ is a differential operator with constant
  coefficients. This product is a distribution on  $({\mathbb
    R}^{2})^{n}$ that acts on a test function $\phi(z_1, z_2, \dots,
  z_{n})$ as
\begin{equation}
  \label{gen1}
\big\langle d\delta(z_{1}-z_{2}) D, \phi\bigr\rangle = \bigl\langle d^{*}_{1} D, \phi(z_{2}, z_{2}, z_{3}, \dots , z_{n})\bigr\rangle 
+ \bigl\langle D, d_{1}^{*} \phi(z_{2}, z_{3}, \dots , z_{n})\bigr\rangle 
\end{equation}
  where $d_{1}^{*}$ is the adjoint differential operator acting on the
  first variable.

  Note also that given a distribution in $n$ variables one can define
  its partial integral in any variable as a distribution in $n-1$
  variables by taking a test function which does not depend on the
  given variable on an interval of radius $R$ and then taking $R$ to
  infinity. If the limit exists it is a distribution in $n-1$
  variables.

  We will put square brackets around the regularised functions to
  denote the corresponding distributions. Using \eqref{gen1} we
  define
\begin{align}
  \label{def1}
& \biggl[ \frac{1}{(z_{1}-w)^{2}(\bar w - \bar z_2)^{2}}\biggr] = 
\partial_{w}\bar \partial_{w} \partial_{z_1}\bar \partial_{z_2} \ln|z_1-w|^{2} \ln |z_{2}-w|^{2} \nonumber \\
& + \frac{\pi}{\bar z_{2} - \bar w} \partial_{w}\delta(z_1-w)  + \frac{\pi}{z_{1}-w} \bar \partial_{w}\delta(z_2-w) 
- \pi^{2}\delta(z_{1}-w)\delta(z_{2}-w) 
\end{align}
  Here the first term on the right hand side is defined as a
  distributional derivative of $\ln|z_1-w|^{2} \ln |z_{2}-w|^{2}$
  which is a locally integrable function; the next two terms are well
  defined by virtue of the above general remark because
  $\frac{\pi}{\bar z_{2} - \bar w}$ and $\frac{\pi}{z_{1}-w}$ are
  locally integrable. Analogously we define
\begin{align}
  \label{def11}
& \biggl[ \frac{1}{(z_1-w)^{2}(\bar z_2 - \bar w)}\biggr] =\partial_{w}\partial_{z_1}\bar \partial_{z_{2}} 
 \ln|z_1-w|^{2} \ln |z_{2}-w|^{2} + \frac{\pi}{z_1-w}\delta(z_2-w) \, , \nonumber \\
 & 
\biggl[ \frac{1}{(z_1-w)(\bar z_2 - \bar w)^{2}}\biggr] =\bar \partial_{\bar w}\partial_{z_1}\bar \partial_{z_{2}} 
 \ln|z_1-w|^{2} \ln |z_{2}-w|^{2} + \frac{\pi}{\bar z_2-\bar w}\delta(z_1-w) \, .
\end{align}

  We can easily take the integrals of the distributions defined in
  \eqref{def1}, \eqref{def11} to obtain
\begin{equation}
  \label{int1}
\int\!\! d^{2}w \biggl[ \frac{1}{(z_{1}-w)^{2}(\bar w - \bar z_2)^{2}}\biggr] =  \pi^{2} \delta(z_{1}-z_{2}) \, ,
\end{equation}
\begin{equation} 
\label{int2} \int\!\! d^{2}w  \biggl[ \frac{1}{(z_1-w)^{2}(\bar z_2 - \bar w)}\biggr] = \frac{\pi}{z_1-z_2} \, ,
\end{equation}
\begin{equation}
  \label{int3}
\int\!\! d^{2} w \biggl[ \frac{1}{(z_1-w)(\bar z_2 - \bar w)^{2}}\biggr] = \frac{\pi}{\bar z_2 - \bar z_1} \,. 
\end{equation}
  For reference we also include here another useful integral 
\begin{equation}
  \label{int4}
\int\limits_{|w|\le R}\!\! d^{2}w\ \frac{1}{(z_1-w)(\bar z_2 - \bar w) } = -\pi \ln |z_1 - z_2|^{2} 
+ \pi \ln(R^{2} -z_1\bar z_2)\, . 
\end{equation}

  We can now define the regularised rational functions \eqref{1},
  \eqref{2} recursively. Take for definiteness \eqref{1}. We can
  define the corresponding distribution as
\begin{align}\label{D2n_main}
& D_{2k}\equiv \biggl[  \frac{1}{(z_{1}-w_{1})^{2}} \frac{1}{(\bar w_{1}-\bar w_{2})^{2}} \frac{1}{ (w_{2}- w_{3})^{2}} \cdot \dots \cdot 
\frac{1}{(w_{2k}- z_{2})^{2}} 
\biggr] \nonumber \\
& =\partial_{z_1}\partial_{z_2} (\partial_{1}\bar \partial_{1}) \cdot \dots \cdot 
(\partial_{2k}\bar \partial_{2k}) \ln|z_{1}-w_{1}|^{2} \ln|w_{2}-w_{3}|^{2} \cdot \dots \cdot 
\ln|w_{2k}-z_{2}|^{2} - C_{2k}
\end{align}
  where $C_{2k}$ are terms each of the form
  $d\delta(w_{i_{1}}-w_{i_{2}}) D$ where $D$ is a distributional
  regularisation of the rational functions in a smaller number of
  variables, all containing first or second powers in the denominator,
  and $d$ is some differential operator with constant
  coefficients. The precise form of $C_{2k}$ is worked out by
  differentiating the product of logarithms and using
\begin{equation}
\partial\bar \partial \ln |z|^{2} = -\pi \delta(z) \, . 
\end{equation}
  Thus starting with \eqref{def1}, \eqref{def11}  we can build all the
  required distributions recursively. It is also clear that this
  definition also defines regularisations of a more general class of
  functions of the form
\begin{equation}
  \label{1a}
\frac{1}{(z_{1}-w_{1})^{p_1}} \frac{1}{(\bar w_{1}-\bar w_{2})^{p_2}} \frac{1}{ (w_{2}- w_{3})^{p_3}} \cdot \dots \cdot 
\frac{1}{(w_{2k}- z_{2})^{p_{2k+1}}} \, , 
\end{equation}
\begin{equation}
  \label{2a}
\frac{1}{(z_{1}-w_{1})^{q_1}} \frac{1}{(\bar w_{1}-\bar w_{2})^{q_2}} \frac{1}{ (w_{2}- w_{3})^{q_3}} \cdot \dots \cdot 
\frac{1}{(\bar w_{2k+1}- \bar z_{2})^{q_{2k+2}}} \, . 
\end{equation}
  where $p_{i}=1,2$ and $q_{j}=1,2$. 

  Next we would like to prove that the distributions $D_{2k}$ satisfy
  the property
\begin{equation}
  \label{property}
\int\!\! dw_{n} D_{2k} = \pi^{2} \delta(w_{n-1} - w_{n-2}) D_{2k-1} \, .
\end{equation}
  This property in particular accounts for the values of  integrals
  3a, 4a in \cite{Moore:1993zc}. Naively \eqref{property} is obtained
  by using integral \eqref{int1} for a partial integral of
  $D_{2k}$. It is not clear however that this is consistent with the
  recursive definition of $D_{2k}$. We now explain how one proves this
  rigorously. Denote for brevity
\begin{equation}
\Delta_{i}\equiv \partial_{i}\bar \partial_{i}
\end{equation}
  then the total derivative used to define $D_{2k}$ can be written as
\begin{equation}
  \label{d2n}
\Delta_{n}\Delta_{n-1}\Delta_{n+1} \Bigl[ R_{1}\ln|w_{n-1}-w_{n}|^{2} \ln |w_{n} - w_{n+1}|^{2} R_{2} \Bigr]
\end{equation}
  where 
\begin{equation}
R_{1} = \partial_{1} \Delta_{1} \dots \Delta_{n-2} \ln |z_{1}-w_{1}|^{2}\ln \dots \ln|w_{n-2}-w_{n-1}|^{2}  
\end{equation}
  and 
\begin{equation}
R_{2} = \partial_{2} \Delta_{n+2} \dots \Delta_{2k} \ln
|w_{n+1}-w_{n+2}|^{2} \dots  \ln|w_{2k}-z_{2}|^{2}\ .
\end{equation}
   Using the representation \eqref{d2n} we can recast  \eqref{D2n_main} into the following form
\begin{align}
 & D_{2k} = \Bigl[{\rm Sm}(\bar \partial_{n-1} R_{1})\Bigl( \frac{\pi}{\bar w_{n+1} - \bar w_{n}} 
 \partial_{n}\delta(w_{n-1}-w_n)  + \frac{\pi}{w_{n-1}-w_{n}} \bar \partial_{n}\delta(w_{n+1}-w_{n}) 
\nonumber \\
& - \pi^{2}\delta(w_{n-1}-w_{n})\delta(w_{n}-w_{n+1})\Bigr)    {\rm Sm}( \partial_{n+1} R_{2})\Bigr] + 
 \Delta_{n}(...)\, .
\end{align}
  Here the square brackets make all rational functions appearing in
  this expression into a distribution as recursively defined
  above. ${\rm Sm}(\bar \partial_{n-1} R_{1})$ and ${\rm Sm}(
  \partial_{n+1} R_{2})$ stand for the rational functions obtained by
  taking the derivatives of the products of logarithms at finite
  separation. The total derivatives $\Delta_{n}(...)$ drop out when
  taking the integral over $w_{n}$. This is  ensured by the absence of
  IR divergences.

  We can now take the integral over $w_{n}$ and we are left with the
  desired result \eqref{property} by means of the following lemma

\noindent \underline{{\bf Lemma}}
\begin{equation}
  \label{diff}
\partial_{n} \Bigl[{\rm Sm}(\bar \partial_{n-1} R_{1}) \frac{\pi}{\bar w_{n+1} - \bar w_{n}} 
{\rm Sm}( \partial_{n+1} R_{2}) \Bigr] = -\pi^{2} \delta(w_{n+1}-w_{n})\Bigl[ {\rm Sm}(\bar \partial_{n-1} R_{1})
 {\rm Sm}( \partial_{n+1} R_{2}) \Bigr]
\end{equation}
  This lemma is proven by induction in $k$ - the length of the
  chain. Using the definition we write the left hand side as 
\begin{align}
&\partial_{n} \Bigl( -\partial_{z_{1}}\partial_{z_{2}}\Delta_{1}\dots \Delta_{n-2}
\bar \partial_{n-1}\bar \partial_{n}\partial_{n+1}\Delta_{n+2} \dots \Delta_{2k}\nonumber \\
&\ln|z_1 - w_1|^2 \dots 
 \ln|w_{n-2}-w_{n-1}|^{2}\ln|w_{n+1}-w_{n}|^{2} \dots  \ln|w_{2k}-z_2|^2\nonumber \\
 & + \pi \delta(w_{n}-w_{n+1})(\dots ) - C_{k,n} \Bigr)
\end{align}
  where $C_{k,n}$ are contact terms all containing a factor of
  $1/({\bar w_{n+1} - \bar w_{n}})$. By  induction the analogue of
  formula \eqref{diff} holds for the distributions in $C_{k,n}$
  because those terms contain a shorter `smooth' part. For the first
  two terms the desired identity holds by definition of the
  distributional derivative.
 
  Using \eqref{property} repeatedly we obtain  integrals 3a and 4a in
  \cite{Moore:1993zc} used in summing up the perturbative series to
  obtain the dressed contractions given in \eqref{dr_conts}. Formula
  \eqref{property} easily extends to more general distributions
  regularising \eqref{1a}, \eqref{2a}.

%************************************************************************
%************************************************************************
%************************************************************************
\section{\label{ap:1storder}First order calculations}

  Section \ref{sc:OPEs} was concerned with the determination of the
  OPE between the currents $J^a(z,\bz)$ and $\bJ^a(z,\bz)$ in deformed
  WZW models. For the sake of clarity, some of the more technical
  calculations have been omitted in the main text. For completeness
  they are summarised in this appendix.

\paragraph{Determination of $t$.}

  In order to determine this coefficient we will put
  $X=:\!J^c\bJ^d\!:$. The calculation of the coefficients $A(X)$ and
  $B_i(X)$ is straightforward but lengthy in this case, and we only
  report the main steps. In the calculation of $A(X)$ we will encounter
  the four-point function of $J$. The quadratic singularities drop out
  after integration. The simple poles can be determined by moving
  $J^e$ to the left and performing the contraction with $J^a$, $J^b$
  and $J^c$.  The last contribution  drops out after
  integration while the rest yields
\begin{align}
  A(X)
  &= \frac{k(-1)^{e(a+b+c+d)}
        \kappa_{fe}\kappa^{df}}{(z_1-z_2)(z_2-\xi)(\xi-z_1)}
        \biggl[\frac{{f^{ea}}_gf^{gbc}}{\bz_1-\bxi}
        +\frac{{f^{eb}}_gf^{agc}(-1)^{ae}}{\bz_2-\bxi}\biggr]\, .
\end{align}
  We then split the bracket into its symmetric and its
  antisymmetric part. After applying the Jacobi identity we  find
\begin{align}
  A(X)
  &= \frac{k(-1)^{d(a+b+c)}}{(z_1-z_2)(z_2-\xi)(\xi-z_1)}
        \biggl\{
        \frac{1}{2}\Bigl[{f^{da}}_gf^{gbc}+{f^{db}}_gf^{agc}(-1)^{ad}\Bigr]
        \biggl[\frac{1}{\bz_1-\bxi}+\frac{1}{\bz_2-\bxi}\biggr]\\[2mm]  
  &\qquad+\frac{1}{2}\Bigl[{f^{da}}_gf^{gbc}-{f^{db}}_gf^{agc}(-1)^{ad}\Bigr]
        \biggl[\frac{1}{\bz_1-\bxi}-\frac{1}{\bz_2-\bxi}\biggr]  
        \biggr\}\\[2mm]
  &= \frac{k(-1)^{d(a+b+c)}}{(z_1-z_2)(z_2-\xi)(\xi-z_1)}
        \biggl\{
        -\frac{1}{2}(-1)^{d(a+b)}\,{f^{ab}}_gf^{gdc}
        \biggl[\frac{1}{\bz_1-\bxi}+\frac{1}{\bz_2-\bxi}\biggr]\\[2mm]  
  &\qquad-\frac{1}{2}\Bigl[{f^{da}}_gf^{gbc}-{f^{db}}_gf^{agc}(-1)^{ad}\Bigr]
        \,\frac{\bz_1-\bz_2}{(\bxi-\bz_1)(\bxi-\bz_2)}
        \biggr\}\, .
\end{align}
  Finally, we expand this expression up to terms involving $\xi$ and
  $\bxi$ to the fourth inverse power. After some elementary algebra
  one obtains
\begin{align}
  A(X)
  &= \frac{1}{2}\,\frac{k\,{f^{ab}}_gf^{gcd}}{z_1-z_2}
        \frac{1}{\xi^2\bxi}
        \biggl[2+\frac{2(z_1+z_2)}{\xi}+\frac{\bz_1+\bz_2}{\bxi}+\cdots\biggr]\\[2mm]  
  &\qquad-\frac{k}{2}(-1)^{d(b+c)}\frac{\bz_1-\bz_2}{z_1-z_2}\frac{1}{\xi^2\bxi^2}
        \Bigl[{f^{ad}}_gf^{gbc}+{f^{db}}_gf^{agc}\Bigr]\Bigl[1+\cdots\Bigr]\, .
\end{align}
  Fortunately, the remaining terms are easier to determine. The first
  coefficient $B_1(X)$ vanishes due to the integration. For the second
  coefficient a simple calculation yields
\begin{align}
  B_2(X)
  &= -\frac{i{f^{ab}}_g}{z_1-z_2}(-1)^{de}\kappa_{fe}
        \frac{1}{\pi k}\int\!\! d^2z\,\frac{ikf^{gce}}{(z_2-\xi)(\xi-z)(z-z_2)}\,\frac{k\kappa^{df}}{(\bxi-\bz)^2}\\[2mm]
  &= \frac{i{f^{ab}}_g}{z_1-z_2}
        \frac{ikf^{gcd}}{(z_2-\xi)^2(\bz_2-\bxi)}\, .
\end{align}
  Upon expansion we immediately find
\begin{align}
  B_2
  &= \frac{k{f^{ab}}_gf^{gcd}}{z_1-z_2}\,
        \frac{1}{\xi^2\bxi}
        \biggl[1+\frac{2z_2}{\xi}+\frac{3z_2^2}{\xi^2}+\cdots\biggr]
        \biggl[1+\frac{\bz_2}{\bxi}+\frac{\bz_2^2}{\bxi^2}+\cdots\biggr]\, .
\end{align}
  Finally, we can recycle the knowledge previously obtained about $A$
  in order to determine the last coefficient $B_3(X)=\lim_{:z_1\to
    z_2:}A$. We only need to expand the term $1/(\xi-z_1)$ in a
  geometric series in $z_1-z_2$ in order to find
\begin{align}
  B_3(X)
  &= \frac{k}{(\xi-z_2)^3}\,{f^{ab}}_gf^{gcd}
        \frac{1}{\bxi-\bz_1}
   =\frac{k}{\xi^3\bxi}
        \,{f^{ab}}_gf^{gcd}
        \Bigl[1+\cdots\Bigr]\, .
\end{align}
  When adding up these contributions, the terms at orders
  $1/\xi^2\bxi$ and $1/\xi^3\bxi$ drop out. The remaining term can be
  simplified using the Jacobi identity. In the end we  obtain
\begin{align}
  A(X)-B_1(X)-B_2(X)-B_3(X)
  &= \frac{1}{\xi^2\bxi^2}\,\frac{\bz_1-\bz_2}{z_1-z_2}
        \,k
        \biggl\{
        -(-1)^{d(b+c)}{f^{ad}}_gf^{gbc}\biggr\}
        +\cdots\, .
\end{align}
  This result should be compared to the unperturbed correlation
  functions. With the present choice $X=:\! J^c\bJ^d\!:$ the most important
  contribution arises from
\begin{equation}
  \begin{split}
 &-\frac{(\bz_1-\bz_2)^2}{(z_1-z_2)^2}\,\partial_\lambda t^{ab}_{rs}(0)\,
  \Bigl\langle :\!J^r\bJ^s\!:(z_2,\bz_2)J^c(\xi)\bJ^d(\bxi)\Bigr\rangle\\[2mm]
  &\qquad= -\frac{(\bz_1-\bz_2)^2}{(z_1-z_2)^2}\,(-1)^{cs}\partial_\lambda t^{ab}_{rs}(0)\,
        \frac{k\kappa^{rc}}{(z_2-\xi)^2}\,\frac{k\kappa^{sd}}{(\bz_2-\bxi)^2}\\[2mm]
  &\qquad= -\frac{(\bz_1-\bz_2)^2}{(z_1-z_2)^2}\,(-1)^{cs}\partial_\lambda t^{ab}_{rs}(0)\,
        \frac{k^2\kappa^{rc}\kappa^{sd}}{\xi^2\bxi^2}\Bigl[1+\cdots\Bigr]\, .
  \end{split}
\end{equation}
  Comparing the two expressions and solving for $t$ we find
\begin{align}
  t_{cd}^{ab}
  =\frac{\lambda}{k}(-1)^{bd}{f^a}_{dg}{f^{gb}}_c\, .
\end{align}

\paragraph{Determination of $w$.}

  The determination of $w$ mimics the calculation for $g$ above. We
  choose $X=:\! \bJ\bJ(\bxi)\!:$. The coefficients $B_i(X)$ all vanish.  Even
  though $A(X)$ is non-zero, it is obviously antisymmetric in $(ab)$ and
  in $(cd)$. On the other hand such contributions can never arise from
  $w$, which is symmetric in the lower indices. Instead they are
  accounted for by the coefficients $u$ and $v$ that have already
  been determined above. As a consequence we find
\begin{align}
  0
  &= \frac{(\bz_1-\bz_2)^2}{(z_1-z_2)^2}\,\partial_\lambda w_{ef}^{ab}(0)\,
  \Bigl\langle:\! \bJ^e\bJ^f\!:(\bz_2):\! \bJ^c\bJ^d\!:(\bxi)\Bigr\rangle\\[2mm]
  &= \frac{(\bz_1-\bz_2)^2}{(z_1-z_2)^2}\,\partial_\lambda
  w_{ef}^{ab}(0)
        \Biggl\{\frac{k^2\kappa^{ec}\kappa^{fd}(-1)^{ef}}{(\bz_2-\bxi)^4}
        +\frac{k^2\kappa^{ed}\kappa^{fc}}{(\bz_2-\bxi)^4}\\[2mm]
  &\qquad-\frac{2k{f^{ec}}_gf^{fgd}(-1)^{ef}}{(\bz_2-\bxi)^4}
        -\frac{k{f^{ed}}_gf^{fcg}(-1)^{e(f+c)}}{(\bz_2-\bxi)^4}\Biggr\}\\[2mm]
  &= \frac{(\bz_1-\bz_2)^2}{(z_1-z_2)^2}\,\frac{1}{\bxi^4}\,
        \partial_\lambda w_{ef}^{ab}(0)
        \Bigl\{k^2\kappa^{ec}\kappa^{fd}(-1)^{ef}
        +k^2\kappa^{ed}\kappa^{fc}\\[2mm]
  &\qquad-2k{f^{ec}}_gf^{fgd}(-1)^{ef}
        -k{f^{ed}}_gf^{fcg}(-1)^{e(f+c)}\Bigr\}+\cdots\, .
\end{align}
  Just as for $\partial_\lambda g(0)$, we postulate that the solution
  for this coefficient is given by $\partial_\lambda w(0)=0$. In other
  words, also the coefficient $w$ does not receive any correction at
  first order in perturbation theory. This concludes our calculations
  with regard to the OPE of the current $J$ with itself.

\paragraph{Determination of $B_c^{ab}$ and $\tilde{B}_c^{ab}$.}

  For  this case we pick
  $X=\bJ^c$. Following the standard prescription we first evaluate
\begin{align}
  A(X)
  &= -\frac{ikf^{abc}}{(\bz_2-\bxi)^2}\biggl[\frac{1}{z_1-z_2}-\frac{1}{z_1-\xi}\biggr]
   =-ikf^{abc}\biggl[\frac{1}{(z_1-z_2)\bxi^2}+\frac{2\bz_2}{(z_1-z_2)\bxi^3}+\frac{1}{\xi\bxi^2}+\cdots\biggr]
  \, .
\end{align}
  In the next step we take the non-singular limit
\begin{align}
  B(X)
  &= \lim_{:z_1\to z_2:}A
   =\frac{ikf^{abc}}{(\bz_2-\bxi)^2}\,\frac{1}{z_2-\xi}
   =-\frac{ikf^{abc}}{\xi\bxi^2}\Bigl[1+\cdots\Bigr]\, .
\end{align}
  The total contribution is hence given by
\begin{align}
  A(X)-B(X)
  &= -\frac{ikf^{abc}}{(z_1-z_2)\bxi^2}\biggl[1+\frac{2\bz_2}{\bxi}+\cdots\biggr]
  \, .
\end{align}
  The result above has to be compared to
\begin{align}
  -\frac{1}{z_1-z_2}\,\partial_\lambda B^{ab}_d(0)
  \,\Bigl\langle\bJ^d(\bz_2)\bJ^c(\xi)\Bigr\rangle
  &= -\frac{1}{z_1-z_2}\,\frac{k\kappa^{dc}}{(\bz_2-\bxi)^2}\,\partial_\lambda B^{ab}_d(0)\\[2mm]
  &= -\frac{k\kappa^{dc}\partial_\lambda
     B^{ab}_d(0)}{z_1-z_2}\,\frac{1}{\bxi^2}
        \biggl[1+\frac{2\bz_2}{\bxi}+\cdots\biggr]\, .
\end{align}
  A comparison of the leading terms yields
\begin{align}
  B^{ab}_c(\lambda)
  =i\lambda{f^{ab}}_c\, .
\end{align}
  It is obvious that this term already accounts even for subleading
  contributions up to the order considered. In other words, we have
\begin{align}
  A(X)-B(X)+\frac{\partial_\lambda B^{ab}_d(0)}{z_1-z_2}\,\Bigl\langle\bJ^d(\bz_2)\bJ^c(\xi)\Bigr\rangle
  &= 0+\cdots\, .
\end{align}
  On the other hand the same result should be obtained when adding up
  the leading contributions of the following two expressions,
\begin{align}
  -\frac{\bz_1-\bz_2}{z_1-z_2}\,\partial_\lambda\tilde{B}^{ab}_d(0)
  \,\Bigl\langle\bartial\bJ^d(\bz_2)\bJ^c(\xi)\Bigr\rangle
  &= \frac{\bz_1-\bz_2}{z_1-z_2}\,\frac{2k\kappa^{dc}}{(\bz_2-\bxi)^3}\,\partial_\lambda\tilde{B}^{ab}_d(0)\\[2mm]
  &= -\frac{\bz_1-\bz_2}{z_1-z_2}\,\frac{1}{\bxi^3}
        \,2k\kappa^{dc}\partial_\lambda\tilde{B}^{ab}_d(0)\,
        \biggl[1+\frac{3\bz_2}{\bxi}+\cdots\biggr]\\[2mm]
  -\frac{\bz_1-\bz_2}{z_1-z_2}\,\partial_\lambda\hat{B}^{ab}_{de}(0)
  \,\Bigl\langle\bJ^d\bJ^e(\bz_2)\bJ^c(\xi)\Bigr\rangle
  &= \frac{\bz_1-\bz_2}{z_1-z_2}\,\frac{ikf^{dec}}{(\bz_2-\bxi)^3}\,\partial_\lambda\hat{B}^{ab}_{de}(0)
   =0\, .
\end{align}
  The comparison yields
\begin{align}
  \tilde{B}^{ab}_d(\lambda)
  &= 0+\cO(\lambda^2)\, .
\end{align}
  Hence $\tilde{B}^{ab}_d(\lambda)$ remains zero (at least to this
  order), even after the deformation is switched on.

\paragraph{Determination of $\hat{C}_c^{ab}$.}

  This case may be treated using $X=:\! J^cJ^d\! :$. Our first focus rests
  on
\begin{align}
  A(X)
  &= \frac{1}{\pi}(-1)^{b(c+d)}\int\!d^2z\,
        \Bigl\langle J^a(z_1):\! J^cJ^d\! :(\xi)J^b(z)\Bigr\rangle
        \,\frac{1}{(\bz_2-\bz)^2}\, .
\end{align}
  Due to the integration we only need the simple pole contributions
  from the four-point correlator. The latter is a bit difficult to
  deal with since two of the currents have a coinciding argument. We
  can calculate it by expressing the correlator as the non-singular
  limit $:\!w\to\xi\!:$ of the correlator $\bigl\langle
  J^a(z_1)J^c(w)J^d(\xi)J^b(z)\bigr\rangle$. A straightforward but
  lengthy calculation yields
\begin{align}
  \Bigl\langle J^a(z_1):\!J^cJ^d\!:(\xi)J^b(z)\Bigr\rangle
  &= (-1)^{b(a+c+d)}\lim_{:w\to\xi:}
        \frac{k}{(z_1-w)(w-\xi)(\xi-z_1)}\\[2mm]
  &\qquad\Biggl\{
        \frac{{f^{ba}}_gf^{gcd}}{z-z_1}
        +\frac{(-1)^{ab}{f^{bc}}_gf^{agd}}{z-w}
        +\frac{(-1)^{b(a+c)}{f^{bd}}_gf^{acg}}{z-\xi}
        \Biggr\}+\cdots\\[2mm]
  &= \frac{k(-1)^{b(a+c+d)}}{(\xi-z_1)^3}\\[2mm]
  &\qquad\Biggl\{
        \frac{{f^{ba}}_gf^{gcd}}{z-z_1}
        +\frac{(-1)^{ab}{f^{bc}}_gf^{agd}}{z-\xi}
        +\frac{(-1)^{b(a+c)}{f^{bd}}_gf^{acg}}{z-\xi}
        \Biggr\}+\cdots\\[2mm]
  &\qquad-\frac{k(-1)^{b(a+c+d)}}{(\xi-z_1)^2}
        \frac{(-1)^{ab}{f^{bc}}_gf^{agd}}{(z-\xi)^2}\, .
\end{align}
  Upon integration the last term drops out, leaving us with
\begin{align}
  A(X)
  &= -\frac{k(-1)^{ab}}{(\xi-z_1)^3}
        \Biggl\{
        \frac{{f^{ba}}_gf^{gcd}}{\bz_2-\bz_1}
        +\frac{(-1)^{ab}{f^{bc}}_gf^{agd}}{\bz_2-\bxi}
        +\frac{(-1)^{b(a+c)}{f^{bd}}_gf^{acg}}{\bz_2-\bxi}
        \Biggr\}+\cdots\\[2mm]
  &= \frac{k(-1)^{ab}{f^{ba}}_gf^{gcd}}{(\bz_1-\bz_2)(\xi-z_1)^3}
        +\frac{k}{(\xi-z_1)^3(\bxi-\bz_2)}
        \Bigl\{{f^{bc}}_gf^{agd}
        +(-1)^{bc}{f^{bd}}_gf^{acg}\Bigr\}+\cdots\, .
\end{align}
  Expanding $A(X)$ in inverse powers of $\xi$ yields
\begin{align}
  A(X)
  &= -\frac{k{f^{ab}}_gf^{gcd}}{\bz_1-\bz_2}
        \frac{1}{\xi^3}
        \biggl[1+\frac{3z_1}{\xi}+\frac{6z_1^2}{\xi^2}+\cdots\biggr]\\[2mm]
  &\qquad+\frac{k}{\xi^3\bxi}
        \biggl[1+\frac{3z_1}{\xi}+\frac{\bz_2}{\bxi}+\cdots\biggr]
        \Bigl\{{f^{bc}}_gf^{agd}
        +(-1)^{bc}{f^{bd}}_gf^{acg}\Bigr\}+\cdots\, .
\end{align}
  We then evaluate
\begin{align}
  B(X)
  &= \lim_{:z_1\to z_2:}A
   =\frac{k}{\xi^3\bxi}
        \biggl[1+\frac{3z_2}{\xi}+\frac{\bz_2}{\bxi}+\cdots\biggr]
        \Bigl\{{f^{bc}}_gf^{agd}
        +(-1)^{bc}{f^{bd}}_gf^{acg}\Bigr\}+\cdots\, .
\end{align}
  Eventually we arrive at
\begin{align}
  A(X)-B(X)
  &= -\frac{k{f^{ab}}_gf^{gcd}}{\bz_1-\bz_2}
        \frac{1}{\xi^3}
        \biggl[1+\frac{3z_1}{\xi}+\frac{6z_1^2}{\xi^2}+\cdots\biggr]\\[2mm]
  &\qquad+\frac{k}{\xi^4\bxi}
        \Bigl[3(z_1-z_2)+\cdots\Bigr]
        \Bigl\{{f^{bc}}_gf^{agd}
        +(-1)^{bc}{f^{bd}}_gf^{acg}\Bigr\}+\cdots\, .
\end{align}

  The leading term of this contribution can be attributed to the
  coefficient $C_c^{ab}$. Indeed, using the previously obtained result
  for $C_c^{ab}$ we find
\begin{align}
  -\frac{1}{\bz_1-\bz_2}\,\partial_\lambda C_e^{ab}(0)
  \,\Bigl\langle J^e(z_2):\!J^cJ^d\!:(\xi)\Bigr\rangle
  &= -\frac{\partial_\lambda C_e^{ab}(0)}{\bz_1-\bz_2}
       \frac{ikf^{ecd}}{(z_2-\xi)^3}\\[2mm]
  &= \frac{1}{\bz_1-\bz_2}\,ikf^{ecd}\,
        \partial_\lambda C_e^{ab}(0)
        \,\frac{1}{\xi^3}\biggl[1+\frac{3z_2}{\xi}+\frac{6z_2^2}{\xi^2}+\cdots\biggr]\\[2mm]
  &= \frac{k\,{f^{ab}}_ef^{ecd}}{\bz_1-\bz_2}\,
        \,\frac{1}{\xi^3}\biggl[1+\frac{3z_2}{\xi}+\frac{6z_2^2}{\xi^2}+\cdots\biggr]\, .
\end{align}
  Omitting the terms which are non-singular in $z_1-z_2$ one is left
  with
\begin{align}
  A(X)-B(X)+\frac{1}{\bz_1-\bz_2}\,\partial_\lambda C_e^{ab}(0)
  \,\Bigl\langle J^e(z_2):\!J^cJ^d\!:(\xi)\Bigr\rangle
  &=  -\frac{z_1-z_2}{\bz_1-\bz_2}
        \frac{3k{f^{ab}}_gf^{gcd}}{\xi^4}
        \biggl[1+\frac{2(z_1+z_2)}{\xi}+\cdots\biggr]\, .
\end{align}
  The most singular contribution here can now be cancelled by
\begin{align}
 - \frac{z_1-z_2}{\bz_1-\bz_2}\,\partial_\lambda\tilde{C}_e^{ab}(0)
  \,\Bigl\langle\partial J^e(z_2):\!J^cJ^d\!:(\xi)\Bigr\rangle
  &= \frac{z_1-z_2}{\bz_1-\bz_2}
        \,3\partial_\lambda\tilde{C}_e^{ab}(0)\,
        \frac{ikf^{ecd}}{(z_2-\xi)^4}\\[2mm]
  &= \frac{z_1-z_2}{\bz_1-\bz_2}\,3k\,{f^{ab}}_ef^{ecd}
        \,\frac{1}{\xi^4}\biggl[1+\frac{4z_2}{\xi}+\cdots\biggr]\, ,
\end{align}
  where the coefficient $\tilde{C}_c^{ab}$ again has been determined
  previously. Since the contribution
\begin{align}
  -\frac{z_1-z_2}{\bz_1-\bz_2}\,\partial_\lambda\hat{C}_{ef}^{ab}(0)
  \,\Bigl\langle:\!J^eJ^f\!:(z_2):\!J^cJ^d\!:(\xi)\Bigr\rangle
  =-\frac{z_1-z_2}{\bz_1-\bz_2}\,\frac{\partial_\lambda\hat{C}^{ab}_{ef}(0)}{(z_2-\xi)^4}
        \Bigl\{\text{some tensor structure}\Bigr\}
\end{align}
  is expected to arise at the same order, we conclude that
\begin{align}
  \hat{C}_{cd}^{ab}(\lambda)
  =0+\cO(\lambda^2)\, .
\end{align}
  The reasoning is identical to the reasoning leading to the vanishing
  of $\partial_\lambda g_{cd}^{ab}(0)$ and $\partial_\lambda
  w_{cd}^{ab}(0)$.

\paragraph{Determination of $\hat{B}_c^{ab}$.}

  This coefficient may be determined by putting $X=:\! \bJ^c\bJ^d\! :$. We
  first evaluate
\begin{align}
  A(X)
  &= \frac{1}{\pi}(-1)^{a(b+c+d)}\int d^2z\frac{1}{(z_1-z)^2}\Bigl\langle
        \bJ^b(\bz_2):\!\bJ^c\bJ^d\!:(\bxi)\bJ^a(\bz)\Bigr\rangle_0\, .
\end{align}
  As above only the simple poles in the four-point function will
  contribute after integration. Using the standard procedure we can
  hence determine
\begin{align}
  \Bigl\langle \bJ^b(\bz_2):\!\bJ^c\bJ^d\!:(\bxi)\bJ^a(\bz)\Bigr\rangle
  &= \lim_{:\bw\to\bxi:}\Bigl\langle \bJ^b(\bz_2)\bJ^c(\bw)\bJ^d(\bxi)\bJ^a(\bz)\Bigr\rangle\\[2mm]
  &= \frac{k(-1)^{a(b+c+d)}}{(\bxi-\bz_2)^3}\\[2mm]
  &\qquad\Biggl\{
        \frac{{f^{ab}}_gf^{gcd}}{\bz-\bz_2}
        +\frac{(-1)^{ba}{f^{ac}}_gf^{bgd}}{\bz-\bxi}
        +\frac{(-1)^{a(b+c)}{f^{ad}}_gf^{bcg}}{\bz-\bxi}
        \Biggr\}\\[2mm]
  &\qquad-\frac{k(-1)^{a(b+c+d)}}{(\bxi-\bz_2)^2}
        \Biggl\{\frac{(-1)^{ba}{f^{ac}}_gf^{bgd}}{(\bz-\bxi)^2}
        \Biggr\}+\cdots\, .
\end{align}
  Upon integration the last term drops out, resulting in
\begin{align}
  A(X)
  &= -\frac{k}{(\bxi-\bz_2)^3}
        \frac{{f^{ab}}_gf^{gcd}}{z_1-z_2}
        +\frac{k}{(\bxi-\bz_2)^3(\xi-z_1)}
        \Bigl\{
        (-1)^{ba}{f^{ac}}_gf^{bgd}
        +(-1)^{a(b+c)}{f^{ad}}_gf^{bcg}
        \Bigr\}\\[2mm]
  &= -\frac{k{f^{ab}}_gf^{gcd}}{z_1-z_2}
        \biggl[\frac{1}{\bxi^3}+\frac{3\bz_2}{\bxi^4}+\frac{6\bz_2^2}{\bxi^5}+\cdots\biggr]\\[2mm]
  &\qquad+\pi k^2
        \biggl[\frac{1}{\xi\bxi^3}+\frac{3\bz_2}{\xi\bxi^4}+\frac{z_1}{\xi^2\bxi^3}+\cdots\biggr]
        \Bigl\{
        (-1)^{ba}{f^{ac}}_gf^{bgd}
        +(-1)^{a(b+c)}{f^{ad}}_gf^{bcg}
        \Bigr\}\, .
\end{align}
  We then evaluate
\begin{align}
  B(X)
  &= \lim_{:z_1\to z_2:}A
   = \frac{k}{(\bxi-\bz_2)^3(\xi-z_2)}
        \Bigl\{(-1)^{ba}{f^{ac}}_gf^{bgd}
        +(-1)^{a(b+c)}{f^{ad}}_gf^{bcg}
        \Bigr\}\\[2mm]
  &= \frac{k}{\xi\bxi^3}
        \biggl[1+\frac{z_2}{\xi}+\frac{3\bz_2}{\bxi}+\cdots\biggr]
        \Bigl\{(-1)^{ba}{f^{ac}}_gf^{bgd}+(-1)^{a(b+c)}{f^{ad}}_gf^{bcg}\Bigr\}\, .
\end{align}
  Hence the complete contribution is
\begin{align}
  A(X)-B(X)
  &= -\frac{k{f^{ab}}_gf^{gcd}}{z_1-z_2}
        \frac{1}{\bxi^3}
        \biggl[1+\frac{3\bz_2}{\bxi}+\frac{6\bz_2^2}{\bxi^2}+\cdots\biggr]\\[2mm]
  &\qquad+\frac{k}{\xi^2\bxi^3}
        \biggl[z_1-z_2+\cdots\biggr]
        \Bigl\{
        (-1)^{ba}{f^{ac}}_gf^{bgd}
        +(-1)^{a(b+c)}{f^{ad}}_gf^{bcg}
        \Bigr\}\, .
\end{align}
  This expression has to be compared to
\begin{align}
  -\frac{1}{z_1-z_2}\,\partial_\lambda B_e^{ab}(0)\,
  \Bigl\langle\bJ^e(\bz_2):\!\bJ^c\bJ^d\!:(\bxi)\Bigr\rangle
  &= -\frac{1}{z_1-z_2}\,\partial_\lambda B_e^{ab}(0)\,\frac{ikf^{ecd}}{(\bz_2-\bxi)^3}\\[2mm]
  &= \frac{1}{z_1-z_2}\,k
  \,{f^{ab}}_ef^{ecd}\,\frac{1}{\bxi^3}\biggl[1+\frac{3\bz_2}{\bxi}+\frac{6\bz_2^2}{\bxi^2}+\cdots\biggr]\, .
\end{align}
  Consequently we find
\begin{align}
  A(X)-B(X)+\frac{\partial_\lambda B_e^{ab}(0)}{z_1-z_2}\,
  \Bigl\langle\bJ^e(\bz_2):\!\bJ^c\bJ^d\!:(\bxi)\Bigr\rangle
  &= \frac{k(-1)^{ab}}{\xi^2\bxi^3}
        \Bigl[z_1-z_2+\cdots\Bigr]
        \Bigl\{{f^{ac}}_gf^{bgd}+(-1)^{ac}{f^{ad}}_gf^{bcg}\Bigr\}\, .
\end{align}
  We recognise that the quartic contribution expected to arise in
\begin{align}
  -\frac{1}{z_1-z_2}\,\partial_\lambda\tilde{B}_e^{ab}(0)\,
  \Bigl\langle\bartial\bJ^e(\bz_2):\!\bJ^c\bJ^d\!:(\bxi)\Bigr\rangle
  &= \frac{1}{z_1-z_2}\,\partial_\lambda\tilde{B}_e^{ab}(0)
        \,\frac{3ikf^{ecd}}{(\bz_2-\bxi)^4}
   =0\\[2mm]
 - \frac{\bz_1-\bz_2}{z_1-z_2}\,\partial_\lambda\hat{B}^{ab}_{ef}(0)
  \,\Bigl\langle\bJ^e\bJ^f(\bz_2):\!\bJ^c\bJ^d\!:(\bxi)\Bigr\rangle
  &= -\frac{\bz_1-\bz_2}{z_1-z_2}\,\frac{\partial_\lambda\hat{B}^{ab}_{ef}(0)}{(\bz_2-\bxi)^4}
        \Bigl\{\text{some tensor structure}\Bigr\}
\end{align}
  is absent. Consequently we find that
\begin{align}
  \hat{B}(\lambda)
  =0+\cO(\lambda^2)\, .
\end{align}
  This result concludes our calculation of the deformed mixed OPE
  between the currents $J$ and $\bJ$ at the leading order in 
  $\lambda$.

%************************************************************************
%************************************************************************
%************************************************************************
\section{\label{ap:4pt}Computation of four-point functions}

  Here we give the missing details of the computation leading to
  \eqref{4pt_final}. We start with formula \eqref{corr2}. Each term
  $a^{(i)}$ contributes a term denoted $A^{(i)}_{4}$ upon substitution
  into \eqref{step2} so that we have
\begin{equation}
(1-\lambda^{2})A_{4}=A_{4}^{(0)} + A_{4}^{(1)} + A_{4}^{(2)} + A_{4}^{(3)} + A_{4}^{(4)} + A_{4}^{(5)} + A_{4}^{(6)} 
+ A_{4}^{(7)} \, . 
\end{equation}

  To compute $A^{(1)}_{4}$  we need  the integral
\begin{equation}
I_{1}=\int \frac{d^{2}w}{z_1-w}\frac{1}{(z_2-w)^2}\frac{\bar z_{34}}{(z_{34})^{2}(\bar z_3-\bar w)(\bar z_4-\bar w)}
= \frac{\pi}{z_{34}^{2}z_{12}^{2}}\ln\left| \frac{z_{13}z_{24}}{z_{23}z_{14}}\right|^{2} - \frac{\pi}{z_{34}z_{12}z_{23}z_{24}}
\, . 
\end{equation}
  We obtain 
\begin{align}
&A_{4}^{(1)}= \frac{(1-\lambda)}{1-\lambda^{2}} \left(\frac{-\lambda}{k\pi}\right)(-1)^{a(b+c+d)}
\int\!\! \frac{d^{2}w}{z_1-w} i{f^{a}}_{re}(-1)^{b(c+d)}\frac{k\kappa^{be}}{(z_2-w)^{2}}\,\bigl\langle J^{c}(z_3)J^{d}(z_4) \bar J^{r}(\bar w)\bigr\rangle_{\mu} \nonumber \\
& = \frac{(1-\lambda)}{1-\lambda^2}
 \left(\frac{\lambda}{k\pi}\right)(-1)^{a(b+c+d)}k^{2}\kappa^{be}\Bigl(-if^{cdr}\Bigr)\frac{\lambda(1-\lambda)}{(1-\lambda^{2})^{3}}
 \Bigl(i{{f^{a}}_{r}}^{b}(-1)^{b}\Bigr)I_{1} 
 \nonumber \\
 & = -k\frac{\lambda^{2}(1-\lambda)^{2}}{(1-\lambda^2)^{4}}{f^{ab}}_{r}f^{rcd}\biggl[\, 
 \frac{1}{z_{34}^{2}z_{12}^{2}}\ln\left| \frac{z_{13}z_{24}}{z_{23}z_{14}}\right|^{2} - \frac{1}{z_{34}z_{12}z_{23}z_{24}}\,
 \biggr]\, , 
\end{align}
Furthermore a straightforward computation yields 
\begin{align}
&A^{(2)}_{4}=  \frac{\lambda^{2}(1-\lambda)}{1-\lambda^2}
 (-1)^{a(b+c+d)}i{f^{a}}_{re}\int\!\! \frac{d^{2}w}{z_1-w}
 (-1)^{b(c+d+e+r)}\kappa^{rb}\delta(z_2-w)\nonumber \\
& \qquad\qquad\times\bigl\langle J^{c}(z_3)J^{d}(z_4)J^{e}(w)\bigr\rangle_{\lambda}= k\frac{\lambda^{2}(1-\lambda^{3})(1-\lambda)}{(1-\lambda^2)^{4}}\frac{{f^{ab}}_{e}f^{ecd}}{z_{12}z_{23}z_{34}z_{42}} \, . 
\end{align}

  Up to now all formulas were exact. Now we will start using the
  abelian approximation for the remaining correlators entering
  $A_{4}^{(3)}, A_{4}^{(4)}, A_{4}^{(5)}$, $A^{(6)}$, $A^{(7)}$. In
  our approximation we have
\begin{equation}
\bigl\langle J^{c}(z_3)J^{d}(z_4)\, :\!\!J^{s}(w)\bar J^{r}(\bar w)\!\!:\bigr\rangle_{\lambda} = 
-\frac{\pi k^{2}\lambda}{(1-\lambda^2)^2}\biggl[ \kappa^{cr}\kappa^{ds}\frac{\delta(z_3-w)}{(z_4-w)^{2}} 
+ (-1)^{cd}\kappa^{cs}\kappa^{dr}\frac{\delta(z_4-w)}{(z_3-w)^{2}}\biggr]
\end{equation}
  Using this formula we obtain 
\begin{align}
A_{4}^{(3)}&=-k\frac{(1-\lambda)\lambda^2}{(1-\lambda^2)^3}\biggl[ \frac{(-1)^{ab}{f^{ac}}_{e}f^{bed}}{z_{13}z_{23}z_{43}^{2}} 
- \frac{(-1)^{a(b+c)}{f^{ad}}_{e}f^{bce}}{z_{14}z_{24}z_{34}^{2}}\biggr]\, , \\[2mm]
A_{4}^{(4)}&=k\frac{(1-\lambda)\lambda^3}{(1-\lambda^2)^3}\biggl[ \frac{(-1)^{ab}{f^{ac}}_{e}f^{bed}}{z_{14}z_{24}z_{43}^{2}} 
- \frac{(-1)^{a(b+c)}{f^{ad}}_{e}f^{bce}}{z_{13}z_{23}z_{34}^{2}}\biggr]\, .
\end{align}
  To compute the remaining term $A_{4}^{(5)}$ we need to compute the
  correlator
\begin{equation}
\bigl\langle J^{c}(z_3)J^{d}(z_4) :\!J^{e}\bar J^{r}\!:(w,\bar w)\, :\!J^{s}\bar J^{q}\!:(w_2,\bar w_2)\bigr\rangle_{\lambda}
\end{equation}
  in the abelian approximation. There are eight non-vanishing contraction
  schemes contributing to $A_{4}^{(5)}$:\footnote{Some contraction
    schemes drop out upon contraction with the ${f^{a}}_{bc}$ tensors
    present in $A_{4}^{(5)}$.}
\begin{align}
  A&= \bigl\langle \,(J^{c} [J^{d}\, :\!\! J^{e}] \{\bar J^{r}\!\!: \, :\!\! J^{s})\bar J^{q}\}\!\!:\, \bigr\rangle 
\qquad\qquad B=\bigl\langle\, (J^{c} [J^{d}\, :\!\! J^{e}) \{\bar J^{r}\!\!: \, :\!\! J^{s}]\bar J^{q}\}\!\!:\, \bigr\rangle \\
  C&= \bigl\langle\, (J^{c} [J^{d}\, :\!\! J^{e}] \{\bar J^{r}\!\!: \, :\!\! J^{s}\}\bar J^{q})\!\!:\, \bigr\rangle 
\qquad\qquad D=\bigl\langle\,  (J^{c} [J^{d}\, :\!\! J^{e}) \{\bar J^{r}\!\!: \, :\!\! J^{s}\}\bar J^{q}]\!\!:\, \bigr\rangle \\
  E&= \bigl\langle\, (J^{c} [J^{d}\, :\!\! \{J^{e} \bar J^{r}]\!\!: \, :\!\! J^{s})\bar J^{q}\}\!\!:\, \bigr\rangle 
\qquad\qquad F=\bigl\langle\, (J^{c} [J^{d}\, :\!\! \{J^{e} \bar J^{r})\!\!: \, :\!\! J^{s}]\bar J^{q}\}\!\!:\, \bigr\rangle \\
  G&= \bigl\langle\, (J^{c} [J^{d}\, :\!\! \{J^{e} \bar J^{r}]\!\!: \, :\!\! J^{s}\}\bar J^{q})\!\!:\, \bigr\rangle 
\qquad\qquad H=\bigl\langle\, (J^{c} [J^{d}\, :\!\! \{J^{e} \bar
J^{r})\!\!: \, :\!\! J^{s}\}\bar J^{q}]\!\!:\,\bigr\rangle \, ,
\end{align}
where each of the 3 contractions we mark by a pair of brackets: $\{J \, J\}$, $(J\, J)$, $[J, J]$.
The corresponding contributions to $A_{4}^{(5)}$ are 
\begin{align}
  A_{4}^{(5)}
  =\tilde A + \tilde B + \tilde C + \tilde D + \tilde E + \tilde F +
  \tilde G + \tilde H\, .
\end{align}
  Evaluating the integrals in $\tilde C, \tilde D, \tilde E, \tilde F,
  \tilde G, \tilde H$ is straightforward because each integrands contains two delta functions. Evaluating $\tilde A$ involves 
\begin{align}
  I_{2}
  &= \iint\!\! \frac{d^{2}wd^{2}w_{2}}{(z_1-w)(z_2-w_2)(z_4-w)^{2}(z_3-w_2)^{2}(\bar w -\bar w_{2})^{2}}\nonumber \\ 
  &= \frac{\pi^{2}}{z_{14}z_{24}z_{34}^{2}} - \frac{\pi^{2}}{z_{14}^{2}z_{23}^{2}}
\ln\left|\frac{z_{24}z_{13}}{z_{12}z_{34}}\right|^{2} + \frac{\pi^{2}}{z_{14}z_{32}z_{31}z_{34}}\, .
\end{align}
  The integral emerging in the $\tilde B$ contribution is obtained by
  interchanging $z_{4}$ and $z_{3}$. We obtain 
\begin{align}
  \tilde A + \tilde B 
  &= k\frac{\lambda^2(1-\lambda)^2}{(1-\lambda^2)^4}\biggl[ -(-1)^{a(b+c)}{f^{ad}}_{r}f^{bcr}
\biggl( \frac{1}{z_{14}z_{24}z_{34}^{2}} - \frac{1}{z_{14}^{2}z_{23}^{2}}
\ln\left|\frac{z_{24}z_{13}}{z_{12}z_{34}}\right|^{2}
 + \frac{1}{z_{14}z_{32}z_{31}z_{34}}\biggr)\nonumber\\ 
 &\qquad+ (-1)^{ab}{f^{ac}}_{r}f^{brd}\biggl(\frac{1}{z_{13}z_{23}z_{34}^{2}} - \frac{1}{z_{13}^{2}z_{24}^{2}}\ln\left|\frac{z_{23}z_{14}}{z_{12}z_{34}}\right|^{2} 
+\frac{1}{z_{31}z_{42}z_{41}z_{34}}\biggr)\biggr]
\end{align}
  We further obtain 
\begin{align}
  \tilde E + \tilde G
  &= k\frac{\lambda^{4}(1-\lambda)^{2}}{(1-\lambda^{2})^{4}}\frac{(-1)^{a(b+c)}{f^{ad}}_{s}f^{bcs}}{z_{14}z_{23}z_{34}z_{42}}\, ,  \\[2mm]
 \tilde H + \tilde F
  &=  k\frac{\lambda^{4}(1-\lambda)^{2}}{(1-\lambda^{2})^{4}}\frac{(-1)^{ab}{f^{a
 c}}_{s}f^{bsd}}{z_{13}z_{23}z_{34}z_{42}}\, , \\[2mm]
  \tilde C + \tilde D
  &= k\frac{\lambda^{4}(1-\lambda)^{2}}{(1-\lambda^{2})^{4}}\biggl[ 
        \frac{(-1)^{a(b+c)}{f^{ad}}_{s}f^{bcs}}{z_{13}z_{23}z_{43}^{2}} - 
\frac{(-1)^{ab}{f^{ac}}_{s}f^{bsd}}{z_{14}z_{24}z_{43}^{2}}\biggr]\, . 
\end{align}
  Collecting all terms proportional to $1/(z_{14}z_{24}z_{34}^2)$ and
  $1/z_{13}z_{23}z_{34}^2$ from $A_{4}^{(3)}$, $A_{4}^{(4)}$,
  $A_{4}^{(5)}$ we obtain
\begin{align}
 \label{secondp}
& k\frac{\lambda^{3}(1-\lambda)^{2}}{(1-\lambda^2)^4}\Bigl[ (-1)^{ab}{f^{ac}}_{s}f^{bsd} + (-1)^{a(b+c)}{f^{ad}}_{s}f^{bcs}\Bigr]
\biggl( \frac{1}{z_{14}z_{24}z_{43}^{2}} - \frac{1}{z_{13}z_{23}z_{43}^{2}}   \biggr)\nonumber \\
& = -k\frac{\lambda^{3}(1-\lambda)^{2}}{(1-\lambda^2)^4}\Bigl[ (-1)^{ab}{f^{ac}}_{s}f^{bsd} + (-1)^{a(b+c)}{f^{ad}}_{s}f^{bcs}\Bigr]
\biggl( \frac{1}{z_{13}z_{24}z_{23}z_{34}} + \frac{1}{z_{24}z_{34}z_{14}z_{13}}   \biggr) \, .
\end{align}
Here we used the identities
\begin{align}
  \frac{\lambda^{2}(1-\lambda)^{2}}{(1-\lambda^2)^{4}} - \frac{(1-\lambda)\lambda^{2}}{(1-\lambda^{2})^{3}}
  &= -\frac{\lambda^{3}(1-\lambda)^{2}}{(1-\lambda^{2})^{4}} \, , \\[2mm]
  \frac{\lambda^{4}(1-\lambda)^{2}}{(1-\lambda^2)^{4}} -
  \frac{(1-\lambda)\lambda^{3}}{(1-\lambda^{2})^{3}}
  &= -\frac{\lambda^{3}(1-\lambda)^{2}}{(1-\lambda^{2})^{4}} \, .
\end{align}

  We also compute
\begin{align}
A^{(6)}_{4}&= k\frac{\lambda^{2}(1-\lambda)}{(1-\lambda^{2})^{3}}(-1)^{a(b+c)}\frac{{f^{ad}}_{s}f^{bcs}}{z_{23}z_{34}z_{13}z_{14}}\, , \\
A^{(7)}_{4}&= k\frac{\lambda^{2}(1-\lambda)}{(1-\lambda^{2})^{3}}(-1)^{ab}\frac{{f^{ac}}_{s}f^{bsd}}{z_{24}z_{34}z_{13}z_{14}}\, .
\end{align}

  Collecting all terms we obtain 
\begin{equation}
(1-\lambda^{2})A_{4} = A_{4}^{(0)} + L_4 + R_4
\end{equation}
  where the logarithmic part $L_4$ is given by
\begin{align}
  L_{4} 
  &= k\frac{\lambda^{2}(1-\lambda)^{2}}{(1-\lambda^2)^{4}}\biggl[ -{f^{ab}}_{r}f^{rcd}
 \frac{1}{z_{34}^{2}z_{12}^{2}}\ln\left| \frac{z_{13}z_{24}}{z_{23}z_{14}}\right|^{2} \nonumber \\
  &\qquad + (-1)^{a(b+c)}{f^{ad}}_{r}f^{bcr}\frac{1}{z_{14}^{2}z_{23}^{2}}
\ln\left|\frac{z_{24}z_{13}}{z_{12}z_{34}}\right|^{2} 
- (-1)^{ab}{f^{ac}}_{r}f^{brd}\frac{1}{z_{13}^{2}z_{24}^{2}}\ln\left|\frac{z_{23}z_{14}}{z_{12}z_{34}}\right|^{2}\,\biggr]
\end{align}
  and the additional rational part $R_4$ is
\begin{align}
  R_{4}
  &= k\frac{\lambda^{3}(1-\lambda)}{(1-\lambda^2)^3}\biggl[ \frac{{f^{ab}}_{r}f^{rcd}}{z_{12}z_{23}z_{34}z_{42}}
 + \frac{(-1)^{ab}{f^{ac}}_{s}f^{bsd}}{z_{13}z_{23}z_{34}z_{42}} 
 + \frac{(-1)^{a(b+c)}{f^{ad}}_{s}f^{bcs}}{z_{14}z_{23}z_{34}z_{42}} \biggr]
\end{align}
  where the first term in the square brackets contains contributions
  from $A_{4}^{(1)}$ and $A_{4}^{(2)}$. These terms combine together
  with $A^{(0)}$ so that we finally get \eqref{4pt_final}.

%************************************************************************
%************************************************************************
%************************************************************************
\section{\label{ap:GG}The OPE closure in the $G\times G$-preserving deformation}

  In this appendix we give further considerations regarding the issue
  of the current algebra closure for the $G\times G$-preserving
  deformation. This deformation is generated by the operator $:\!
  J^{a}\phi_{ab} \bar J^{b}\!:$ that involves the adjoint
  representation primary field $\phi_{ab}$. Such a field satisfies the
  following OPEs with the WZW currents
\begin{align}\label{Jphi}
  J^{a}(z)\phi_{bc}(w, \bar w)
  & \sim \frac{i{f^{a}}_{bd}{\phi^{d}}_{c}}{z-w} \, , &
  \bar J^{a}(\bar z)\phi_{bc}(w, \bar w)
  & \sim (-1)^{ab}\frac{i{f^{a}}_{cd}{\phi_{b}}^{d}}{\bar z-\bar w} \, .
\end{align}
  In the deformed theory the global symmetry group is $G\times G$. The
  components of the Noether current associated with the left action of
  $G$ on itself can be classically written as $K_{z}= -k\partial
  gg^{-1}$ and $K_{\bar z}=k\bar \partial g g^{-1}$. At the quantum
  level these components reduce to the operators
\begin{align}
  \label{KK}
  K_{z}^{a}(z)
  = J^{a}(z)
  \ ,\qquad\qquad
  K_{\bar z}^{a}
  = \kappa^{ab}:\! {\phi^{a}}_{c}\bar J^{c}\!:(z,\bar z)
\end{align}
  at the WZW point. The Knizhnik-Zamolodchikov equations for the field
  $\phi_{ab}$ read
\begin{align} \label{KZ1}
  \partial\phi_{ab}(z, \bar z)
  &=\frac{i}{k}{f_{aep}}:\! J^{p}{\phi^{e}}_{b}\!:(z, \bar z)\, ,\\
 \label{KZ2}
  \bar{\partial}\phi_{ab}(z, \bar z)
  &=\frac{i}{k}{f_{bep}}(-1)^{ap}:\! \bar J^{p}{\phi_{a}}^{e}\!:(z, \bar z)\, . 
\end{align}
  Equation \eqref{KZ1} implies the Maurer-Cartan equation 
\begin{equation}
  \partial K_{\bar z}^{a}
  = \frac{i}{k}{f^{a}}_{bc}:\! K_{z}^{c}K_{\bar z}^{b}\!: \, . 
\end{equation}
  The assumption of the OPE closure in the current algebra leads to
  the OPE (formula (2.21) of \cite{Ashok:2009xx})
\begin{equation}
  \label{KK2}
  K_{\bar z}^{a}(z, \bar z)K^{b}_{\bar z}(0)
  \ \sim\ \frac{\kappa^{ab}k}{\bar z^{2}} 
-2i{f^{ab}}_{c}\frac{K_{\bar z}^{c}(0)}{\bar z}  + 
i\frac{(z-w)}{(\bar z - \bar w)^{2}}\,{f^{ab}}_{c}J^{c}(w) + \text{less
  singular terms}\, .
\end{equation}
  We can check this OPE using formula \eqref{KK}. The OPE of the
  field $\phi_{ab}$ with itself is not known in detail, however we do
  know the group-theoretic content and can estimate the singularities
  present in the OPE in the large $k$ limit. One finds that possible
  power singularities go as $|z-w|^{-1/k}$ and  thus are very mild
  for large $k$'s. The leading and subleading singularities in the OPE
  of $K_{\bar z}^{a}$ with $K_{\bar z}^{b}$ then come from the
  singularities in the OPEs of $\bar J^{c}$ with themselves and with
  the fields $\phi_{ab}$. We obtain 
\begin{align}
  \label{OPEcomp}
  :\!{\phi^{a}}_{e}\bar J^{e}\!:(z, \bar z) :\!{\phi^{b}}_{p}\bar J^{p}\!:(w, \bar w) 
 &\sim \frac{k}{(\bar z- \bar w)^{2}}\kappa^{pe}(-1)^{eb}:\!{\phi^{a}}_{e}{\phi^{b}}_{p}\!:(w, \bar w) 
 \nonumber \\
 &\qquad-2i\frac{1}{\bar z - \bar w}{f^{p}}_{er}(-1)^{p(p+b)}:\!(:\!\phi^{ar}\bar J^{e}\!:){\phi^{b}}_{p}\!:(w, \bar w) 
 \nonumber \\
 &\qquad + i\frac{(z-w)}{(\bar z - \bar w)^{2}}{f^{a}}_{rs}(-1)^{pe}\kappa^{pe}
 :\!(:\!J^{s}{\phi^{r}}_{e}\!:){\phi^{b}}_{p}\!:(w, \bar w)
 + \cdots\, ,
 \end{align}
  where we used \eqref{eq:CasAdjoint} to get rid of the terms
  containing two factors of the structure constants. Matching the
  singularities in \eqref{OPEcomp} with those in \eqref{KK2} we obtain
  the equations\footnote{An equation  rather similar to our equation (\ref{phi_id1}) also appeared  in 
  \cite{Benichou:2010rk}. See formulae (3.8), (3.9) in that paper. }
\begin{subequations}
\begin{align}
  \label{phi_id1}
   \kappa^{pe}(-1)^{eb}:\!{\phi^{a}}_{e}{\phi^{b}}_{p}\!:(w, \bar w)& = \kappa^{ab}\,{\bf 1} \, \\
  \label{phi_id2}
   {f^{p}}_{er}(-1)^{p(p+e)}:\!(:\!\phi^{ar}\bar J^{e}\!:){\phi^{b}}_{p}\!:(w, \bar w)& = {f^{ab}}_{c}:\! {\phi^{c}}_{e}\bar J^{e}\!:(w, \bar w)\, , \\
  \label{phi_id3}
   {f^{a}}_{rs}(-1)^{pe}\kappa^{pe}
 :\!(:\!J^{s}{\phi^{r}}_{e}\!:){\phi^{b}}_{p}\!:(w, \bar w)& = {f^{ab}}_{c}J^{c}(w) \, .
\end{align}
\end{subequations}
  In equation  \eqref{phi_id2} we can rearrange the normal ordering on
  the left hand side as
\begin{equation}
  \label{rearr}
   {f^{p}}_{er}(-1)^{p(p+b)}:\!(:\!\phi^{ar}\bar J^{e}\!:){\phi^{b}}_{p}\!:(w, \bar w) = {f^{p}}_{er}(-1)^{r(p+b)} :\!(:\!   \!\phi^{ar}   {\phi^{b}}_{p} \!:) \bar J^{e}\! :(w,\bar w ) \,.   
\end{equation}
   This can be done because by virtue of \eqref{eq:CasAdjoint} the
   operator $\bar J^{e}$  has no singularities with the other two
   operators in that expression. This suggests the following relation
\begin{equation}
  \label{phi_id4}
  -{f^{sr}}_{p}(-1)^{rb}:\! {\phi^{a}}_{r}{\phi^{b}}_{s}\!:(w,\bar w) =\ :\! {f^{ab}}_{c}{\phi^{c}}_{p}\!:(w,\bar w) \, .
\end{equation}
  Relations \eqref{phi_id1} and \eqref{phi_id4} have classical
  analogues. The classical analogue of operator $\phi_{ab}$ is the
  matrix $\Ad_g$. This matrix satisfies the equations
  $\Ad_g\Ad_{g^{-1}}=1$ and $\Ad_g[X,Y]=[\Ad_g(X),\Ad_g(Y)]$ which are
  analogous to quantum equations \eqref{phi_id1} and
  \eqref{phi_id4}. On the quantum level, however, their validity is
  far from obvious. It is easy to observe that such equations can only
  hold when the field $\phi_{ab}$ has dimension zero, and thus do not
  hold for the WZW theories built on ordinary semisimple Lie groups or
  supergroups with non-vanishing Killing form. The meaning of relation
  \eqref{phi_id3} is less clear to us.\footnote{ If one could
    rearrange the terms as in \eqref{rearr} equation \eqref{phi_id3}
    would then follow from \eqref{phi_id1}. However  direct
    investigation of singularities shows that such rearrangement is
    not possible in this case.}

  We conclude that  identities \eqref{phi_id1}, \eqref{phi_id2},
  \eqref{phi_id3}  are  non-trivial necessary conditions for the
  closure of the OPEs of the currents $K_{z}$, $K_{\bar z}$. It would
  be interesting to verify these identities directly as this would
  test the bootstrap approach suggested in
  \cite{Ashok:2009xx,Benichou:2010rk}.

%\bibliographystyle{JHEP-TQ}
%\bibliography{bibliography}

\begin{thebibliography}{10}

\bibitem{Efetov1983:MR708812}
K.~B. Efetov, {\it Supersymmetry and theory of disordered metals},  {\rm Adv.
  in Phys.} {\bf 32} (1983) 53--127.

\bibitem{Berkovits:2002zk}
N.~Berkovits, {\it {ICTP} lectures on covariant quantization of the
  superstring},  \href{http://arXiv.org/abs/hep-th/0209059}{{\tt
  hep-th/0209059}}.
%%CITATION = HEP-TH/0209059;%%

\bibitem{Metsaev:1998it}
R.~R. Metsaev and A.~A. Tseytlin, {\it Type {IIB} superstring action in
  {$AdS_5\times S^5$} background},  {\rm Nucl. Phys.} {\bf B533} (1998)
  109--126 [\href{http://arXiv.org/abs/hep-th/9805028}{{\tt hep-th/9805028}}].
%%CITATION = HEP-TH 9805028;%%

\bibitem{Berkovits:1999im}
N.~Berkovits, C.~Vafa and E.~Witten, {\it Conformal field theory of {AdS}
  background with {Ramond-Ramond} flux},  {\rm JHEP} {\bf 03} (1999) 018
  [\href{http://arXiv.org/abs/hep-th/9902098}{{\tt hep-th/9902098}}].
%%CITATION = HEP-TH 9902098;%%

\bibitem{Berkovits:1999zq}
N.~Berkovits, M.~Bershadsky, T.~Hauer, S.~Zhukov and B.~Zwiebach, {\it
  Superstring theory on {$AdS_2\times S^2$} as a coset supermanifold},  {\rm
  Nucl. Phys.} {\bf B567} (2000) 61--86
  [\href{http://arXiv.org/abs/hep-th/9907200}{{\tt hep-th/9907200}}].
%%CITATION = HEP-TH 9907200;%%

\bibitem{Arutyunov:2008if}
G.~Arutyunov and S.~Frolov, {\it Superstrings on {$AdS_4 \times CP^3$} as a
  coset sigma-model},  {\rm JHEP} {\bf 09} (2008) 129
  [\href{http://arXiv.org/abs/0806.4940}{{\tt 0806.4940}}].
%%CITATION = 0806.4940;%%

\bibitem{Stefanski:2008ik}
     B. Stefanski, jr, 
     {\it Green-Schwarz action for Type IIA strings on $AdS_4\times
                  CP^3$}, {\rm Nucl. Phys.}{\bf B808} (2009) 80--87
    [\href{http://arXiv.org/abs/0806.4948}{{\tt 0806.4948}}].
     



\bibitem{Babichenko:2009dk}
A.~Babichenko, B.~Stefanski, Jr. and K.~Zarembo, {\it Integrability and the
  {$AdS_3/CFT_2$} correspondence},  {\rm JHEP} {\bf 03} (2010) 058
  [\href{http://arXiv.org/abs/0912.1723}{{\tt 0912.1723}}].
%%CITATION = 0912.1723;%%

\bibitem{Zarembo:2010sg}
K.~Zarembo, {\it Strings on semisymmetric superspaces},  {\rm JHEP} {\bf 05}
  (2010) 002 [\href{http://arXiv.org/abs/1003.0465}{{\tt 1003.0465}}].
%%CITATION = 1003.0465;%%

\bibitem{Gotz:2006qp}
G.~{G\"otz}, T.~Quella and V.~Schomerus, {\it The {WZNW} model on
  {$PSU(1,1|2)$}},  {\rm JHEP} {\bf 03} (2007) 003
  [\href{http://arXiv.org/abs/hep-th/0610070}{{\tt hep-th/0610070}}].
%%CITATION = HEP-TH 0610070;%%

\bibitem{Bershadsky:1999hk}
M.~Bershadsky, S.~Zhukov and A.~Vaintrob, {\it {$PSL(n|n)$} sigma model as a
  conformal field theory},  {\rm Nucl. Phys.} {\bf B559} (1999) 205--234
  [\href{http://arXiv.org/abs/hep-th/9902180}{{\tt hep-th/9902180}}].
%%CITATION = HEP-TH 9902180;%%

\bibitem{Kagan:2005wt}
D.~Kagan and C.~A.~S. Young, {\it Conformal sigma-models on supercoset
  targets},  {\rm Nucl. Phys.} {\bf B745} (2006) 109--122
  [\href{http://arXiv.org/abs/hep-th/0512250}{{\tt hep-th/0512250}}].
%%CITATION = HEP-TH 0512250;%%

\bibitem{Babichenko:2006uc}
A.~Babichenko, {\it Conformal invariance and quantum integrability of sigma
  models on symmetric superspaces},  {\rm Phys. Lett.} {\bf B648} (2007)
  254--261 [\href{http://arXiv.org/abs/hep-th/0611214}{{\tt hep-th/0611214}}].
%%CITATION = HEP-TH/0611214;%%

\bibitem{Rozansky:1992rx}
L.~Rozansky and H.~Saleur, {\it Quantum field theory for the multivariable
  {Alexander-Conway} polynomial},  {\rm Nucl. Phys.} {\bf B376} (1992)
  461--509.
%%CITATION = NUPHA,B376,461;%%

\bibitem{Maassarani:1996jn}
Z.~Maassarani and D.~Serban, {\it Non-unitary conformal field theory and
  logarithmic operators for disordered systems},  {\rm Nucl. Phys.} {\bf B489}
  (1997) 603--625 [\href{http://arXiv.org/abs/hep-th/9605062}{{\tt
  hep-th/9605062}}].
%%CITATION = HEP-TH 9605062;%%

\bibitem{Schomerus:2005bf}
V.~Schomerus and H.~Saleur, {\it The {$GL(1|1)$} {WZW} model: {From}
  supergeometry to logarithmic {CFT}},  {\rm Nucl. Phys.} {\bf B734} (2006)
  221--245 [\href{http://arXiv.org/abs/hep-th/0510032}{{\tt hep-th/0510032}}].
%%CITATION = HEP-TH 0510032;%%

\bibitem{Saleur:2006tf}
H.~Saleur and V.~Schomerus, {\it On the {$SU(2|1)$} {WZW} model and its
  statistical mechanics applications},  {\rm Nucl. Phys.} {\bf B775} (2007)
  312--340 [\href{http://arXiv.org/abs/hep-th/0611147}{{\tt hep-th/0611147}}].
%%CITATION = HEP-TH 0611147;%%

\bibitem{Quella:2007hr}
T.~Quella and V.~Schomerus, {\it Free fermion resolution of supergroup {WZNW}
  models},  {\rm JHEP} {\bf 09} (2007) 085
  [\href{http://arXiv.org/abs/0706.0744}{{\tt 0706.0744}}].
%%CITATION = 0706.0744;%%

\bibitem{Luscher:1977uq}
M.~{L\"uscher}, {\it Quantum nonlocal charges and absence of particle
  production in the two-dimensional nonlinear sigma model},  {\rm Nucl. Phys.}
  {\bf B135} (1978) 1--19.
%%CITATION = NUPHA,B135,1;%%

\bibitem{Bernard:1990jw}
D.~Bernard, {\it Hidden {Yangians} in 2-d massive current algebras},  {\rm
  Commun. Math. Phys.} {\bf 137} (1991) 191--208.
%%CITATION = CMPHA,137,191;%%

\bibitem{Schwarz:1995td}
J.~H. Schwarz, {\it Classical symmetries of some two-dimensional models},  {\rm
  Nucl. Phys.} {\bf B447} (1995) 137--182
  [\href{http://arXiv.org/abs/hep-th/9503078}{{\tt hep-th/9503078}}].
%%CITATION = HEP-TH/9503078;%%

\bibitem{Lu:2008kb}
H.~Lu, M.~J. Perry, C.~N. Pope and E.~Sezgin, {\it {Kac-Moody} and {Virasoro}
  symmetries of principal chiral sigma models},  {\rm Nucl. Phys.} {\bf B826}
  (2010) 71--86 [\href{http://arXiv.org/abs/0812.2218}{{\tt 0812.2218}}].
%%CITATION = 0812.2218;%%

\bibitem{Ashok:2009xx}
S.~K. Ashok, R.~Benichou and J.~Troost, {\it Conformal current algebra in two
  dimensions},  {\rm JHEP} {\bf 06} (2009) 017
  [\href{http://arXiv.org/abs/0903.4277}{{\tt 0903.4277}}].
%%CITATION = 0903.4277;%%

\bibitem{Ashok:2009jw}
S.~K. Ashok, R.~Benichou and J.~Troost, {\it Asymptotic symmetries of string
  theory on {$AdS_3\times S^3$} with {Ramond-Ramond} fluxes},  {\rm JHEP} {\bf
  10} (2009) 051 [\href{http://arXiv.org/abs/0907.1242}{{\tt 0907.1242}}].
%%CITATION = 0907.1242;%%

\bibitem{Benichou:2010rk}
R.~Benichou and J.~Troost, {\it The conformal current algebra on supergroups
  with applications to the spectrum and integrability},  {\rm JHEP} {\bf 04}
  (2010) 121 [\href{http://arXiv.org/abs/1002.3712}{{\tt 1002.3712}}].
%%CITATION = 1002.3712;%%

\bibitem{Candu:2008yw}
C.~Candu and H.~Saleur, {\it A lattice approach to the conformal
  {$OSp(2S+2|2S)$} supercoset sigma model. {Part II:} {The} boundary spectrum},
   {\rm Nucl. Phys.} {\bf B808} (2009) 487--524
  [\href{http://arXiv.org/abs/0801.0444}{{\tt 0801.0444}}].
%%CITATION = 0801.0444;%%

\bibitem{Mitev:2008yt}
V.~Mitev, T.~Quella and V.~Schomerus, {\it Principal chiral model on
  superspheres},  {\rm JHEP} {\bf 11} (2008) 086
  [\href{http://arXiv.org/abs/0809.1046}{{\tt 0809.1046}}].
%%CITATION = 0809.1046;%%

\bibitem{Quella:2007sg}
T.~Quella, V.~Schomerus and T.~Creutzig, {\it Boundary spectra in superspace
  sigma models},  {\rm JHEP} {\bf 10} (2008) 024
  [\href{http://arXiv.org/abs/0712.3549}{{\tt 0712.3549}}].
%%CITATION = 0712.3549;%%

\bibitem{Obuse:2008nc}
H.~Obuse, A.~R. Subramaniam, A.~Furusaki, I.~A. Gruzberg and A.~W.~W. Ludwig,
  {\it Boundary multifractality at the integer quantum {Hall} plateau
  transition: Implications for the critical theory},  {\rm Phys. Rev. Lett.}
  {\bf 101} (2008) 116802 [\href{http://arXiv.org/abs/0804.2409}{{\tt
  0804.2409}}].
%%CITATION = 0804.2409;%%

\bibitem{Cardy:1989da}
J.~L. Cardy, {\it Conformal invariance and statistical mechanics},  in {\rm
  Fields, strings and critical phenomena} (E.~Br{\'e}zin and J.~Zinn-Justin,
  eds.), Les Houches Summer School, 1988.

\bibitem{Moore:1993zc}
G.~W. Moore, {\it Finite in all directions},
  \href{http://arXiv.org/abs/hep-th/9305139}{{\tt hep-th/9305139}}.
%%CITATION = HEP-TH/9305139;%%

\bibitem{Guida:1995kc}
R.~Guida and N.~Magnoli, {\it {All order I.R. finite expansion for short
  distance behavior of massless theories perturbed by a relevant operator}},
  {\rm Nucl. Phys.} {\bf B471} (1996) 361--388
  [\href{http://arXiv.org/abs/hep-th/9511209}{{\tt hep-th/9511209}}].
%%CITATION = HEP-TH/9511209;%%

\bibitem{Faddeev:1987ph}
L.~D. Faddeev and L.~A. Takhtajan, {\rm Hamiltonian methods in the theory of
  solitons}.
\newblock Springer, 1987.

\bibitem{AKTQ2}
A.~Konechny and T.~Quella, {\it Bulk anomalous dimensions in superspace
  {$\sigma$}-models},  {\rm work in progress}.

\bibitem{Evers:2008}
F.~Evers, A.~Mildenberger and A.~D. Mirlin, {\it Multifractality at the quantum
  {Hall} transition: Beyond the parabolic paradigm},  {\rm Phys. Rev. Lett.}
  {\bf 101} (2008) 116803 [\href{http://arXiv.org/abs/0804.2334}{{\tt
  0804.2334}}].

\bibitem{Benichou:2010ts}
R.~Benichou, {\it Fusion of line operators in conformal sigma-models on
  supergroups, and the {Hirota} equation},
  \href{http://arXiv.org/abs/1011.3158}{{\tt 1011.3158}}.
%%CITATION = 1011.3158;%%

\bibitem{Kac:1977em}
V.~G. Kac, {\it Lie superalgebras},  {\rm Adv. Math.} {\bf 26} (1977) 8--96.
%%CITATION = ADMTA,26,8;%%

\bibitem{Frappat:1996pb}
L.~Frappat, P.~Sorba and A.~Sciarrino, {\rm Dictionary on {Lie} algebras and
  superalgebras}.
\newblock Academic Press Inc., San Diego, CA, 2000.
\newblock Extended and corrected version of the E-print [hep-th/9607161].
%%CITATION = HEP-TH 9607161;%%

\end{thebibliography}
\def\cprime{$'$} \def\cprime{$'$}
\providecommand{\href}[2]{#2}\begingroup\raggedright\endgroup

\end{document}